\documentclass{iopart}

\usepackage{iopams}
\usepackage{graphicx}

\usepackage{mymacros}

\begin{document}



\title[Continued-fraction methods for the Wigner Caldeira--Leggett
master equation]{
Caldeira--Leggett quantum master equation in Wigner phase space:
continued-fraction solution and application to  Brownian
motion in periodic potentials
}

\author{
J. L. Garc\'{\i}a-Palacios and D. Zueco
}
\address{%
Dep.\ de F\'{\i}sica de la Materia Condensada e
Instituto de Ciencia de Materiales de Arag\'on,
C.S.I.C.--Universidad de Zaragoza,
E-50009 Zaragoza, Spain
}


\date{\today}

\begin{abstract}
The continued-fraction method to solve classical Fokker--Planck
equations has been adapted to tackle quantum master equations of the
Caldeira--Leggett type.
This can be done taking advantage of the phase-space (Wigner)
representation of the quantum density matrix.
The approach differs from those in which some continued-fraction
expression is found for a certain quantity, in that the full solution
of the master equation is obtained by continued-fraction methods.
This allows to study in detail the effects of the environment
(fluctuations and dissipation) on several classes of nonlinear quantum
systems.
We apply the method to the canonical problem of quantum Brownian
motion in periodic potentials both for cosine and ratchet potentials
(lacking inversion symmetry).
\end{abstract}




\section{
Introduction
}
\label{intro}

The phase-space formulation of quantum mechanics has received a
renewed attention in the last decades \cite{hiletal84,lee95,schleich}
because it allows to employ notions and tools of classical physics in
the quantum realm.
The central object in this approach is {\em Wigner's function}
%
\begin{equation}
\label{wigner:def}
\Wf(\x,\p)
=
\frac{1}{2\pi\hbar}
\int
\drm y\,
\e^{-\iu \p\,y/\hbar}
\varrho(\x+\case{1}{2} y,\x-\case{1}{2} y)
\end{equation}
which is merely a phase-space $(\x,\p)$ representation of the density
matrix $\varrho(\x,\x')=\langle\x|\hat{\varrho}|\x'\rangle$.
In a closed system, the dynamical equation for the Wigner function is
equivalent to {\em Schr{\"o}dinger's (or von Neumann) equation\/}
while, in the $\hbar\to0$ limit, it recovers {\em Liouville's
equation\/} for the phase-space distribution in classical mechanics.

Another remarkable property of the Wignerian representation is the
{\em average property}.
The expectation value of a quantum operator $\hat{A}$ is obtained
from the corresponding classical variable $A(\x,\p)$ (related with
$\hat{A}$ by Weyl's rule \cite[Ch.~1]{balescu}) by a prescription
analogous to the classical one, namely
%
\begin{equation}
\label{average}
\big\langle A\big\rangle
\equiv
\mathrm{Tr}(\hat{\varrho}\,\hat{A})
=
\int\!\drm\x\drm\p\,
\Wf(\x,\p)\,
A(\x,\p)
\;.
\end{equation}
Thus, in spite of not being necessarily positive, the Wigner function
can be considered as a sort of quantum-mechanical ``distribution''.
Besides, the marginal distributions of $\Wf$ give the true quantum
probabilities of $\x$, as
$P(\x)
\equiv
\langle\x|\hat{\varrho}|\x\rangle=\int\!\drm\p\,\Wf(\x,\p)$,
or of $\p$, as
$P(\p)
\equiv
\langle\p|\hat{\varrho}|\p\rangle=\int\!\drm\x\,\Wf(\x,\p)$.
Therefore, the Wigner formalism provides a natural quantum-classical
connection.
This has been specially exploited in the search for quantum analogues
of various classical effects (e.g., chaos) or in the problem of the
fuzzy borders between the quantum and classical worlds \cite{zur91}.

In contact with the surrounding medium, the (open) system experiences
dissipation, fluctuations, and decoherence
\cite{grasching88,weiss,brepet2002}.
Typically one is interested in a system consisting of a few relevant
degrees of freedom coupled to its environment, or bath, which has a
very large number of degrees of freedom (e.g., phonons, photons,
nuclear spins, etc.).
The system is not necessarily microscopic, but it can be a mesoscopic
system described by some collective variables which under appropriate
conditions can behave quantum mechanically
\cite{guibascal98,walletal2003}.
Because of the genericity of these situations and the fundamental
issues raised (e.g., approach to thermal equilibrium, dissipation in
tunnelling and coherence, measurement problem), the study of {\em
quantum dissipative systems\/} is of interest in several areas of
physics and chemistry \cite{kohmartan97}.

The dynamics of an open system can in many cases be formulated in
terms of a {\em quantum master equation\/} for the (reduced) density
matrix.
In the Wigner representation this can be compactly written as
$\dptT\Wf=\LFP\Wf$, with $\LFP$ a certain evolution operator.
Taking the classical limit, the master equation for a particle of mass
$\sM$ in a potential $\V(\x)$ goes over the Klein--Kramers equation
\cite{kle21,kra40}
%
\begin{equation}
\label{KK}
\dpt
\Wf
=
\big[
-
(\p/\sM)\,
\px
+
\V'\,
\pp
+
\gamma\,
\pp
\big(
\p
+
\sM\kT
\pp
\big)
\big]
\Wf
\end{equation}
which is simply a {\em Fokker--Planck equation\/} in phase space.
The first two terms (Poisson bracket) generate the classical
reversible evolution (Liouville equation).
The last term (``collision operator'') accounts for irreversible
effects due to the coupling to the environment (dissipation and
fluctuations).
The damping parameter $\gamma$ measures the strengh of the coupling to
the bath, which is at temperature $T$.
It is worth recalling that phase-space dynamics also includes problems
reducible to ``mechanical'' analogues: Josephson junctions, certain
electrical circuits, chemical reactions, etc.

A suitable non-perturbative technique to solve classical
Fokker--Planck equations of systems with few degrees of freedom is the
{\em continued-fraction method\/} \cite{risken}.
This is a special case of the expansion into complete sets (Grad's)
method to solve kinetic equations in statistical mechanics
\cite{balescu2}.
In brief, the distribution is expanded into a series of orthogonal
polynomials and, as a result, the kinetic equation is transformed into
an {\em infinite set of coupled equations\/} for the expansion
coefficients $\ec_{\ir}$.
Approximate solutions are obtained by truncating the hierarchy of
equations at various levels.
To get manageable expressions, however, the truncation needs to be
performed at a low level.
In the continued-fraction method, instead of truncating directly, one
seeks for a basis in which the range of index coupling is short
(ideally, the equation for $\ec_{\ir}$ involves $\ec_{\ir-1}$,
$\ec_{\ir}$, and $\ec_{\ir+1}$).
Then, the (differential) recurrence relations among the $\ec_{\ir}$
can be solved by iterating a simple algorithm, the structure of which
is like that of a continued fraction (\ref{app:RR-CF}).
For the Klein--Kramers equation~(\ref{KK}) the tridiagonal chain of
coupled equations for the $\ec_{\ir}$ is called the {\em Brinkman
hierarchy} \cite{bri56I}.
It has been solved by continued-fraction methods to study classical
particles subjected to dissipation and fluctuations \cite{risken}.
The method has been also extended to rotational Brownian motion
problems involving classical spins and dipoles (vd.\ \cite{kalcof97}
and references therein).

To deal with quantum dissipative systems, however, is a more delicate
task \cite{weiss}.
To begin with, phenomenological quantisation of dissipative systems
poses fundamental problems (e.g., with the uncertainty and
superposition principles).
A rigorous route is to model the bath in a simple way, quantise it
together with the system, and eventually trace over the environmental
variables.
However, the resulting theoretical frameworks are technically
involved.
{\em Quantum master equations\/}, except for simple cases, are
difficult to solve.
An alternative is provided by {\em quantum state diffusion\/} methods
(vd.\ \cite{sto2003}), where stochastic evolution equations for state
vectors in Hilbert space are introduced as computational tools.
Nevertheless, its implementation seems to be restricted to systems
with discrete spectrum (e.g., oscillators, 2-state systems, etc.).
Similarly, {\em  quantum Langevin equations\/}  \cite{forlewoco88} are
of limited use beyond nearly harmonic systems.
For an arbitrary system, there exist exact {\em path-integral\/}
expressions for the evolution of the density matrix (involving the
so-called {\em influence functional\/} that incorporates environmental
effects) \cite{grasching88,weiss}.
However, those expressions are difficult to evaluate, even
numerically, because the propagating function is highly oscillatory,
rendering numerical methods unstable at long times.
Finally, {\em quantum Monte Carlo\/} simulations can in principle be
used.
Nevertheless, in spite of ongoing progress (vd.\ \cite{sto2003}), they
are computationally complex and suffer from the (dynamical) sign
problem.

This situation strongly motivates the development of {\em alternative
methods\/} for quantum dissipative systems.
Inspired on their suitability for classical systems,
continued-fraction techniques have been developed for a number of
problems.
Shibata and co-workers \cite{shi80,shiuch93} applied them to solve a
c-number quantum Fokker--Planck equation for a spin in a dissipative
environment.
Vogel and Risken \cite{vogris88} employed continued-fraction
methods to solve master equations in quantum nonlinear optics (vd.\
also \cite{Qopts-matlab}).
Recently \cite{gar2004} this method has been extended to study genuine
phase-space problems within the Wigner formalism.
A generalisation of the Brinkman hierarchy for quantum master
equations of the Caldeira--Leggett type was presented.
It was shown that the continued-fraction method for the classical
problem can, in principle, be adapted to solve this hierarchy,
yielding a promising technique to study several classes of nonlinear
quantum systems subjected to environmental effects.

In this article we first present the details necessary for the
derivation of the quantum hierarchies and discuss their solution by
continued fractions.
The approach is then implemented for the problem of quantum Brownian
motion in periodic potentials (a demanding problem with (partly)
continuous spectrum).
Both the cosine and simple ratchet potentials (lacking inversion
symmetry) are considered.
For the former a number of previous results are recovered (so testing
the method) and extended, while the physical interpretations are
revisited within the Wigner formalism.
For particles in ratchet potentials we study the effects of finite
damping (inertia) on the rectified velocities.
The phenomenology is interpreted in terms of the interplay of thermal
hopping, overbarrier wave reflection, and tunnelling.
Results for non-equilibrium dynamics under oscillating forcing are
also discussed, which show the combined effect of quantum phenomena,
thermal activation, and dissipation in a nonlinear system.
A number of technical issues are consigned to the appendices.


\section{
Quantum master equations in phase space
}

We start from the following generic form for the quantum master
equation in the Wigner representation
\cite{calleg83pa,tanwol91,hupazzha92,anahal95}:
%
\begin{equation}
\label{WKK:Dxp}
\fl
\dpt
\Wf
=
\Big[
-
\frac{\p}{\sM}
\px
+
\V'
\pp
+
\pp
\big(
\gamma\,\p
+
\Dpp
\pp
\big)
+
\Dxp\pxp
+
\sum_{\iq=1}^{\infty}
\frac{(\iu\hbar/2)^{2\iq}}{(2\iq\!+\!1)!}
\V^{(2\iq+1)}
\pp^{(2\iq+1)}
\Big]
\Wf
\;.
\;
\end{equation}
The first three terms, identifying $\Dpp=\gamma\sM\kT$, give the
classical Klein--Kramers equation~(\ref{KK}).
The mixed diffusion term $\Dxp\pxp\Wf$ is heuristically related with
the colour of the quantum noise \cite{coh97}.
The series of derivatives $\pp^{(2\iq+1)}\Wf$ (Wigner--Moyal term)
gives the quantum contribution to the unitary evolution of the closed
system.
Conditions under which this series can be truncated are sometimes
discussed, to avoid dealing with an infinite-order partial
differential equation.
However, we shall fully keep this term because it can be specially
important in nonlinear systems \cite{zurpaz94}.

Conditions of validity for the master-equation description are
discussed in \cite{kargra97,kohdithan97,ank2003qme}.
Equations of the type of (\ref{WKK:Dxp}) can be derived modelling the
particle surroundings as a {\em bath of oscillators\/} representing
the normal modes of the environment.
However, a number of assumptions are typically involved like
semiclassical or high-temperature bath (rendering the kernels of the
influence functional local in time), weak system-bath coupling
(Born--Markov approximations), etc.
Then, one would heuristically expect equation (\ref{WKK:Dxp}) being
valid for small enough $\hbar\gamma/\kT$.
Notwithstanding this, the {\em structure\/} of the equation seems to
be quite generic.
Thus, a quantum master equation recently derived for strong coupling
\cite{ank2003epl}, and valid for all $\hbar\gamma/\kT$, involves terms
similar to those in (\ref{WKK:Dxp}) [with $\x$-dependent coefficients
and renormalised $\V(\x)$].
Besides, the phase-space representation of the celebrated Lindblad
master equation is obtained by simply adding to (\ref{WKK:Dxp})
terms of the form $\px(\x\,\Wf)$ and $\px^{2}\Wf$ \cite{isasansch96}.
Most of these terms can be readily incorporated in the treatment below
(indeed, the term $\Dxp\pxp\Wf$, absent in the original
Caldeira--Leggett equation, is included to illustrate this).
The same, in principle, would apply to possible extensions of the
quantum master equation~(\ref{WKK:Dxp}).


\section{
Preliminary manipulations of the master equation
}
\label{sec:WKK:scaled}

First it is convenient to introduce appropriate scaled units.
These are based on a characteristic length $\xo$ (e.g., the distance
between the minima in a double-well oscillator, the period in a
periodic potential) and a characteristic energy $\Eo$ (e.g., a barrier
height).
Then one scales frequencies by $\wo=(\Eo/\sM\xo^{2})^{1/2}$, time by
$\wo^{-1}$, forces by $F_{0}=\Eo/\xo$, the action by $\So=\Eo/\wo$,
etc.
On the other hand, $\Dpp$ ($=\gamma\sM\kT$ in a high-$T$ bath) is
handled as an effective temperature $\kT_{\eff}=\Dpp/\gamma\sM$ and we
introduce some convenient thermal rescalings via
$\omega_{\Th}=(\kT_{\eff}/\sM\xo^{2})^{1/2}$.
Thus, in the equations $\p$ is scaled by $\sM\xo\omega_{\Th}$,
the potential and $\Dxp$ appear divided by $\kT_{\eff}$, $\gamma$
enters in the combination $\gammaT=\gamma/\omega_{\Th}$, time is
multiplied by $\omega_{\Th}$, while the thermal de Broglie wave length
$\ldB=\hbar/(4\sM\kT_{\eff})^{1/2}$ enters divided by $\xo$.

Omitting any marks for the scaled quantities (but keeping in mind
specially the thermal rescaling of $t$, $\V(\x)$ and $\p$), the master
equation is simply written as
\begin{equation}
\label{WKK:scaled}
\fl
\partial_{\tT}
\Wf
=
\Big[
-\p\,
\px
+
\bV'\,
\pp
+
\gammaT\,
\pp\,
\big(\p+\pp\big)
+
\bDxp
\pxp
+
{\textstyle\sum}_{\iq=1}^{\infty}
\qcoef^{(\iq)}
\,
\bV^{(2\iq+1)}(\x)
\,
\pp^{(2\iq+1)}
\Big]
\Wf
\quad
\end{equation}
with the coefficient in the quantum sum given by
%
\begin{equation}
\label{kappa:s}
\qcoef^{(\iq)}
=
(-1)^{\iq}\ldB^{2\iq}\big/(2\iq+1)!
\;.
\end{equation}
Planck's constant $\hbar$ is introduced in terms of the characteristic
action $\So=\Eo/\wo$ via the quantum parameter $\kondobar$ (denoted
$\bar{K}$ in \cite{gar2004})
%
\begin{equation}
\label{kondos}
\hbar/\So=2\pi/\kondobar
\;,
\qquad
\ldB
=
\pi\gammaT/\kondo
\;,
\qquad
\kondo=(\gamma/\wo)\kondobar
\;.
\end{equation}
The second relation gives $\ldB$ in terms of the Kondo parameter
$\kondo$, related by the third one with our $\kondobar$.
The classical limit is approached as $\hbar/\So\to0$, i.e., letting
$\kondobar\to\infty$.

Let us now introduce a splitting of the evolution operator $\LFP$ that
will facilitate the calculation of the matrix elements required when
expanding $\Wf(\x,\p)$ into complete sets.
On inspecting equations~(\ref{WKK:scaled}) and~(\ref{kappa:s}), we see
that extending the quantum sum down to $\iq=0$, we obtain the part
$\bV'\,\pp$ of the classical term.
Thus, we can decompose $\LFP$ into the irreversible, kinetic, and
potential parts:
\begin{equation}
\label{WKK:Lir:Lkin:Lv}
\fl
\partial_{\tT}
\Wf
=
\left(
\Lir
+
\Lkin
+
\Lv
\right)
\Wf
\;,
\qquad
\left\{
\begin{array}{lcl}
\Lir
&=&
\gammaT\,
\pp\,
\big(\p+\pp\big)
\\[0.5ex]
\Lkin
&=&
-\big(\p-\bDxp\pp\big)\,
\px
\\[0.5ex]
\Lv
&=&
\sumiqo
\qcoef^{(\iq)}
\,
\bV^{(2\iq+1)}
\,
\pp^{(2\iq+1)}
\end{array}
\right.
\;.
\end{equation}
This natural splitting has the advantage of dealing with $\V(\x)$ (the
most problem dependent part) in a unified way.
Besides, we have grouped the term $\pp\px$ with the kinetic term
$\p\,\px$ because they are structurally similar.


\section{
Derivation of the quantum hierarchies
}
\label{sec:hierarchies}

In the expansion into complete sets approach to solve kinetic
equations, the distribution is expanded in an orthonormal basis
$\{\psi_{\ip}\}$ and the coupled equations for the coefficients
$\ec_{\ip}$ derived \cite[p.~175]{balescu2}.
For the Klein--Kramers equation~(\ref{KK}) one uses Hermite functions
of $\p$ as basis \cite[Sec~4.4]{galpasI}
%
\begin{equation}
\label{hermite}
\psi_{\ip}(\p)
=
r_{\ip}
\,
\e^{-\p^{2}/4}
H_{\ip}(\p/\sqrt{2})
\;,
\qquad
r_{\ip}
=
\big[(2\pi)^{1/2}2^{\ip}\ip!\big]^{-1/2}
\;.
\end{equation}
One starts with the expansion in the momentum because, while the $\p$
dependence of the Hamiltonian is fully specified (kinetic energy
$\p^{2}/2\sM$), the potential part $\V(\x)$ depends on the problem.
Thus, explicit manipulations can be done on the
parts involving $\p$, which are valid for any system.
On the other hand, the Hermite basis has a number of advantages
\cite{risken}; not the least is the handling of the derivatives $\pp$
in the dynamical equation by means of the associated creation and
annihilation operators $\bd=\pp+\case{1}{2}\p$ and
$\bp=-\pp+\case{1}{2}\p$ (this will prove very convenient in the
quantum case).

For the Klein--Kramers equation the resulting equations for the
expansion coefficients $\ec_{\ip}$ are called the {\em Brinkman
hierarchy\/} \cite{risken,bri56I}.
This plays an important r\^{o}le in classical systems, both for
analytical treatments---derivations of $1/\gamma$ expansions, etc.---
and to develope efficient non-perturbative numerical methods.
In fact, it has the structure of a $3$-term recurrence relation which,
after expansion in the position basis, can be solved by continued
fractions.
In what follows we shall proceed in the quantum case following as
closely as possible the steps of the classical limit.


\subsection{
Expansion in the momentum basis
}
\label{sec:p-expansion}

We start expanding the Wigner function as
%
\begin{equation}
\label{W:expansion}
\Wf(\x,\p)
=
\Wo
\sum_{\ip}
\ec_{\ip}(\x)
\psi_{\ip}(\p)
\;,
\qquad
\Wo
=
\frac{\e^{-\etab\,\p^{2}/2}}{(2\pi)^{1/4}}\,
\e^{-\bVo(\x)}
\end{equation}
where we have extracted the Boltzmann-type factor $\Wo$.
This involves the auxiliary parameter $0\leq\etab\leq1/2$ and
potential $\bVo(\x)$, which is frequently chosen proportional to
$\bV(\x)$, i.e., $\bVo=\epspot\bV$.
The results should not depend on the $\etab$ or $\epspot$ used, but
these may (i) put parts of the evolution operator $\LFP$ in
self-adjoint form and/or (ii) improve the stability and convergence of
the eventual numerical implementation.

From the orthonormality of the $\psi_{\ip}(\p)$, the expansion
``coefficients'' can be written as
$\ec_{\ip}=\int\!\drm\p\,\psi_{\ip}\sum_{\ipp}\ec_{\ipp}\psi_{\ipp}
=
\int\!\drm\p\,\psi_{\ip}\,(\Wf/\Wo)$.
Then, differentiating with respect to $t$ and using the evolution
equation $\dptT\Wf=\LFP\Wf$, we get the dynamical equations for
$\ec_{\ip}$
%
\begin{equation}
\label{dcdt:Qtot:Lbar}
\fl
\dptT
\ec_{\ip}
=
\sum_{\ipp}
\Q_{\ip\ipp}
\ec_{\ipp}
\;,
\qquad
\Q_{\ip\ipp}
=
\int\!\drm \p\,
\psi_{\ip}
\LFPb
\psi_{\ipp}
\;,
\qquad
\LFPb
=
\Wo^{-1}
\LFP
\Wo
\;.
\end{equation}
To get the matrix elements $\Q_{\ip\ipp}$, which are still operators
on the $\x$-dependence of $\ec_{\ipp}$, one needs the
$\Wo^{-1}(\,\cdot\,)\Wo$ transform of $\LFP$.
This can be split as $\LFPb=\Lirb+\Lkinb+\Lvb$, with $\Lir$, $\Lkin$,
and $\Lv$ given by (\ref{WKK:Lir:Lkin:Lv}).
The transformation of these operators is done in \ref{app:Lb}, taking
advantage of results for {\em normal ordering\/} of $\bd$ and $\bp$
(which we use to express the $\pp$ and $\p$'s in $\LFP$).
Then, with the $\Lirb$, $\Lkinb$, and $\Lvb$ obtained we compute their
matrix elements between the $\psi_{\ip}$, getting the $\Q_{\ip\ipp}$.
The details of this part of the calculation are given in
\ref{app:Qnm}.

Introducing the results obtained for $\Q_{\ip\ipp}$ into
$\dptT\ec_{\ip}=\sum_{\ipp}\Q_{\ip\ipp}\ec_{\ipp}$ one gets the
following {\em quantum (Brinkman) hierarchy}
%
\begin{eqnarray}
\label{QBH}
\hspace*{-3.em}
-\dptT
\ec_{\ip}
&=&
\sum_{\iq=0}^{[(\ip-1)/2]}
\,
\big[
\G_{\ip}^{\iq,-}
\,
\bV^{(2\iq+1)}
\big]
\,
\ec_{\ip-(2\iq+1)}
+
\sqrt{(\ip-1)\ip}\;
\gammamm
\,
\ec_{\ip-2}
\nonumber\\
& &
{}+
\sqrt{\ip}\;
\Dm
\ec_{\ip-1}
+
\gammao_{\ip}
\,
\ec_{\ip}
+
\sqrt{\ip+1}\;
\Dp
\ec_{\ip+1}
\\
& &
{}+
\sqrt{(\ip+1)(\ip+2)}\;
\gammapp
\,
\ec_{\ip+2}
+
\sum_{\iq=0}^{\infty}
\big[
\G_{\ip+(2\iq+1)}^{\iq,+}
\,
\bV^{(2\iq+1)}
\big]
\,
\ec_{\ip+(2\iq+1)}
\nonumber
\;.
\end{eqnarray}
The auxiliary (damping) parameters introduced read
%
\begin{equation}
\label{gammas}
\gamma_{\pm}
=
\gammaT\,
\etapm(1-\etapm)
\;,
\qquad
\gammao_{\ip}
=
\gammaT
\left[
2\ip(\etab-\etapro)
-
\etap^{2}
\right]
\end{equation}
which involve the following $\etab$-related parameters (note the sign
exchange)
\begin{equation}
\label{etam:etap}
\etapm
=
\etab
\mp
\case{1}{2}
\;,
\qquad
\etapro
=
\etam\etap
\;.
\end{equation}
In equation (\ref{QBH}), the operators on the $\x$ dependences are
%
\begin{eqnarray}
\label{DmDp}
\fl
\Dpm
&=
\dplmi
\big(\px-\bVo'\big)
\;,
\qquad
\dplmi
=
1+\etapm\bDxp
\;,
\qquad
\dlpxx
=
\etapro\ldB^{2}\px^{2}
\;,
\\
\label{Gamma:qcoeff}
\fl
\G_{\ip}^{\iq,\pm}
&=
\etapm^{2\iq+1}
\qcoef^{(\iq)}_{\ip}
\,
\e^{-\dlpxx/2}
\kum\big(-\ipp,2\iq+2\,;\dlpxx\big)
\;,
\qquad
\qcoef^{(\iq)}_{\ip}
=
\qcoef^{(\iq)}
\sqrt{\ip!/\ipp!}
\;.
\end{eqnarray}
Here $\ipp=\ip-(2\iq+1)$, $\kum(a,c\,;z)$ is the confluent
hypergeometric (Kummer) function \cite[Ch.~13.6]{arfken}, and
$\qcoef^{(\iq)}$ is given by equation (\ref{kappa:s}).
In is important to note that the $\G_{\ip}^{\iq}$ act only on $\V(\x)$
not on the $\ec_{\ipp}(\x)$ (\ref{app:Lb}).

The quantum hierarchy (\ref{QBH}) is equivalent to the
Caldeira--Leggett master equation (\ref{WKK:Dxp}).
Previously, several hierarchies had been derived
\cite{plorho85,lilhafher89,joh98,burmol2002}, but the treatment was
for closed (Hamiltonian) systems \cite{plorho85,lilhafher89,joh98} or
involved unsuitable bases like $\{\p^{\ip}\}$
\cite{plorho85,joh98,burmol2002}.
In contrast, our hierarchy of {\em Hermitian moments\/} constitutes a
direct quantum generalisation of the Brinkman hierarchy.
To verify this, note that in the classical case only the $\iq=0$ terms
survive in the sums of (\ref{QBH}) (\ref{app:Qnm}).
Then, setting the usual $\etab=1/2$, equation~(\ref{QBH}) reduces to
%
\begin{equation}
\label{brinkman}
\dptT
\ec_{\ip}
=
-
\Big[
\sqrt{\ip}\;
\big(
\Dm
+
\bV'
\big)\,
\ec_{\ip-1}
+
\ip\gammaT
\,
\ec_{\ip}
+
\sqrt{\ip+1}\;
\Dp
\ec_{\ip+1}
\Big]
\end{equation}
which is indeed the celebrated hierarchy associated to the
Klein--Kramers equation.

For potentials obeying $\V^{(\iq)}\equiv0$, $\forall\iq\ge S$, the
quantum hierarchy (\ref{QBH}) is a recurrence relation with coupling
to a finite number of terms.
Examples include the harmonic oscillator
$\V=\case{1}{2}\sM\wo^{2}\x^{2}$ and the double-well (Duffing)
oscillator $\V=-\case{1}{2}a\,\x^{2}+\frac{1}{4} b\,\x^{4}$.
Finite-coupling recurrences are always reducible to $3$-term vector
recurrences (\ref{app:RR-CF}) and solvable by continued fractions, as
in the classical case.
However, for non-polynomial $\V(\x)$ this approach is prevented by the
infinite coupling range in the index $\ip$, consequence of the
infinite series of $\p$-derivatives $\V^{(2\iq+1)}\,\pp^{(2\iq+1)}$ in
the Wigner--Moyal term.
In what follows we shall exploit the expansion in the position basis
to show how this problem can be circumvented.


\subsection{
Expansion in the position basis
}
\label{sec:x-expansion}

Recall that the coefficients $\ec_{\ip}$ are still functions of
$\x$.
Inserting their expansion in an orthonormal basis $\{u_{\ix}(\x)\}$
into the quantum hierarchy (\ref{dcdt:Qtot:Lbar}), we get the generic
form
%
\begin{equation}
\label{dcdt:generic}
\ec_{\ip}(\x)
=
\sum_{\ix}
\ec_{\ip}^{\ix}
u_{\ix}(\x)
\quad
\leadsto
\quad
\frac{\drm}{\drm\tT}
\ec_{\ip}^{\ix}
=
\sum_{\ipp}
\sum_{\ixp}
\big(\Q_{\ip\ipp}\big)_{\ix\ixp}
\ec_{\ipp}^{\ixp}
\end{equation}
with matrix elements
$(B)_{\ix\ixp}
=
\int\!\drm\x\,u_{\ix}^{\ast}\,\hat{B}(\x,\px)\,u_{\ixp}$.
To put (\ref{QBH}) into the above form we just need to (i) substitute
the operators (and $\x$-dependent coefficients) by their matrix
elements $(B)_{\ix\ixp}$, (ii) attach position superscripts to the
$\ec_{\ip}$, and, finally, (iii) sum over them.
On so doing we get the {\em expanded quantum hierarchy}
%
\begin{eqnarray}
\label{QBH:pre-matrix:I}
\fl
{}-\frac{\drm}{\drm\tT}
\ec_{\ip}^{\ix}
&=
\sumiqo
\sum_{\ixp}
\big[
\G_{\ip}^{\iq,-}
\bV^{(2\iq+1)}
\big]_{\ix\ixp}\,
\ec_{\ip-(2\iq+1)}^{\ixp}
+
\sqrt{(\ip-1)\ip}
\sum_{\ixp}
\gammamm
\,
\delta_{\ix\ixp}
\,
\ec_{\ip-2}^{\ixp}
\nonumber\\
\fl
&
{}+
\sqrt{\ip}
\sum_{\ixp}
\Dm_{\ix\ixp}\,
\ec_{\ip-1}^{\ixp}
+
\sum_{\ixp}
\gammao_{\ip}\,
\delta_{\ix\ixp}
\,
\ec_{\ip}^{\ixp}
+
\sqrt{\ip+1}
\sum_{\ixp}
\Dp_{\ix\ixp}\,
\ec_{\ip+1}^{\ixp}
\\
\fl
&
{}+
\sqrt{(\ip+1)(\ip+2)}
\sum_{\ixp}
\gammapp
\,
\delta_{\ix\ixp}
\,
\ec_{\ip+2}^{\ixp}
+
\sumiqo
\sum_{\ixp}
\big[
\G_{\ip+(2\iq+1)}^{\iq,+}
\bV^{(2\iq+1)}
\big]_{\ix\ixp}\,
\ec_{\ip+(2\iq+1)}^{\ixp}
\nonumber
\;.
\end{eqnarray}
Considering that $\Dpm=\dplmi(\px-\bVo')$ and $\bVo=\epspot\bV$, we
see that to get all matrix elements involved we just need those of
$\px$ and of $\bV^{(2\iq+1)}$, namely
%
\begin{equation}
\label{matr-elem:der:V}
\big(\px\big)_{\ix\ixp}
=
\int\!\drm\x\,
u_{\ix}^{\ast}\,
\px\,
u_{\ixp}
\;,
\qquad
\bV^{(2\iq+1)}_{\ix\ixp}
=
\int\!\drm\x\,
u_{\ix}^{\ast}\,
\bV^{(2\iq+1)}\,
u_{\ixp}
\end{equation}
because the $\G_{\ip}^{\iq}$ only increase the order of the
potential derivatives.
%
%

The expanded hierarchy has the form of a system of ordinary
differential equations,
$\dot{\ec}_{\ip}^{\ix}
=
\sum_{\ipp\ixp}\Qlett_{\ip\ipp}^{\ix\ixp}
\,
\ec_{\ipp}^{\ixp}$.
If $\Ntr$ and $\Atr$ are some large truncation indices for the bases
$\{\psi_{\ip}(\p)\}$ and $\{u_{\ix}(\x)\}$, we have a problem with
$(\Ntr\times\Atr)$ equations.
One may be tempted to try direct Runge--Kutta integration or numerical
diagonalization of the associated
$(\Ntr\times\Atr)\times(\Ntr\times\Atr)$ matrix.
However, the dimensions involved are typically very large, so that it
is worth to pursue a continued-fraction treatment.%
\footnote{
Dynamical equations for the Wigner function have anyway been tackled
by purely numerical methods, like (pseudo-) spectral methods for
partial differential equations (see \cite{hugmensch98jpa,monpaz2001}
and references therein), or grid discretization usually followed by
Runge--Kutta integration \cite{tanwol92}.
}

As mentioned before the $\p$-recurrence has a finite coupling range
for polynomial potentials.
For other potentials, with an appropriate choice of the basis
$\{u_{\ix}(\x)\}$, the matrix elements (\ref{matr-elem:der:V}) may
vanish when the second index $\ixp$ lays at a certain ``distance'' of
the first $\ix$.
Then, the expanded hierarchy would be a recurrence relation in the
position indices with a finite coupling range, also amenable for a
continued-fraction treatment.
To proceed in these cases, we just need to transform the two-index
extended hierarchies into ordinary recurrence problems (i.e., with one
index).


\section{
Matrix quantum hierarchies with one-index recurrences
}
\label{sec:MQBH}


In order to apply the continued-fraction method when finite coupling
range in one of the indices is attained, we need to convert first the
two-index recurrences into ordinary recursions.
In the resulting one-index recurrences the coefficients will be
matrices acting on appropriate vectors formed with the
$\ec_{\ip}^{\ix}$.


\subsection{
Matrix quantum hierarchy: $\p$-recurrence
}

The generic expanded form (\ref{dcdt:generic}) can be converted into
an {\em one-index\/} recurrence relation in the momentum index
introducing the following vectors and matrices
%
\begin{equation}
\label{dcdt:vectors:matrices:p}
\fl
\frac{\drm}{\drm\tT}
\mc_{\ip}
=
\sum_{\ipp}
\mQ_{\ip\ipp}
\mc_{\ipp},
\;\;
\mc_{\ip}
=
\left(
\begin{array}{c}
\!\!
\ec_{\ip}^{-\Atr}
\!\!
\\
\vdots
\\
\!\!
\ec_{\ip}^{\Atr}
\!\!
\end{array}
\right)
\;\;
\mQ_{\ip\ipp}
=
\left(
\begin{array}{ccc}
\!
\big(\Q_{\ip\ipp}\big)_{-\Atr,-\Atr}
\!
&
\!\!\!\!
\cdots
\!\!\!\!
&
\!
\big(\Q_{\ip\ipp}\big)_{-\Atr,\Atr}
\!
\\
\vdots
&
\!\!\!\!
\ddots
\!\!\!\!
&
\vdots
\\
\!
\big(\Q_{\ip\ipp}\big)_{\Atr,-\Atr}
\!
&
\!\!\!\!
\cdots
\!\!\!\!
&
\!
\big(\Q_{\ip\ipp}\big)_{\Atr,\Atr}
\!
\end{array}
\right)
\quad
\end{equation}
where
$(\Q_{\ip\ipp})_{\ix\ixp}
=
\int\!\drm\x\,u_{\ix}^{\ast}\,\Q_{\ip\ipp}\,u_{\ixp}$.
That is, for a fixed $\ip$ one forms the column vector with all the
$\ec_{\ip}^{\ix}$ with $\ix=-\Atr,\dots,\Atr$ (this index may take
positive and negative values).
Similarly, for given $(\ip,\ipp)$ we build the matrix
$\Qlett_{\ip\ipp}^{\ix\ixp}\equiv(\Q_{\ip\ipp})_{\ix\ixp}$ with all
indices $\ix,\ixp$.

Before giving explicit expressions for the matrix coefficients we
introduce the following notation:
$\mQ_{\ip}^{\pm\iq}=\mQ_{\ip,\ip\pm\iq}$,
$\mQ_{\ip}^{\pm\pm}=\mQ_{\ip,\ip\pm2}$,
$\mQ_{\ip}^{\pm}=\mQ_{\ip,\ip\pm1}$,
and
$\mQ_{\ip}=\mQ_{\ip,\ip}$.
Then, the matrix quantum hierarchy (\ref{dcdt:vectors:matrices:p}) can
be written as
%
\begin{eqnarray}
\label{QBH:matrix:p}
\frac{\drm}{\drm\tT}
\mc_{\ip}
&=&
{\textstyle\sum}_{\iq\ge1}
\mQ_{\ip}^{-(2\iq+1)}
\mc_{\ip-(2\iq+1)}
+
\mQ_{\ip}^{--}
\mc_{\ip-2}
\nonumber\\[-0.5ex]
& &
{}+
\mQ_{\ip}^{-}
\mc_{\ip-1}
+
\mQ_{\ip}
\mc_{\ip}
+
\mQ_{\ip}^{+}
\mc_{\ip+1}
\\[+0.5ex]
& &
{}+
\mQ_{\ip}^{++}
\mc_{\ip+2}
+
{\textstyle\sum}_{\iq\ge1}
\mQ_{\ip}^{+(2\iq+1)}
\mc_{\ip+(2\iq+1)}
\nonumber
\;.
\end{eqnarray}
We have accounted for $\mQ_{\ip}^{\pm(2\iq)}\equiv0$, $\forall\iq\ge2$
[vd.\ equation~(\ref{QBH:pre-matrix:I})] and incorporated the $\iq=0$
terms into the central $\mc_{\ip\pm1}$ ones.
The matrix coefficients explicitly read
%
\begin{equation}
\label{Q:matr-elem:p:I}
\left\{
\begin{array}{lcr}
\big(
\mQ_{\ip}^{-(2\iq+1)}
\big)_{\ix\ixp}
&=&
-
\big[
\G_{\ip}^{\iq,-}
\bV^{(2\iq+1)}
\big]_{\ix\ixp}\,
\\
\big(
\mQ_{\ip}^{--}
\big)_{\ix\ixp}
&=&
-\sqrt{(\ip-1)\ip}\;
\gammamm\,
\delta_{\ix\ixp}
\\
\big(
\mQ_{\ip}^{-}
\big)_{\ix\ixp}
&=&
-
\sqrt{\ip}\,
\Dm_{\ix\ixp}
-
\big[
\G_{\ip}^{0,-}
\bV'
\big]_{\ix\ixp}\,
\\
\big(
\mQ_{\ip}
\big)_{\ix\ixp}
&=&
-
\gammao_{\ip}\,\delta_{\ix\ixp}
\\
\big(
\mQ_{\ip}^{+}
\big)_{\ix\ixp}
&=&
-
\sqrt{\ip+1}\,
\Dp_{\ix\ixp}
-
\big[
\G_{\ip+1}^{0,+}
\bV'
\big]_{\ix\ixp}\,
\\
\big(
\mQ_{\ip}^{++}
\big)_{\ix\ixp}
&=&
-\sqrt{(\ip+1)(\ip+2)}\;
\gammapp\,
\delta_{\ix\ixp}
\\
\big(
\mQ_{\ip}^{+(2\iq+1)}
\big)_{\ix\ixp}
&=&
-
\big[
\G_{\ip+(2\iq+1)}^{\iq,+}
\bV^{(2\iq+1)}
\big]_{\ix\ixp}\,
\end{array}
\right.
\end{equation}
For finite coupling range in $\ip$ (e.g., in polynomial potentials),
the hierarchy (\ref{QBH:matrix:p}) is a recurrence relation to which
continued-fraction methods can now be applied.
This amounts to convert the underlying large
$(\Ntr\times\Atr)\times(\Ntr\times\Atr)$ problem into $\Ntr$ problems
with associated $\Atr\times\Atr$ matrices [or $\Atr\to(2\Atr+1)$].
As the recursion ``coefficients'' are matrices, one would need {\em
matrix continued-fraction\/} methods (\ref{app:RR-CF}).


\subsection{
Matrix quantum hierarchy: $\x$-recurrence
}

When the $\p$-recurrence has an infinite coupling range (in periodic
potentials, the Morse potential, etc.), we can convert the $2$-index
hierarchy (\ref{dcdt:generic}) into matrix form, but introducing
recurrences in the position index $\ix$
\begin{equation}
\label{dcdt:vectors:matrices:x}
\fl
\frac{\drm}{\drm\tT}
\mc_{\ix}
=
\sum_{\ixp}
\mQ_{\ix\ixp}
\mc_{\ixp},
\quad
\mc_{\ix}
=
\left(
\begin{array}{c}
\ec_{0}^{\ix}
\\
\vdots
\\
\ec_{\Ntr}^{\ix}
\end{array}
\right)
\quad
\mQ_{\ix\ixp}
=
\left(
\begin{array}{ccc}
\big(\Q_{00}\big)_{\ix\ixp}
&
\cdots
&
\big(\Q_{0\Ntr}\big)_{\ix\ixp}
\\
\vdots
&
\ddots
&
\vdots
\\
\big(\Q_{\Ntr0}\big)_{\ix\ixp}
&
\cdots
&
\big(\Q_{\Ntr\Ntr}\big)_{\ix\ixp}
\end{array}
\right)
\end{equation}
That is,
$(\mQ_{\ix\ixp})_{\ip\ipp}
=
(\Q_{\ip\ipp})_{\ix\ixp}$,
with
$B_{\ix\ixp}
=
\int\!\drm\x\,
u_{\ix}^{\ast}\,
\hat{B}\,
u_{\ixp}$.
Thus, for a fixed $\ix$ one constructs the column vector with the
$\ec_{\ip}^{\ix}$ for all $\ip=0,\dots,\Ntr$, and for given
$(\ix,\ixp)$, the matrix
$\Qlett_{\ip\ipp}^{\ix\ixp}\equiv(\Q_{\ip\ipp})_{\ix\ixp}$ with all
$\ip,\ipp$.
Comparing the matrix in (\ref{dcdt:vectors:matrices:x}) with
equation~(\ref{Q:matr-elem:p:I}) for $(\mQ_{\ip}^{\pm\iq})_{\ix\ixp}$,
we can construct $\mQ_{\ix\ixp}$ diagonal by diagonal
%
\begin{equation}
\label{Q:matr-elem:x:I}
\left\{
\begin{array}{lcr}
\big(
\mQ_{\ix\ixp}
\big)_{\ip,\ip-(2\iq+1)}
&=&
-
\big[
\G_{\ip}^{\iq,-}
\bV^{(2\iq+1)}
\big]_{\ix\ixp}\,
\\
\big(
\mQ_{\ix\ixp}
\big)_{\ip,\ip-2}
&=&
-\sqrt{(\ip-1)\ip}\;
\gammamm\,
\delta_{\ix\ixp}
\\
\big(
\mQ_{\ix\ixp}
\big)_{\ip,\ip-1}
&=&
-
\sqrt{\ip}\,
\Dm_{\ix\ixp}
-
\big[
\G_{\ip}^{0,-}
\bV'
\big]_{\ix\ixp}\,
\\
\big(
\mQ_{\ix\ixp}
\big)_{\ip,\ip}
&=&
-
\gammao_{\ip}\,\delta_{\ix\ixp}
\\
\big(
\mQ_{\ix\ixp}
\big)_{\ip,\ip+1}
&=&
-
\sqrt{\ip+1}\,
\Dp_{\ix\ixp}
-
\big[
\G_{\ip+1}^{0,+}
\bV'
\big]_{\ix\ixp}\,
\\
\big(
\mQ_{\ix\ixp}
\big)_{\ip,\ip+2}
&=&
-\sqrt{(\ip+1)(\ip+2)}\;
\gammapp\,
\delta_{\ix\ixp}
\\
\big(
\mQ_{\ix\ixp}
\big)_{\ip,\ip+(2\iq+1)}
&=&
-
\big[
\G_{\ip+(2\iq+1)}^{\iq,+}
\bV^{(2\iq+1)}
\big]_{\ix\ixp}\,
\end{array}
\right.
\end{equation}
all other diagonals being zero
$(\mQ_{\ix\ixp})_{\ip,\ip\mp2\iq}\equiv0$, $\forall\iq\ge2$.

To recognise better the structure of these matrices, we decompose them
into the ``free'' and potential contributions
$\mQ_{\ix\ixp}=\mQ_{\ix\ixp}^{\rm f}+\mQ_{\ix\ixp}^{\rm v}$.
For $\mQ_{\ix\ixp}^{\rm f}$, including the kinetic and irreversible
terms, we have the pentadiagonal structure (tridiagonal for
$\etab=1/2$)
\begin{equation*}
\fl
\mQ_{\ix\ixp}^{\rm f}
=
-
\left(
\begin{array}{ccccccc}
\gammao_{0}
\delta_{\ix\ixp}
\!\!&\!\!
\sqrt{1}
\Dp_{\ix\ixp}
\!\!&\!\!
\sqrt{1\!\cdot\!2}
\gammapp
\delta_{\ix\ixp}
\!\!&\!\!
0
\!\!&\!\!
0
\!\!&\!\!
0
\!\!&\!\!
\ddots
\\
\sqrt{1}
\Dm_{\ix\ixp}
\!\!&\!\!
\gammao_{1}
\delta_{\ix\ixp}
\!\!&\!\!
\sqrt{2}
\Dp_{\ix\ixp}
\!\!&\!\!
\sqrt{2\!\cdot\!3}
\gammapp
\delta_{\ix\ixp}
\!\!&\!\!
0
\!\!&\!\!
0
\!\!&\!\!
\ddots
\\
\sqrt{1\!\cdot\!2}
\gammamm
\delta_{\ix\ixp}
\!\!&\!\!
\sqrt{2}
\Dm_{\ix\ixp}
\!\!&\!\!
\gammao_{2}
\delta_{\ix\ixp}
\!\!&\!\!
\sqrt{3}
\Dp_{\ix\ixp}
\!\!&\!\!
\sqrt{3\!\cdot\!4}
\gammapp
\delta_{\ix\ixp}
\!\!&\!\!
0
\!\!&\!\!
\ddots
\\
0
\!\!&\!\!
\sqrt{2\!\cdot\!3}
\gammamm
\delta_{\ix\ixp}
\!\!&\!\!
\sqrt{3}
\Dm_{\ix\ixp}
\!\!&\!\!
\gammao_{3}
\delta_{\ix\ixp}
\!\!&\!\!
\sqrt{4}
\Dp_{\ix\ixp}
\!\!&\!\!
\sqrt{4\!\cdot\!5}
\gammapp
\delta_{\ix\ixp}
\!\!&\!\!
\ddots
\\
0
\!\!&\!\!
0
\!\!&\!\!
\sqrt{3\!\cdot\!4}
\gammamm
\delta_{\ix\ixp}
\!\!&\!\!
\sqrt{4}
\Dm_{\ix\ixp}
\!\!&\!\!
\gammao_{4}
\delta_{\ix\ixp}
\!\!&\!\!
\sqrt{5}
\Dp_{\ix\ixp}
\!\!&\!\!
\ddots
\\
0
\!\!&\!\!
0
\!\!&\!\!
0
\!\!&\!\!
\sqrt{4\!\cdot\!5}
\gammamm
\delta_{\ix\ixp}
\!\!&\!\!
\sqrt{5}
\Dm_{\ix\ixp}
\!\!&\!\!
\gammao_{5}
\delta_{\ix\ixp}
\!\!&\!\!
\ddots
\\
\ddots
\!\!&\!\!
\ddots
\!\!&\!\!
\ddots
\!\!&\!\!
\ddots
\!\!&\!\!
\ddots
\!\!&\!\!
\ddots
\!\!&\!\!
\ddots
\end{array}
\right)
\end{equation*}
The part due to $\V(\x)$, on the other hand, has the following
alternate dense structure, reflecting the odd powers of $\pp$ in the
Wigner--Moyal term:
\begin{equation*}
\fl
\mQ_{\ix\ixp}^{\rm v}
=
-
\left(
\begin{array}{ccccccc}
0
\!\!\!\!&\!\!\!\!
[\G_{1}^{0,+}
\bV']_{\ix\ixp}
\!\!\!\!&\!\!\!\!
0
\!\!\!\!&\!\!\!\!
[\G_{3}^{1,+}
\bV^{(3)}]_{\ix\ixp}
\!\!\!\!&\!\!\!\!
0
\!\!\!\!&\!\!\!\!
[\G_{5}^{2,+}
\bV^{(5)}]_{\ix\ixp}
\!\!\!\!&\!\!
\ddots
\\[0.ex]
[\G_{1}^{0,-}
\bV']_{\ix\ixp}
\!\!\!\!&\!\!\!\!
0
\!\!\!\!&\!\!\!\!
[\G_{2}^{0,+}
\bV']_{\ix\ixp}
\!\!\!\!&\!\!\!\!
0
\!\!\!\!&\!\!\!\!
[\G_{4}^{1,+}
\bV^{(3)}]_{\ix\ixp}
\!\!\!\!&\!\!\!\!
0
\!\!\!\!&\!\!
\ddots
\\[0.ex]
0
\!\!\!\!&\!\!\!\!
[\G_{2}^{0,-}
\bV']_{\ix\ixp}
\!\!\!\!&\!\!\!\!
0
\!\!\!\!&\!\!\!\!
[\G_{3}^{0,+}
\bV']_{\ix\ixp}
\!\!\!\!&\!\!\!\!
0
\!\!\!\!&\!\!\!\!
[\G_{5}^{1,+}
\bV^{(3)}]_{\ix\ixp}
\!\!\!\!&\!\!
\ddots
\\[0.ex]
[\G_{3}^{1,-}
\bV^{(3)}]_{\ix\ixp}
\!\!\!\!&\!\!\!\!
0
\!\!\!\!&\!\!\!\!
[\G_{3}^{0,-}
\bV']_{\ix\ixp}
\!\!\!\!&\!\!\!\!
0
\!\!\!\!&\!\!\!\!
[\G_{4}^{0,+}
\bV']_{\ix\ixp}
\!\!\!\!&\!\!\!\!
0
\!\!\!\!&\!\!
\ddots
\\[0.ex]
0
\!\!\!\!&\!\!\!\!
[\G_{4}^{1,-}
\bV^{(3)}]_{\ix\ixp}
\!\!\!\!&\!\!\!\!
0
\!\!\!\!&\!\!\!\!
[\G_{4}^{0,-}
\bV']_{\ix\ixp}
\!\!\!\!&\!\!\!\!
0
\!\!\!\!&\!\!\!\!
[\G_{5}^{0,+}
\bV']_{\ix\ixp}
\!\!\!\!&\!\!
\ddots
\\[0.ex]
[\G_{5}^{2,-}
\bV^{(5)}]_{\ix\ixp}
\!\!\!\!&\!\!\!\!
0
\!\!\!\!&\!\!\!\!
[\G_{5}^{1,-}
\bV^{(3)}]_{\ix\ixp}
\!\!\!\!&\!\!\!\!
0
\!\!\!\!&\!\!\!\!
[\G_{5}^{0,-}
\bV']_{\ix\ixp}
\!\!\!\!&\!\!\!\!
0
\!\!\!\!&\!\!
\ddots
\\[0.ex]
\ddots
\!\!\!\!&\!\!\!\!
\ddots
\!\!\!\!&\!\!\!\!
\ddots
\!\!\!\!&\!\!\!\!
\ddots
\!\!\!\!&\!\!\!\!
\ddots
\!\!\!\!&\!\!\!\!
\ddots
\!\!\!\!&\!\!
\ddots
\end{array}
\right)
\end{equation*}
%
With an appropriate choice of the basis $\{u_{\ix}(\x)\}$ the matrix
elements $(B)_{\ix\ixp}$ may couple only a finite number of basis
functions.
Then, the recurrence~(\ref{dcdt:vectors:matrices:x}) would be solvable
by continued fractions when the $\ip$-recurrence is infinite and this
approach is seemingly precluded.
Even for finite coupling in $\ip$, the $\ix$-recurrence may be
preferable in certain situations (e.g., at low $T$), as indicated by
Risken for a cosine potential in the classical limit
\cite[Sec.~11.5.6]{risken}).
In these cases, one has been able to reduce the underlying large
$(\Ntr\times\Atr)\times(\Ntr\times\Atr)$ problem to $\Atr$ problems
with associated $\Ntr\times\Ntr$ matrices.


\section{
Density matrix, observables, and marginal distributions
}
\label{sec:obs}

Once the $\dot{\mc}_{\ip}=\sum_{\ipp}\mQ_{\ip\ipp}\mc_{\ipp}$ or
$\dot{\mc}_{\ix}=\sum_{\ixp}\mQ_{\ix\ixp}\mc_{\ixp}$ are solved, we
can reconstruct the Wigner function from its expansion coefficients
$\ec_{\ip}^{\ix}$ in the double basis
$\{u_{\ix}(\x)\,\psi_{\ip}(\p)\}$
\begin{equation}
\label{W:expansion:xp}
\Wf(\x,\p)
=
\Wo(\x,\p)
\sum_{\ip,\ix}
\ec_{\ip}^{\ix}\;
u_{\ix}(\x)\,\psi_{\ip}(\p)
\end{equation}
obtaining the {\em full solution of the quantum master equation}.
If preferred, one can switch to the familiar density matrix
$\hat{\varrho}$ by the inverse of transformation (\ref{wigner:def})
%
\begin{equation}
\label{rho:wigner}
\varrho(\x,\x')
=
\int
\drm\p\,
\e^{\iu \p(\x-\x')/\hbar}\,
\Wf\big(\case{1}{2}(\x+\x'),\p\big)
\;.
\end{equation}
We can get any observable inserting the above $\Wf(\x,\p)$ in
equation~(\ref{average}) for the averages.
Nevertheless, common observables can many times be extracted {\em
directly\/} from the $\ec_{\ip}^{\ix}$.

Let us illustrate this with transport observables, which are
characterised by the averages $\llangle\p^{\ell}\rrangle$.
For an arbitrary function of $\p$ only, we get from
(\ref{W:expansion:xp})
%
\begin{equation}
\fl
\llangle f\rrangle
=
\int\!\drm\x\drm\p\,\Wf(\x,\p)\,f(\p)
=
\sum_{\ip,\ix}
\ec_{\ip}^{\ix}
\!
\int\!\drm\x\,
\e^{-\bVo(\x)}
u_{\ix}(\x)
\!
\int\!\drm\p\,
r_{0}
\e^{-\etab\,\p^{2}/2}
f(\p)\,
\psi_{\ip}(\p)
\end{equation}
where we have used $\Wo=r_{0}\,\e^{-\etab\,\p^{2}/2}\e^{-\bVo}$, with
$r_{0}=1/(2\pi)^{1/4}$ [equations~(\ref{hermite}) and
(\ref{W:expansion})].
Next we introduce the auxiliary integrals (for their calculation vd.\
\ref{app:auxint})
%
\begin{equation}
\label{Ial:Knl}
\Ial_{\ix}
=
\int\!\drm\x\,
\e^{-\bVo(\x)}
u_{\ix}(\x)
\;,
\qquad
\Knl_{\ip}^{(\ell)}
=
\int\!\drm\p\,
r_{0}\,
\e^{-\etab\,\p^{2}/2}
\,
\p^{\ell}
\,
\psi_{\ip}(\p)
\;.
\end{equation}
Then, the moments (case $f=\p^{\ell}$) can be written in terms of
the expansion coefficients as
$\langle\p^{\ell}\rangle
=
\sum_{\ip}
\Knl_{\ip}^{(\ell)}\,\big(\sum_{\ix}\ec_{\ip}^{\ix}\,\Ial_{\ix}\big)$.
Taking into account that $H_{\ip}$ has the parity of $\ip$ and
carrying this to $\Knl_{\ip}^{(\ell)}$, we get for the first two
%
\begin{equation}
\label{J:Ial:Knl}
\fl
\llangle\,\p\,\rrangle
=
\sum_{\ip}
\Knl_{2\ip+1}^{(1)}
\;
\Big(
\sum_{\ix}
\ec_{2\ip+1}^{\ix}
\,
\Ial_{\ix}
\Big)
\;,
\qquad
\llangle\p^{2}\rrangle
=
\sum_{\ip}
\Knl_{2\ip}^{(2)}
\;
\Big(
\sum_{\ix}
\ec_{2\ip}^{\ix}
\,
\Ial_{\ix}
\Big)
\;.
\end{equation}
The first moment is the {\em current\/} and the second characterises
{\em fluctuations\/} around it.
The zeroth moment corresponds to the {\em normalisation\/} of $\Wf$,
and imposses on the $\ec_{\ip}^{\ix}$ the condition
$\langle\p^{0}\rangle=\sum_{\ip}\Knl_{2\ip}^{(0)}
\big(\sum_{\ix}\ec_{2\ip}^{\ix}\,\Ial_{\ix}\big)=1$.

Finally, the quantum probabilities of $\x$ and $\p$, given by the
marginal distributions $P(\x)=\int\!\drm\p\,\Wf(\x,\p)$ and
$P(\p)=\int\!\drm\x\,\Wf(\x,\p)$, can be expressed in a similar
fashion
\begin{eqnarray}
\label{Wx}
P(\x)
&=&
\e^{-\epspot\bV(\x)}
\sum_{\ix}
u_{\ix}(\x)
\,
\Big(
\sum_{\ip}
\ec_{\ip}^{\ix}
\,
\Knl_{\ip}^{(0)}
\Big)
\\
\label{Wp}
P(\p)
&=&
r_{0}\,\e^{-\etab\,\p^{2}/2}
\sum_{\ip}
\psi_{\ip}(\p)
\,
\Big(
\sum_{\ix}
\ec_{\ip}^{\ix}
\,
\Ial_{\ix}
\Big)
\;.
\end{eqnarray}


\section{
Periodic potentials: matrix elements and limit cases
}
\label{sec:Vper}

Henceforth we shall apply the method described to incorporate
fluctuations and dissipation in the problem of {\em quantum transport
in periodic potentials}.
%
The simplest model consists of a particle evolving in a cosine
potential subjected to external forces (tilted ``washboard''
potential), which is also a paradigm of {\em quantum Brownian
motion\/} \cite{sch83,guihakmur85,fiszwe85}.
Others include potentials lacking inversion symmetry (ratchets), which
have been used to model rectification of current and {\em directional
motion\/} in several systems \cite{rei2002}.
These problems are demanding because periodic potentials have (partly)
continuous spectra \cite{glukolkor2002}, rendering inappropriate
methods devised for systems with discrete levels.
In this section we compute the matrix coefficients of our recurrences
for arbitrary periodic $\V(\x)$.
With them we can implement the continued-fraction method to solve
the quantum master equation, which we first check by regaining the
classical dissipative and quantum Hamiltonian limits.


\subsection{
Matrix elements
}

In periodic potentials the $\p$-recurrence cannot be used because of
its infinite coupling range (the derivatives of $\V(\x)$ neither
vanish, nor decrease, as they are essentially the potential itself).
However, one can guess that the $\x$-recurrence may have a coupling
range related with the number of harmonics in $\V(\x)$.
To exploit this we introduce plane waves as basis functions and the
Fourier expansion of the potential derivative:
%
\begin{equation}
\label{Vper}
u_{\ix}(\x)
=
\frac{\e^{\iu\ix\x}}{\sqrt{2\pi}}
\;,
\qquad
\V'(\x)
=
\sum_{\ix}
\V_{\ix}'\,
\e^{\iu\ix\x}
\;.
\end{equation}
To preserve the periodicity of $\Wf$, one extracts the external force
$\bF$ from $\V(\x)$ including it in a generalised
$\Dpm=\dplmi(\px-\bVo')-\etapm\bF$; cf.\ equation~(\ref{DmDp}).
Then, the auxiliary potential $\bVo$ is set proportional to the
periodic part $\bVo=\epspot\bV$.

To compute the matrix elements in $\mQ_{\ix\ixp}$, we only need to do
the integrals (\ref{matr-elem:der:V}) with plane waves for $u_{\ix}$.
Using $\px u_{\ix}=\iu\ix\,\e^{\iu\ix\x}/\sqrt{2\pi}$, the integral
$\int_{0}^{2\pi}\!\drm\x\,\e^{\iu(\ixp-\ix)\x}=2\pi\delta_{\ix\ixp}$,
and the expansion $\bV^{(\iq)}=\sum_{r}\bV_{r}^{(\iq)}\e^{\iu r\x}$,
we find
\begin{equation}
(\px)_{\ix\ixp}=\iu\ix\,\delta_{\ix\ixp}
\;,
\qquad
\big[\bV^{(\iq)}(\x)\big]_{\ix\ixp}=\V_{\ix-\ixp}^{(\iq)}
\;.
\end{equation}
From these elements and
$\V_{\qab}^{(2\iq+1)}=(-1)^{\iq}\qab^{2\iq}\V_{\qab}'$
we get
\begin{equation}
\label{DmDpVper:matr-elem}
\fl
\begin{array}{ccclccl}
\big[
\G_{\ip}^{\iq,\pm}
\bV^{(2\iq+1)}
\big]_{\ix\ixp}
&=&
0
&
\delta_{\ix\ixp}
&-&
\etapm^{2\iq+1}
\Gs_{\ip}^{\iq}(-\etapro \qab^{2}\ldB^{2})
&
\V_{\ix-\ixp}'
\\
\Dpm_{\ix\ixp}
&=&
\big(\iu\,\dplmi\ix-\etapm\bF\big)
&
\delta_{\ix\ixp}
&-&
\dplmi\epspot
&
\bV_{\ix-\ixp}'
\end{array}
\end{equation}
with $\qab=\ix-\ixp$, $\ipp=\ip-(2\iq+1)$ and [cf.\ equation
(\ref{Gamma:qcoeff})]
%
\begin{equation}
\label{Gns}
\fl
\Gs_{\ip}^{\iq}(z)
=
|\qcoef^{(\iq)}_{\ip}(\qab)|
\,
\e^{-z/2}
\,
\kum(-\ipp,2\iq+2\,;z)
\;,
\qquad
|\qcoef^{(\iq)}_{\ip}(\qab)|
=
\frac{(\qab\ldB)^{2\iq}}{(2\iq+1)!}
\sqrt{\frac{\ip!}{\ipp!}}
\;.
\end{equation}
That is, $\qcoef^{(\iq)}_{\ip}(\qab)$ is the coefficient
in (\ref{Gamma:qcoeff}) with $\ldB\to\qab\ldB$.
Inserting equation~(\ref{DmDpVper:matr-elem}) into the general
matrices $\mQ_{\ix\ixp}$ of section~\ref{sec:MQBH} we explicitly get
the matrix coefficients of our recurrences (\ref{app:matrices}).
For the cosine potential
%
\begin{equation}
\label{Vcos}
\V(\x)
=
-\Vo
\cos\x
\end{equation}
the coupling range is $1$, because $\V_{\qab}'=0$ for $|\qab|>1$ and
we have a $3$-term recurrence
$\dot{\mc}_{\ix}
=
\mQ_{\ix,\ix-1}\mc_{\ix-1}
+
\mQ_{\ix,\ix}\mc_{\ix}
+
\mQ_{\ix,\ix+1}\mc_{\ix+1}$.
For a $2$-harmonic ratchet potential
%
\begin{equation}
\label{Vrat}
\V(\x)
=
-\Vo
\big[
\sin\x
+
(\rat/2)
\sin(2\x)
\big]
\end{equation}
the range is $2$ ($5$-term recurrence) because $\V_{\qab}'=0$ for
$|\qab|=|\ix-\ixp|>2$, and so on.

As we restricted $0\leq\etab\leq1/2$, we have
$\etapro=\etam\etap=\etab^{2}-1/4\leq0$, so that
$z=-\etapro\qab^{2}\ldB^{2}\geq0$.
Thus, the argument of $\Gs_{\ip}^{\iq}$ is positive and $\exp(-z/2)$
can act as a regularisation factor [an advantage of allowing
$\etab\neq1/2$ in (\ref{W:expansion})].
This factor reduces the weight of the off-diagonal terms inside
$\mQ_{\ix\ixp}$, enhancing the numerical stability when going into the
deep quantum regime.
Then we typically use $\etab\sim0$--$0.05$, while for classical
calculations $\etab=1/2$ performs better.
Concerning the auxiliary potential $\bVo$, for the problems considered
below we use $\epspot=0$.
Convergence is achieved with $\Ntr\sim100$ Hermite functions and
$\Atr\sim25-50$ plane waves [then $\Ntr\times(2\Atr+1)\sim10^{4}$].
Finally, a word on the scaled quantities
(section~\ref{sec:WKK:scaled}).
To get period $2\pi$, we have set $\xo=\period/2\pi$ with $\period$
the original period.
The characteristic energy $\Eo$ is given by the potential amplitude
$\Eo=\Vo$; then $\kT$ is scaled by $\Vo$ and the other characteristic
quantities are: action, $\So=\xo\sqrt{\sM\Vo}$, frequency,
$\wo^{2}=\Vo/\sM\xo^{2}$, and force, $F_{0}=\Vo/\xo$.
\begin{figure}
\includegraphics[width=19.em]{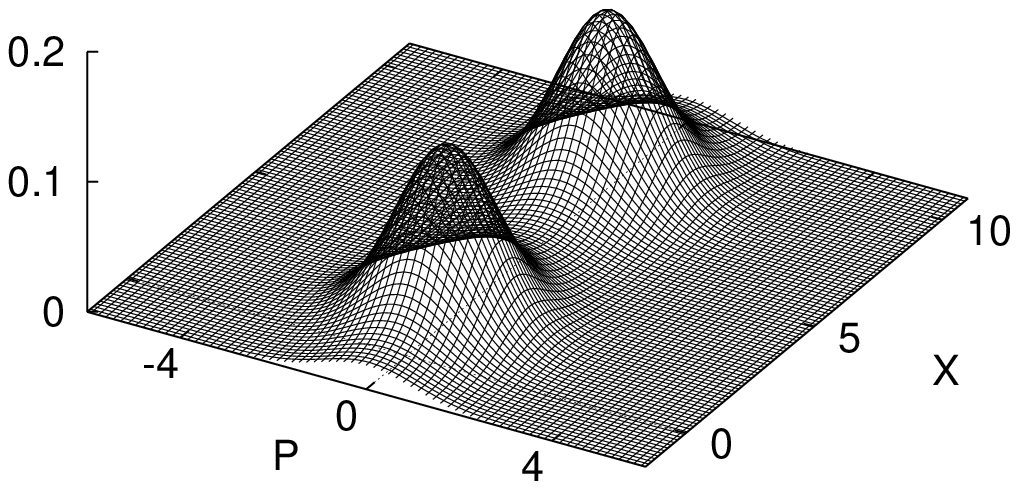}
\includegraphics[width=19.em]{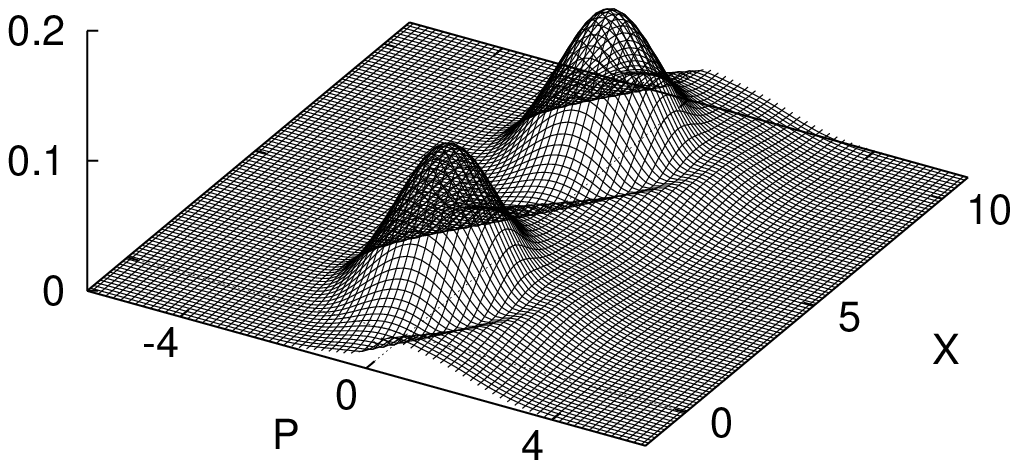}
\includegraphics[width=19.em]{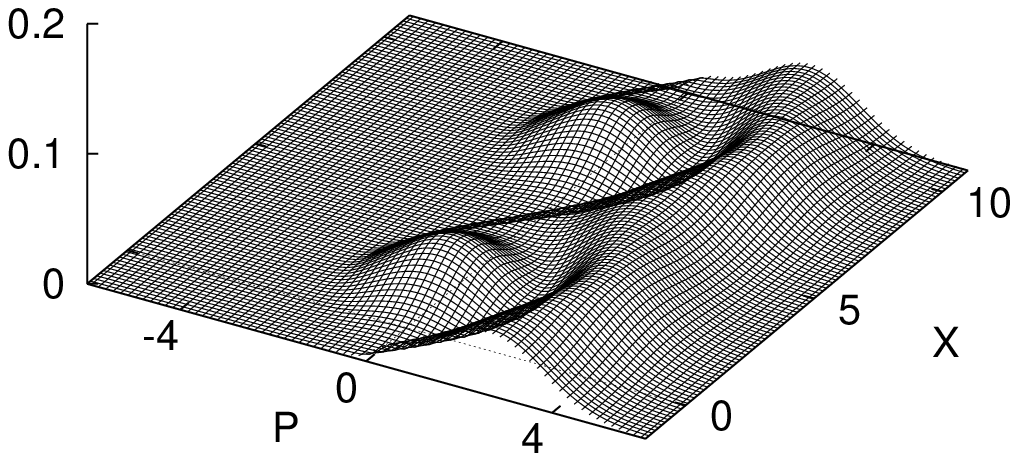}
\includegraphics[width=19.em]{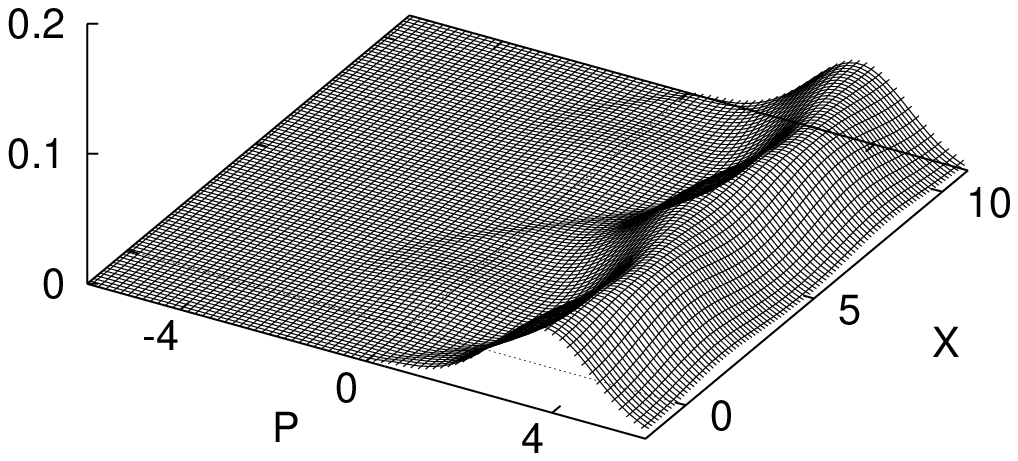}
\caption{
Wigner functions in the classical limit ($\kondobar=1000$) for a
particle in a tilted sinusoidal potential (two periods are displayed
along $\x$).
Here $\gamma=0.05$, and $T=1$, while $F=0$, $0.075$, $0.15$, and $0.2$
(left to right, top to bottom).
}
\label{fig:risken}
\end{figure}


\subsection{
Classical limit
}
\label{sec:risken}

Before going into the quantum regime, let us check that for small
enough $\hbar/\So=2\pi/\kondobar$ we recover the classical results.
Recall that in the deterministic limit a particle in a periodic
potential has two critical forces $F_{1}$ (retrapping force) and
$F_{3}$ (force at which the barrier disappears).
For $F<F_{1}$ the attractors of the dynamical system are solutions
``locked'' around the potential minima, whereas for $F>F_{3}$ the
``running'' solution is an attractor globally stable.
In the range $F_{1}<F<F_{3}$ the system exhibits bistability.
Finally, $F_{1}$ decreases with $\gamma$ [for the cosine potential
$F_{1}\sim(4/\pi)\Vo^{1/2}\gamma$], whereas $F_{1}$ tends to $F_{3}$
as the damping increases, narrowing the bistability range.
\begin{figure}
\includegraphics[width=14.em,angle=-90]{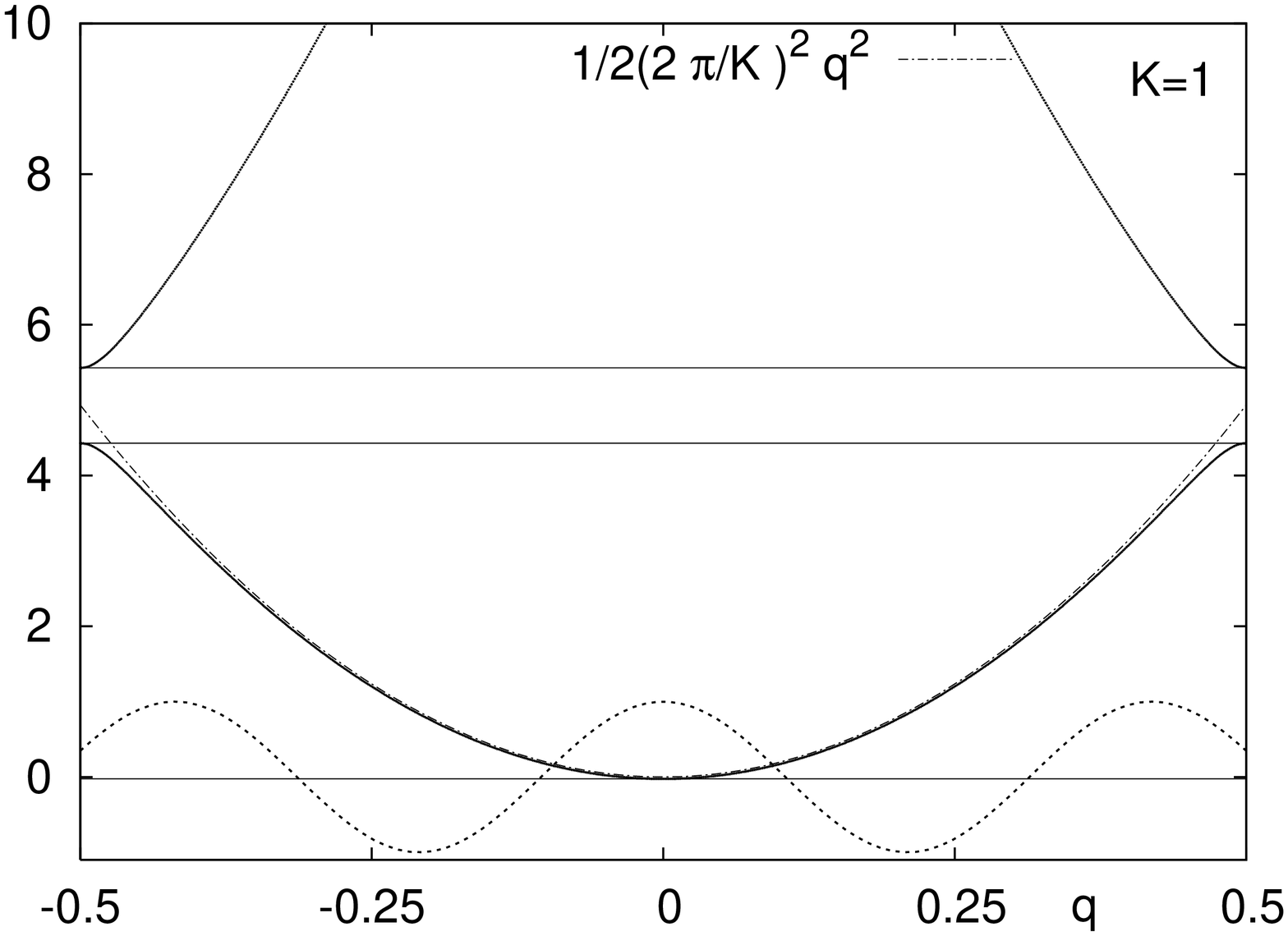}
\includegraphics[width=14.em,angle=-90]{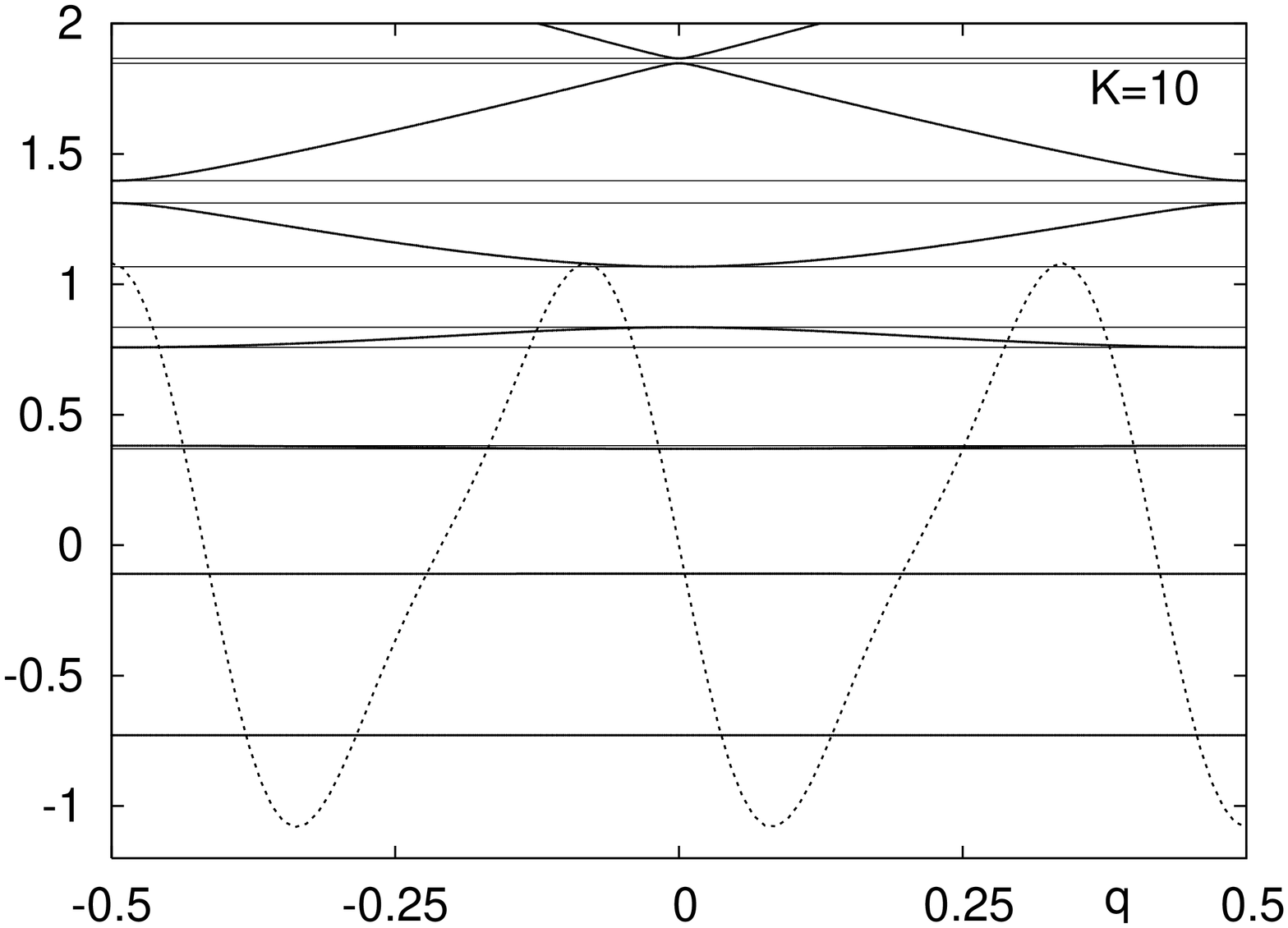}
\includegraphics[width=14.em,angle=-90]{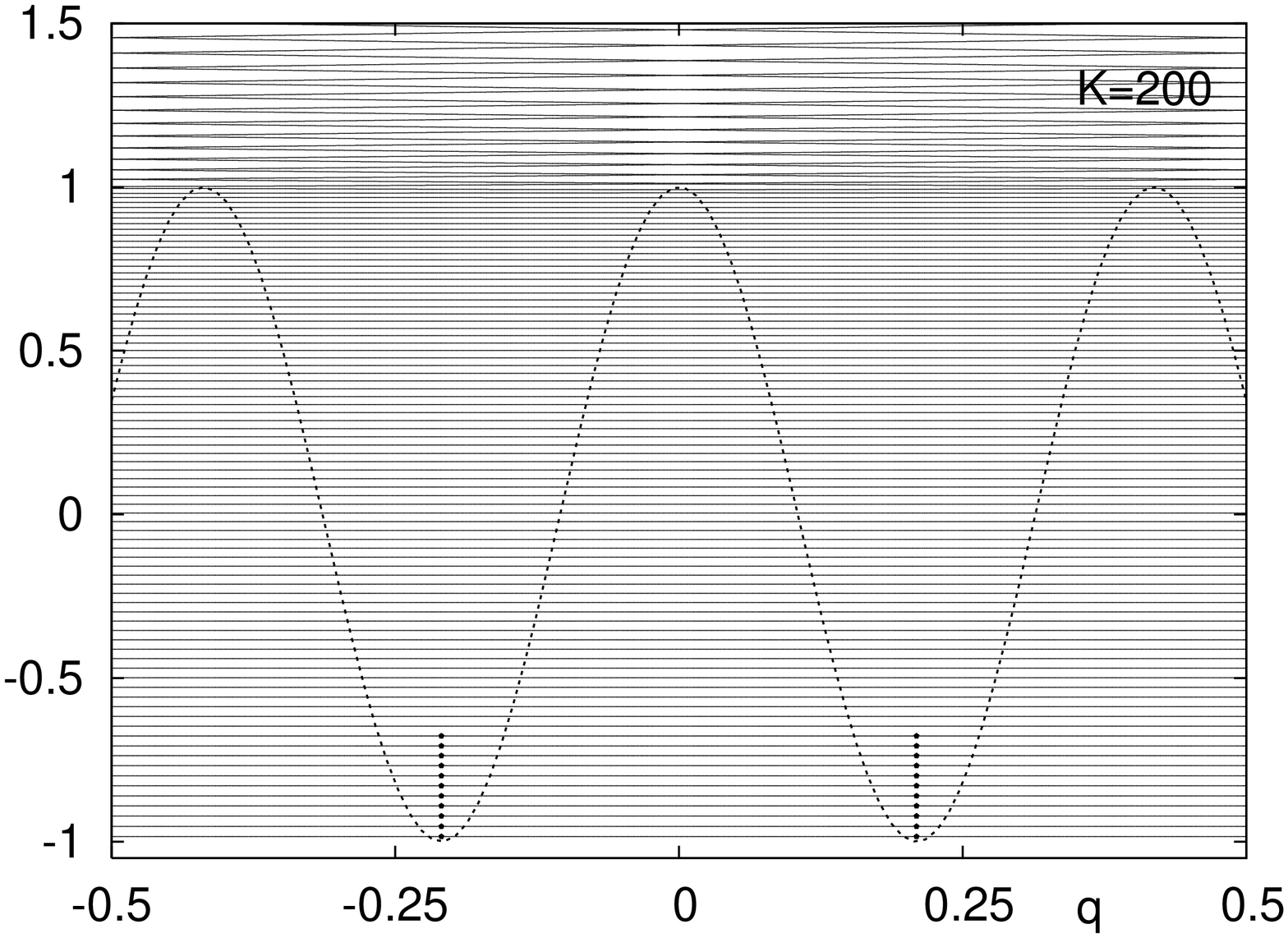}
\includegraphics[width=14.em,angle=-90]{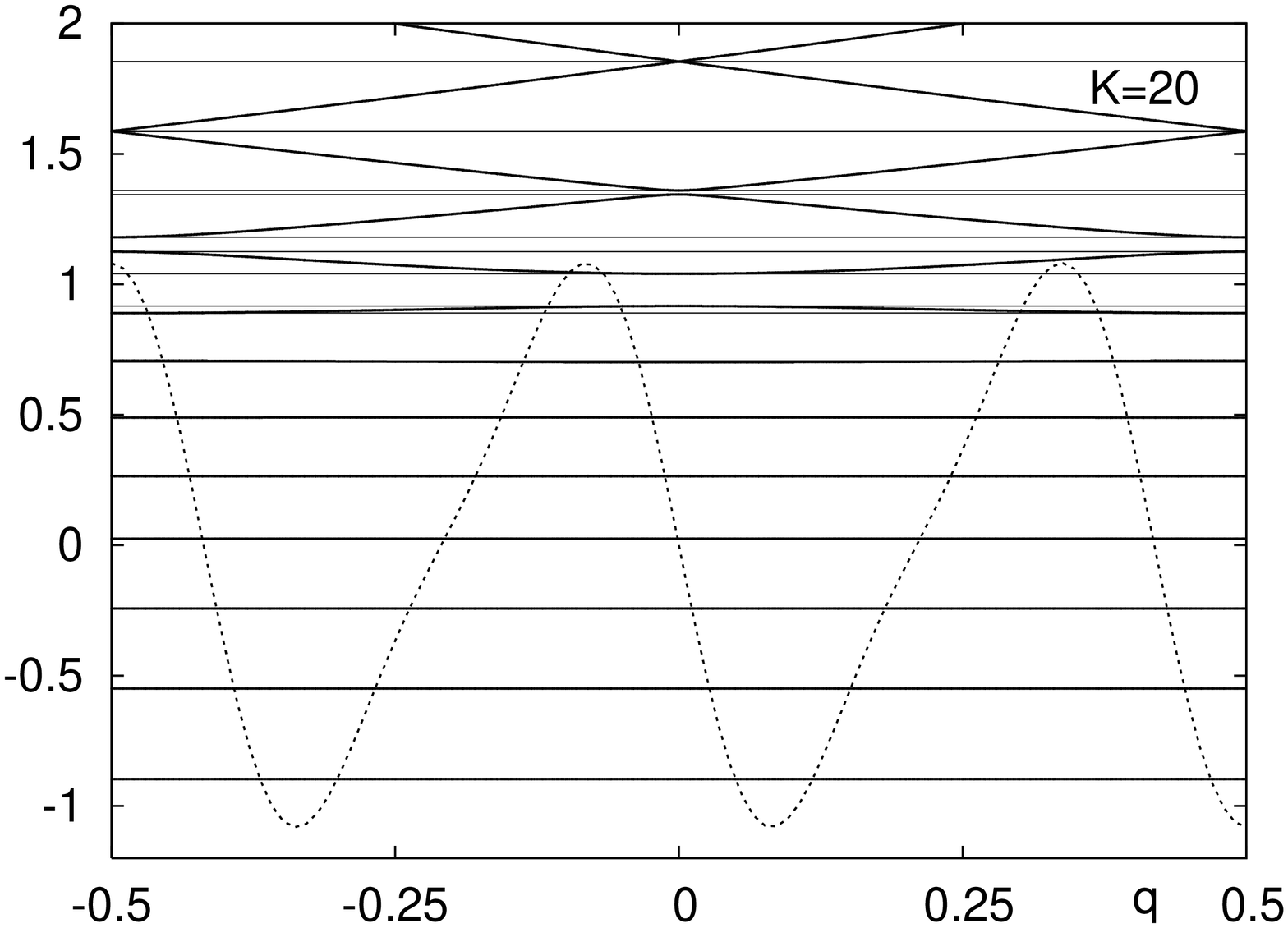}
\caption{
Energy bands for various $\kondobar$ [reduced Kondo parameter
(\ref{kondos})].
The potential profile is plotted to show the number of bands below the
barrier.
Left panels: cosine potential for $\kondobar=1$ (as in
figures~\ref{fig:kandemir} and~\ref{fig:cheleb92}), with the
free-particle parabolic relation $E\propto\q^{2}$, and for
$\kondobar=200$ (figure \ref{fig:suscep-w}), together with the first
levels of the associated anharmonic oscillator [dots;
equation~(\ref{pert-fey})].
Right panels: ratchet potential (\ref{Vrat}) with $\rat=0.44$ for
$\kondobar=10$ and $\kondobar=20$, as in figure~\ref{fig:Q-ratchet}.
}
\label{fig:bands}
\end{figure}

Thermal fluctuations are incorporated by the Klein--Kramers equation,
which for classical particles in periodic potentials was solved by
continued fractions in
\cite{risken,jun93,feretal93,asamazbou99epjb} (using the
$\p$-recurrence).
Here we solve the Caldeira--Leggett quantum master equation
(\ref{WKK:Dxp}) using a large $\kondobar$.
As shown before, when $\V(\x)=-\Vo\cos(\x)$ we have a $3$-term
recurrence which, for the static response, can be written as
\begin{equation}
\label{RR:3:cosine}
\mQ_{\ix}^{-}\mc_{\ix-1}
+
\mQ_{\ix}\mc_{\ix}
+
\mQ_{\ix}^{+}\mc_{\ix+1}
=
0
\end{equation}
with $\mQ_{\ix}^{-}=\mQ_{\ix,\ix-1}$, $\mQ_{\ix}=\mQ_{\ix,\ix}$, and
$\mQ_{\ix}^{+}=\mQ_{\ix,\ix+1}$.
Solving it with the continued-fraction method we reobtain Risken's
impressive graphs for the classical distribution $\Wf(\x,\p)$
\cite[Ch.~11.5]{risken}.
Some of them are displayed in figure~\ref{fig:risken} for weak damping
and various external forces (here $F_{1}\simeq0.064$ and $F_{3}=1$).
At low $F$ the distribution, always periodic along $\x$, has maxima at
the potential minima, while the profile in momenta is a Maxwellian
envelope $\sim\exp(-\p^{2}/2)$.
As $F$ is increased we see the evolution from these {\em locked\/}
solutions ($\llangle\,\p\,\rrangle\simeq0$) to bistability between the
locked and {\em running\/} solutions, and eventually purely running
states $\Wf\sim\exp[-(\p-F/\gamma)^{2}/2]$ are obtained at large
forces.


\subsection{
Hamiltonian quantum case
}


\subsubsection{
Bands and quantum parameters.
}

Having checked the retrieval of the classical results, we now enter
the quantum regime by decreasing $\kondobar\sim\So/\hbar$.
Firstly, it will be useful to get a more quantitative feeling of the
relation of our parameter $\kondobar$ and ``how quantal is the
system''.
To this end we study the bands of the corresponding Hamiltonian
problem (without force).
Inserting the {\em Bloch\/} function
$\Psi_{m\q}(\x)\propto\exp(\iu\q\x)U_{m\q}(\x)$ in the Schr\"odinger
equation, we arrive at the following eigenvalue problem (in scaled
units; section~\ref{sec:WKK:scaled})
\begin{equation}
\label{schrodinger:bloch}
\Big[
-\frac{1}{2}
\Big(
\frac{2\pi}{\kondobar}
\Big)^{2}
\big(
\iu\,\q+\partial_{\x}
\big)^{2}
+
\V(\x)
\Big]
U(\x)
=
E\,
U(\x)
\;,
\quad
-\half\leq \q\leq\half
\;.
\end{equation}
Expanding $U(\x)$ in plane waves (\ref{Vper}) and truncating
$|\ix|$ at some large $\Atr$ one gets a $(2\Atr+1)\times(2\Atr+1)$
matrix which is diagonalised numerically.
Various bands $E(\q)$ pertaining to problems discussed below are
displayed in figure~\ref{fig:bands}.
We see that the rule of thumb is that $\kondobar/2$ is of the order of
the number of bands laying below the barrier.
Finally, the quantum parameter used by Kandemir \cite{kan98} is
$(\kondobar/\pi)^{2}$, while Chen and Lebowitz employed
$\Omega_{q}=\pi/\kondobar$ \cite{cheleb92b,cheleb92c}.
Besides, for a Josephson junction problem we have
$\kondobar=\pi(E_{\rm J}/2E_{\rm C})^{1/2}$, with $E_{\rm J}$ and
$E_{\rm C}$ the Josephson and Coulomb energies.
\begin{figure}
\includegraphics[width=19.em]{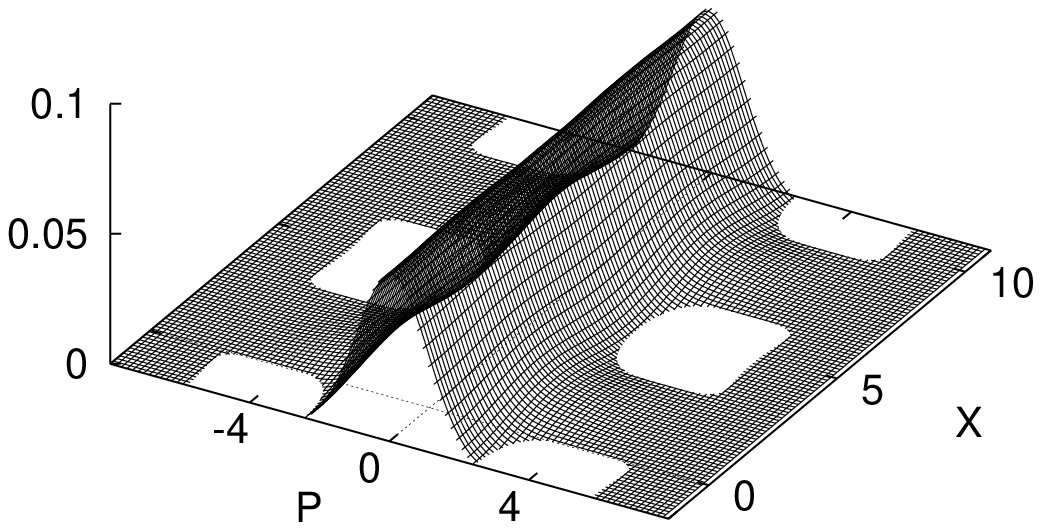}
\includegraphics[width=19.em]{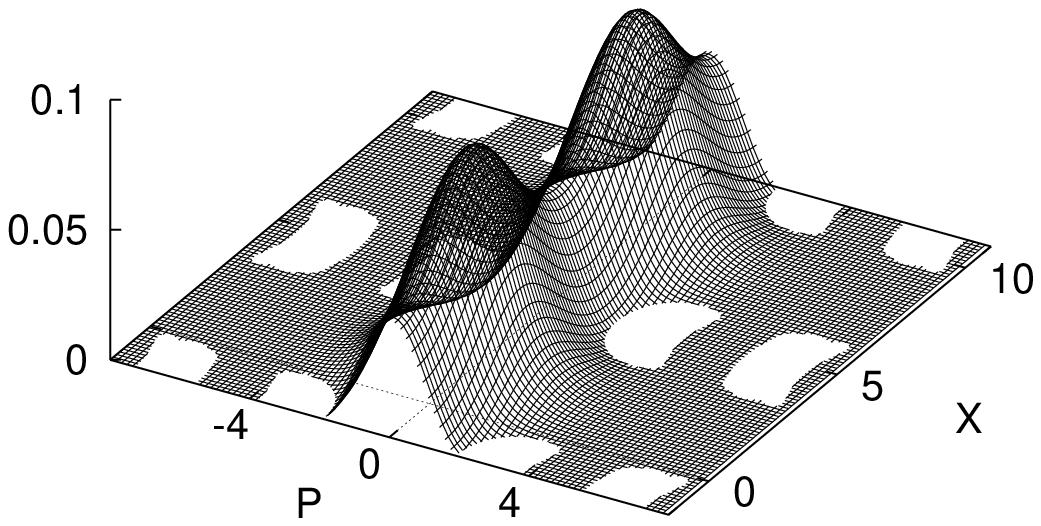}
\caption{
Wigner functions in the absence of dissipation for a particle in a
sinusoidal potential (two periods are displayed).
The reduced Kondo parameter is $\kondobar=1$ (left) and $2$ (right).
The ``islands'' correspond to zones of negative $\Wf$.
}
\label{fig:kandemir}
\end{figure}


\subsubsection{
Wignerian representation.
}

Let us continue for a while in the Hamiltonian case to familiarise
ourselves with the structure of the Schr\"odinger eigenstates in a
phase-space representation.
Kandemir \cite{kan98} obtained analytical approximations for the
eigenfunctions of a particle in a cosine potential in terms of Mathieu
functions.
They were represented through the Wigner function associated to the
density matrix of pure states
$\varrho(\x,\x')=\Psi(\x)\Psi^{\ast}(\x')$.
(In \cite{kan98} various Dirac deltas occurring were regularised with
Gaussians $\delta(\p)\sim(k/\pi)\exp(-k^{2}\p^{2})$ and in the plots
$k=1$ was used; we attain the same setting $T=1$ since our $\p$ is
thermally rescaled.)

Using a very small damping ($\gamma=10^{-6}$) in the Caldeira--Leggett
master equation we reobtain Kandemir's Hamiltonian results for the
ground state (figure~\ref{fig:kandemir}).
We see that at low $\kondobar$ the state is extended or delocalised,
becoming more localised at the potential wells when
$\kondobar\propto\So/\hbar$ is increased.
According to \cite{kan98}, the regions of negative $\Wf$ are
associated to Brillouin zone boundaries.
We have represented a larger range of $\p$ to see how these $\Wf<0$
islands are repeated in the $\p$ direction ($\Wf$ can be as small as
$\Wf\sim-10^{-9}$ bordering the limits of our numerical accuracy).

To see how these structures arise, note that for small $\kondobar$ the
potential can be treated perturbatively
[equation~(\ref{schrodinger:bloch})].
Then, the wave functions have the form
$\Psi_{\q}(\x)
\propto
\e^{\iu\q\x}(1+\lambda\,\e^{+\iu\x}+\lambda^{\ast}\e^{-\iu\x})$,
with $\lambda=\iu(\kondobar/2\pi)^{2}$.
The first term is a pure plane wave, and the rest first order
corrections due to the sinusoidal $\V(\x)$.
Performing the Wigner transform (\ref{wigner:def}) of $\Psi_{\q}(\x)$,
we get for the ground state $\q=0$ (cf.\ \cite[p.~314]{ozoalm98})
%
\begin{eqnarray}
\label{wigner:turkeys}
\Wf(\x,\p)
&=
\delta(\p)
+
|\lambda|^{2}
\big[
\delta(\p-1)
+
\delta(\p+1)
\big]
\nonumber\\
&-
2|\lambda|^{2}
\cos(2\x)\,
\delta(\p)
+
2\iu\lambda
\sin(\x)
\big[
\delta(\p-\half)
+
\delta(\p+\half)
\big]
\;.
\end{eqnarray}
The first line gives the momentum localisation at $\p=0$ (unperturbed)
and at $\p=\pm1$.
The first term in the second line arises from the interference of
$\e^{+\iu\x}$ and $\e^{-\iu\x}$, and is responsible for the weak
modulation of the height of the $\p=0$ bell along $\x$.
This modulation is more intense for larger $\lambda$ (i.e., larger
$\kondobar$).
The last term accounts for the interference between the plane wave
centred at $\p=0$ with those at $\p=\pm1$, and can produce the
negative islands.
Therefore, the above simple functional form captures most features of
the Wigner funtions of figure~\ref{fig:kandemir}.


\section{
Periodic potentials: quantum dissipative case
}

After checking the connection with the classical and Hamiltonian
quantum limits, we finally include dissipation and temperature in
the quantum case.
Here examples of particles in cosine and ratchet potentials will be
discussed.


\subsection{
Quantum Brownian motion in a cosine potential
}
\label{cosQBM}

Transport properties of weakly damped particles in a cosine potential
were studied by Chen and Lebowitz \cite{cheleb92b,cheleb92c}.
They started from the system-plus-bath density matrix, traced out the
bath variables, and performed a perturbative calculation in the
potential height (followed by a resummation) to get $\Ppro$.
For low forces a free-particle like behaviour $\Ppro\propto F/\gamma$
was obtained.
Increasing $F$, the wave-vector associated to $\p$ approaches the
first zone boundary (figure~\ref{fig:bands}).
There, while Landau--Zener tunnelling can bring the particle to the
next band, Bragg scattering reduces the velocity ($\propto\partial
E/\partial\q$).
Eventually, at larger forces, $\p$ corresponding to states inside the
next band become favoured, and the free-particle behaviour is
progressively recovered.
\begin{figure}
\includegraphics[width=19.em]{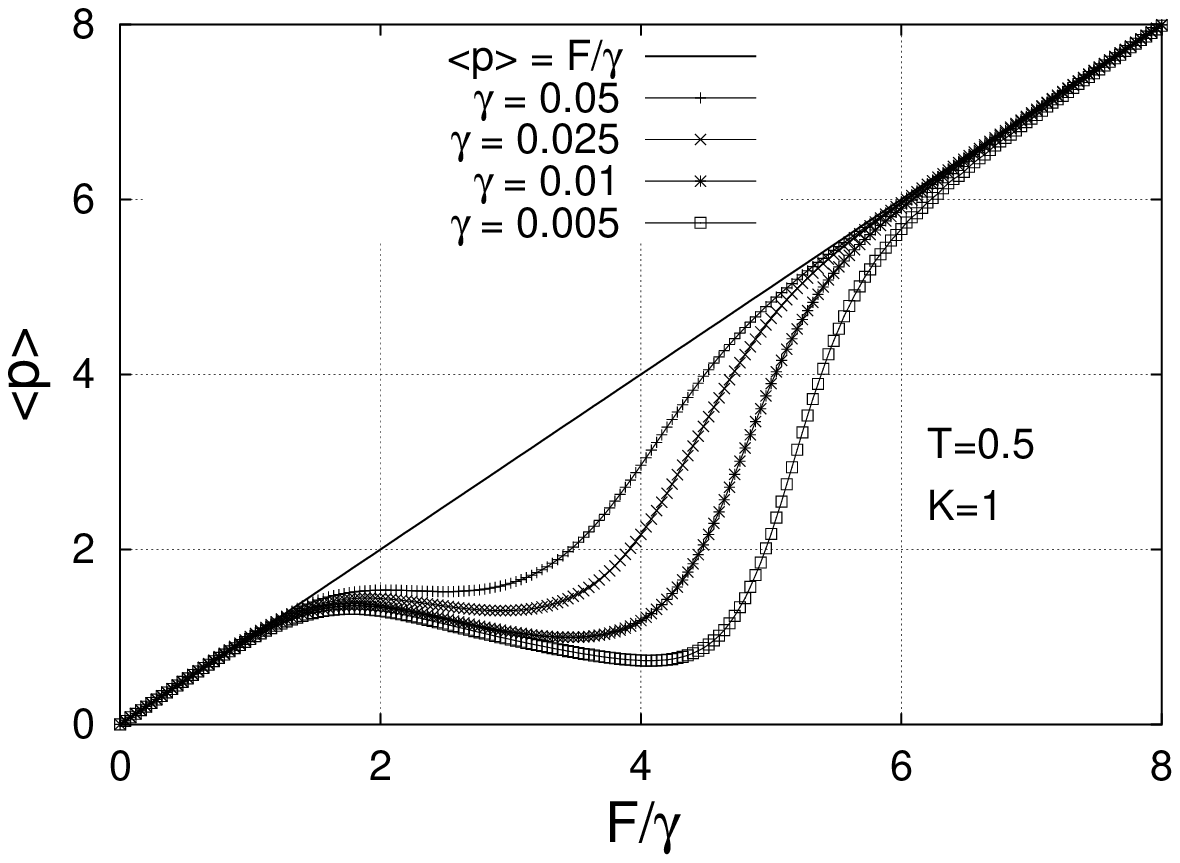}
\includegraphics[width=19.em]{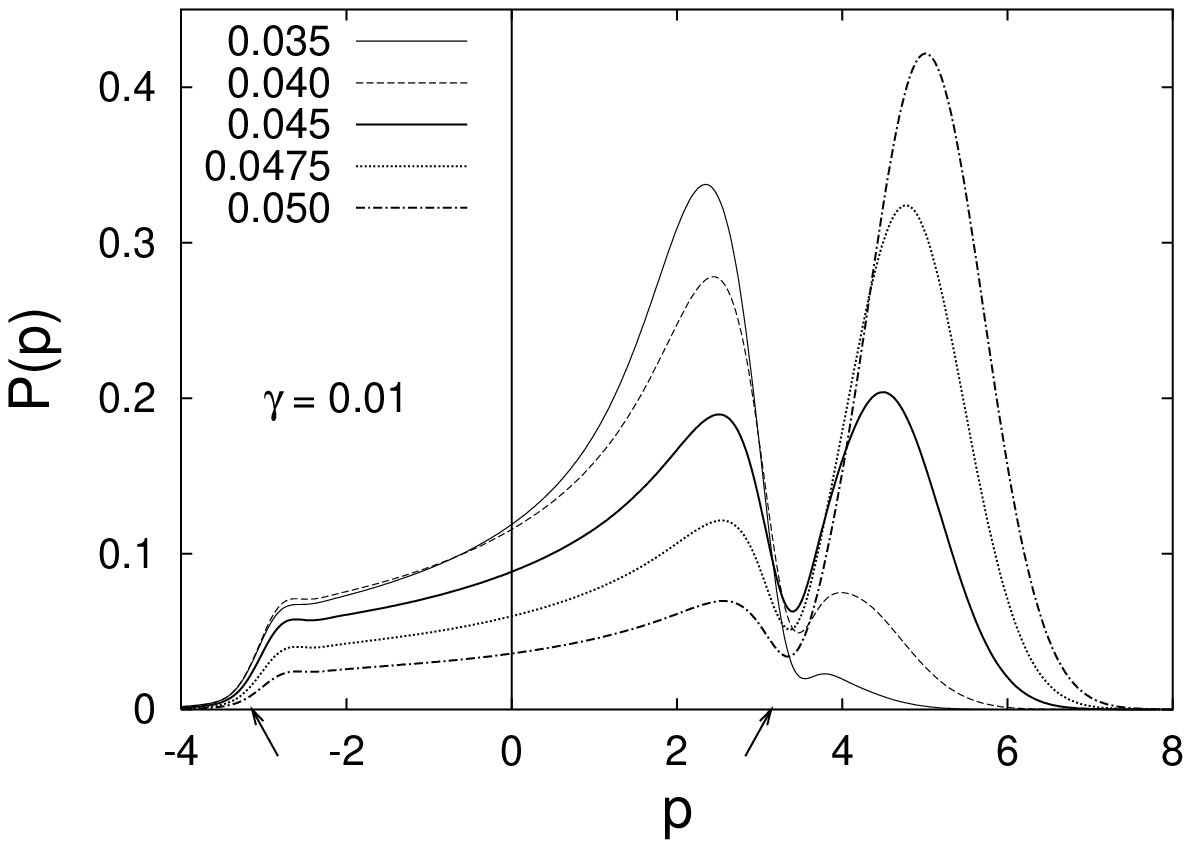}
\includegraphics[width=19.em]{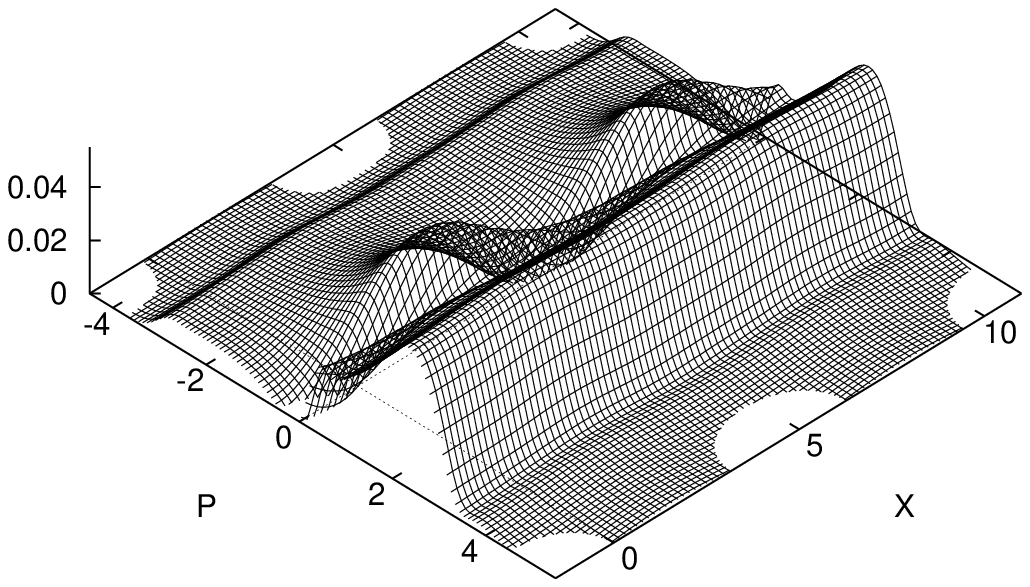}
\includegraphics[width=19.em]{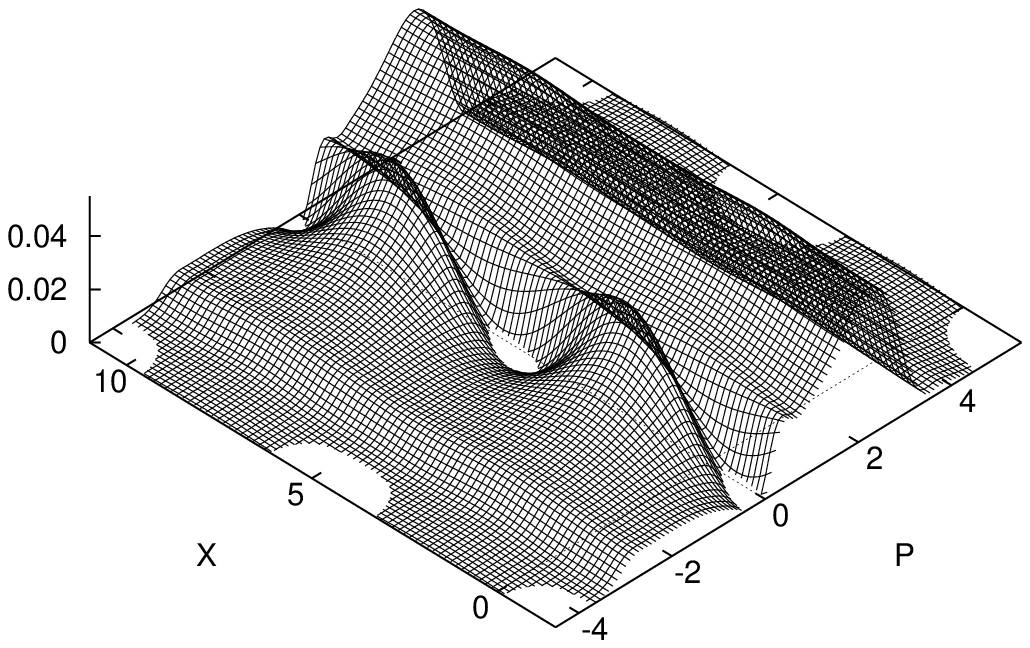}
\caption{
Upper panels.
Left: $\Ppro$ vs.\ $F$ for various dampings at
$T=0.5$ and $\kondobar=1$.
Right: marginal distribution of momenta
$P(p)=\int\!\drm\x\,\Wf(\x,\p)$ for $\gamma=0.01$ at $F/\gamma=3.5$,
$4$, $4.5$, $4.75$, and $5$.
The arrows mark the zone boundaries $|\p|=\pi/\kondobar$.
Bottom panels: two views of $\Wf(\x,\p)$ for $\gamma=0.01$ and
$F/\gamma=3.5$.
}
\label{fig:cheleb92}
\end{figure}

Since high $\kT/\hbar\gamma$ approximations were involved in their
calculations, we set $\Dpp=\gamma\sM\kT$ and $\Dxp=0$ in the master
equation~(\ref{WKK:Dxp}).
Solving it with the continued-fraction method we reobtain the effect
just described (figure~\ref{fig:cheleb92}; here we work in the regime
$\hbar\gamma/\kT\sim0.1$).
Note that this quantum slowing effect is reduced as $\gamma$
increases, since the coupling to the bath ``broadens'' the effective
energy levels.
%
%
This broadening makes less relevant the presence of the band gap
(``bridges'' it) and brings the curves progressively closer to the
free-damped-particle behaviour $\Ppro\propto F/\gamma$.

With our method we get, in addition, the Wigner function (i.e., the
full density matrix).
The distribution of velocities (\ref{Wp}),
$P(\p)=\int\!\drm\x\,\Wf(\x,\p)$, when the curves start to rise again
($F/\gamma\sim4.5$), shows two peaks separated by a minimum
corresponding to the wave vector of the first zone boundary.
Loosely speaking the system shows coexistence of {\em two} quantum
``running'' solutions.
Classical running solutions have a wiggling structure along $\x$, as
the particle slows down near the potential maxima
(figure~\ref{fig:risken}).
The full $\Wf(\x,\p)$ shows that the quantum running solutions have
a nearly straight structure (substrate insensitive).
We also recognise the familiar crescent islands of negative $\Wf$ at
large $\p$ (there $\Wf\sim-0.0003$).
Around $\p\sim0$ a complex structure is developed along $\x$ (``locked
part''), with the maxima deformed and the minima becoming slightly
negative ($\Wf\sim-0.005$).
These features may reflect some interference between the locked and
the nearby running solution.
However, simply adding a plane wave $\e^{+\iu\p_{0}\x}$ to the
$\Psi_{\q}(\x)|_{\q=0}$ used to analyse the Wigner functions of the
Hamiltonial problem [equation~(\ref{wigner:turkeys})] we have not been
able to reproduce such structure.
Anyway, it is erased when integrating over $\x$ to get $P(\p)$.


\subsection{
Quantum Brownian motion in ratchet potentials
}
\label{ratQBM}

We now turn our attention to periodic potentials without spatial
inversion symmetry.
This feature combined with out-of-equilibrium conditions can give rise
to directional motion with net unbiased driving ({\em ratchet
effect}).
This directional motion has been invoked to explain the behaviour of
molecular motors and, with the emergence of nanoscience, opens the way
to build mesoscopic devices and engines based on it.
The theoretical work has concentrated on the classical behaviour and
in the high-damping regime \cite{rei2002}.
The few quantum studies also addressed strong system-bath coupling in
two situations: semiclassical \cite{reigrihan97} and a extreme quantum
case disregarding thermal effects \cite{grietal2002}.
An exception to the large $\gamma$ studies is \cite{schvin2002}, but
there the substrate potential (and the force) were treated
perturbatively.
Here, we shall solve the quantum master equation (\ref{WKK:Dxp}) with
the (non-perturbative) continued-fraction method for particles in
ratchet potentials, taking into account finite damping ($\sim$
non-negligible inertia).

For the ratchet potential (\ref{Vrat}), the Fourier coefficients of
$\V'(\x)$ are $\V_{\pm1}'=-\Vo/2\kT$ and $\V_{\pm2}'=-\rat\Vo/2\kT$
(recall the thermal scaling of $\V$; section \ref{sec:WKK:scaled}).
We use $\rat=0.44$, which smooths a small shoulder that this potential
exhibits on the ``easy'' side (figure~\ref{fig:bands}).
As discussed before, the range of index coupling is $2$, which gives a
$5$-term recurrence.
For the stationary response it can be written as [cf.\
equation~(\ref{RR:3:cosine})]
\begin{equation}
\label{RR:5:ratchet}
\mQ_{\ix}^{--}\mc_{\ix-2}
+
\mQ_{\ix}^{-}\mc_{\ix-1}
+
\mQ_{\ix}\mc_{\ix}
+
\mQ_{\ix}^{+}\mc_{\ix+1}
+
\mQ_{\ix}^{++}\mc_{\ix+2}
=
0
\end{equation}
which can be folded onto a canonical $3$-term recurrence by
introducing appropriate (block) vectors and matrices
(\ref{app:RR-CF}).

The potential considered is the minimal extension of the cosine
potential lacking inversion symmetry, so that the average velocities
can be different for positive and negative forces (vd.\ figure
\ref{fig:cl-ratchet}).
We excite the system with a square-wave force switching alternatively
between $\pm F$, and compute the ``rectified'' current of particles
\begin{equation}
\label{rectified}
\gamma
\Ppro_{\rm r}
= 
\gamma\Ppro_{+F}
+
\gamma\Ppro_{-F}
\;.
\end{equation}
Here $\Ppro_{\pm F}$ are the correponding stationary velocities since
we consider adiabatic conditions ($\omega\to0$).
In what follows we briefly investigate the classical limit
(understudied for finite damping) and then proceed to study quantum
corrections.


\subsubsection{
Classical case.
}

\begin{figure}
\includegraphics[width=19.em]{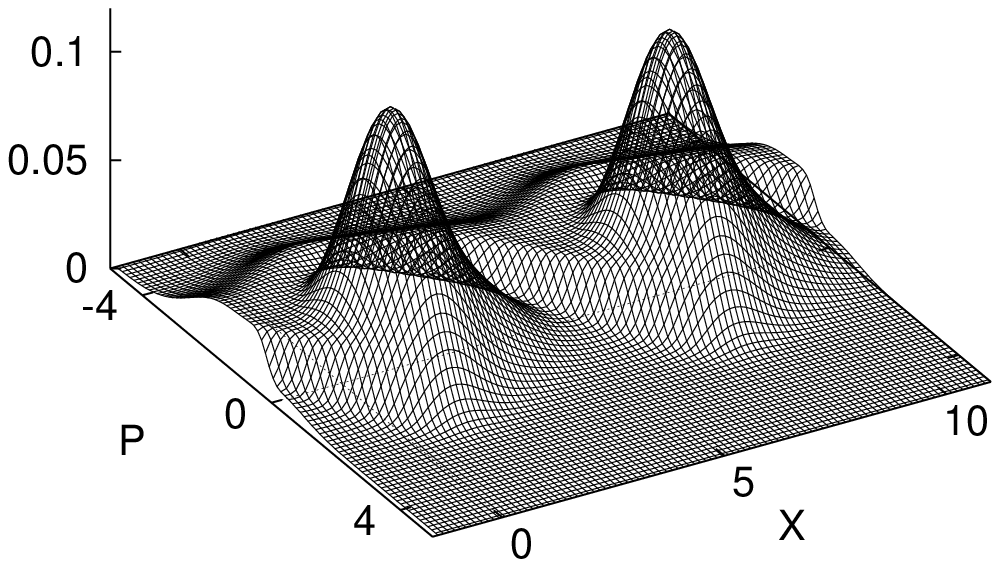}
\includegraphics[width=19.em]{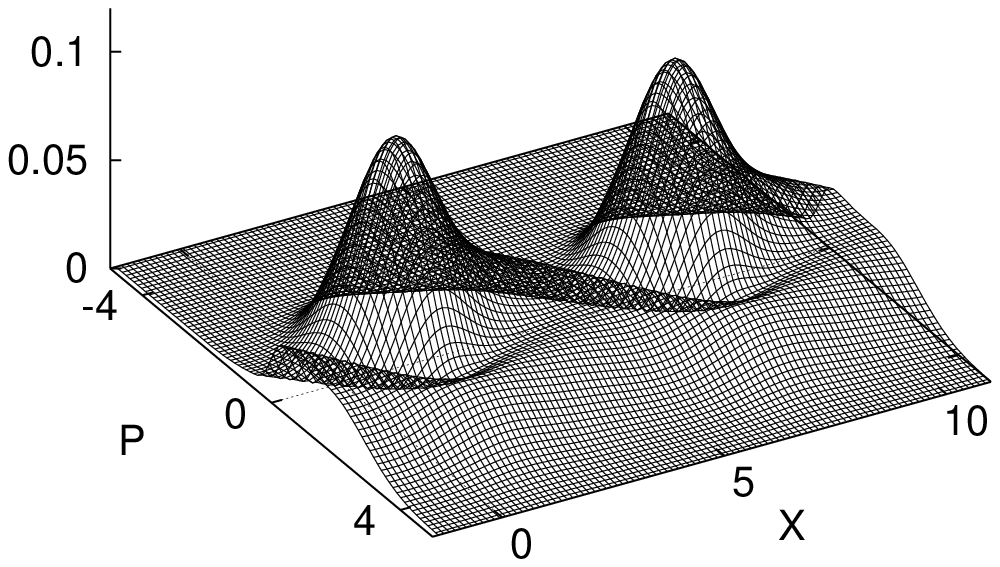}
\includegraphics[width=13.5em,angle=-90]{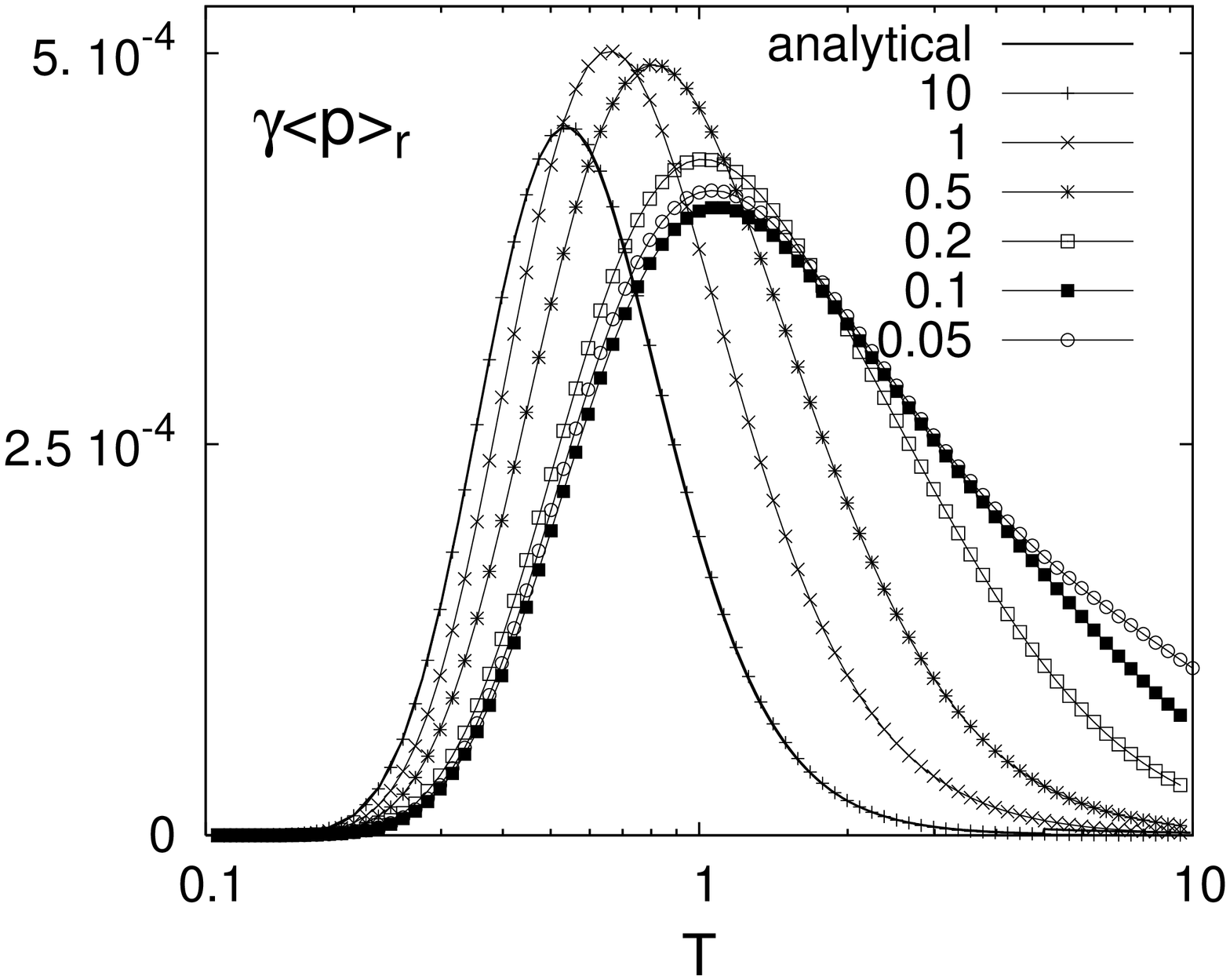}
\includegraphics[width=13.5em,angle=-90]{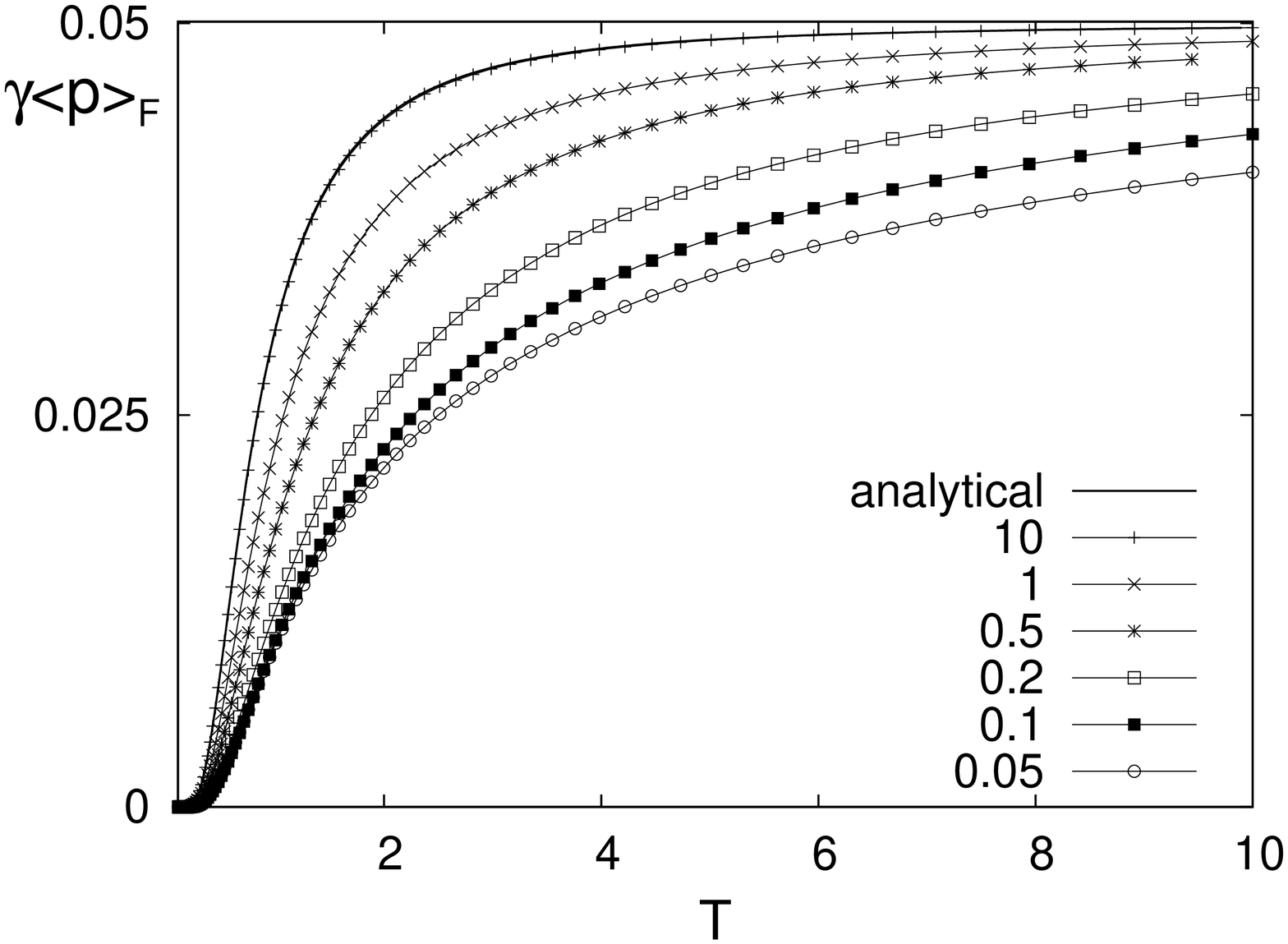}
\caption{
Top panels: Wigner functions for a ratchet potential in the classical
limit ($\kondobar=1000$) with $\gamma=0.05$ and $T=1$.
Two forces are used $F=-0.1$ (left) and $F=0.1$ (right), to show the
lack of symmetry in the response (running part more developed to the
easy side) and the locked regions reflecting the potential profile.
Bottom left: Rectified classical velocity vs.\ temperature for various
values of the damping and force $|F|=0.05$.
Right: $\gamma\Ppro$ vs.\ $T$ for the positive force.
}
\label{fig:cl-ratchet}
\end{figure}

Figure~\ref{fig:cl-ratchet} displays $\Ppro_{\rm r}$ as a function of
$T$, showing the appearance of non-zero rectified velocities (ratchet
effect).
The rectification is optimum at some intermediate temperature (at too
low $T$ there is hardly a response, and at too high $T$ the potential
is irrelevant, so its asymmetry plays no r\^{o}le in either case).
At $\gamma=10$ the results coindice with those obtainable from the
analytical $\Ppro$ for overdamped classical particles
\cite[Ch.~11.3]{risken} (vd.\ also \cite{rei2002,barhankis94})
\begin{equation}
\label{risken:eq}
\fl
\gamma
\Ppro
=
\frac{
2\pi T\,
(1-\e^{-2\pi F/T})
}
{
\int_{0}^{2\pi} \!\! \drm x \,
\e^{-\phi(x)}
\!
\int_{0}^{2\pi} \!\! \drm y \,
\e^{\phi(y)}
-
(1-\e^{-2\pi F/T})
\int_{0}^{2\pi} \!\! \drm x \,
\e^{-\phi(x)}
\!
\int_{0}^{x}    \!\! \drm y \,
\e^{\phi(y)}
}
\end{equation}
with $\phi(\x) = [V(\x)-F\x]/T$.
Inertia, however, broadens the $\Ppro_{\rm r}$ curves and shifts the
maxima to higher $T$ (for the lowest $\gamma=0.05$ the maximum moves
to a slightly lower $T$; vd.\ infra).
This broadening is accompanied by a slower decrease of $\Ppro_{\rm r}$
with $T$.
This is consistent with the asymptotic behaviour of $\Ppro_{\rm r}$
when $T\to\infty$ \cite{schvin2002}.
For overdamped particles it goes as $\Ppro_{\rm r}\sim T^{-4}$, while
for weak damping it decreases only with $\Ppro_{\rm r}\sim
T^{-(17/6)}$ ($17/6\lesssim3$).
We have checked that our results recover these dependences (curves not
shown due to the smallness of the asymptotic $\Ppro_{\rm
r}\sim10^{-6}-10^{-7}$).

In our case $F \lesssim \gamma\sim F_{1}$ (the ``retrapping'' force;
section~\ref{sec:risken}), so that at $T=0$ the attractors are locked
solutions.
Then, the absolute $\gamma\Ppro$ vs.\ $T$ curves start from $\Ppro=0$
at $T=0$, depart from zero as $T$ is increased and evolve towards
$\gamma\Ppro=F$ for $T\to\infty$ (free damped particle).
In the overdamped case this evolution is nearly a step, whereas the
slope of the initial raising decreases with decreasing $\gamma$ (if
$F<F_{1}$).
To understand this, let us assimilate $\Ppro$ to the escape rate
$\Gamma$ (modified by some mean free path), so that
$\gamma\Ppro\sim\gamma\Gamma$.
Kramers' theory \cite{mel91} shows that $\gamma\Gamma$ increases
monotonously with $\gamma$.
Therefore, the $\gamma\Ppro$ vs.\ $T$ curves for finite $\gamma$ must
lay ordered below the overdamped result.
On the other hand, the maximum rectification occurs in the temperature
region where $\gamma\Ppro$ transits from $0$ to $F$.
This intermediate region is narrower (and the slopes steeper) the
higher the damping is (figure~\ref{fig:cl-ratchet}), while the curves
start to raise from zero at a lower $T$.
Therefore, increasing the damping the $\gamma\Ppro_{\rm r}$ curves
will be narrower and the maxima shift to lower $T$, as observed.
Eventually, this also allows to explain the anomalous curve showing
the return ($\gamma=0.05$).
Here $\gamma\sim F$, so that at $T=0$ the system starts close to the
zone of bistability.
Then the particle can perform incursions into the running solution,
increasing the initial slope of $\gamma\Ppro_{\pm F}$ but now when the
damping is lowered, shifting the maximum to a lower $T$.


\subsubsection{
Quantum corrections.
}

\begin{figure}
\includegraphics[width=20.em]{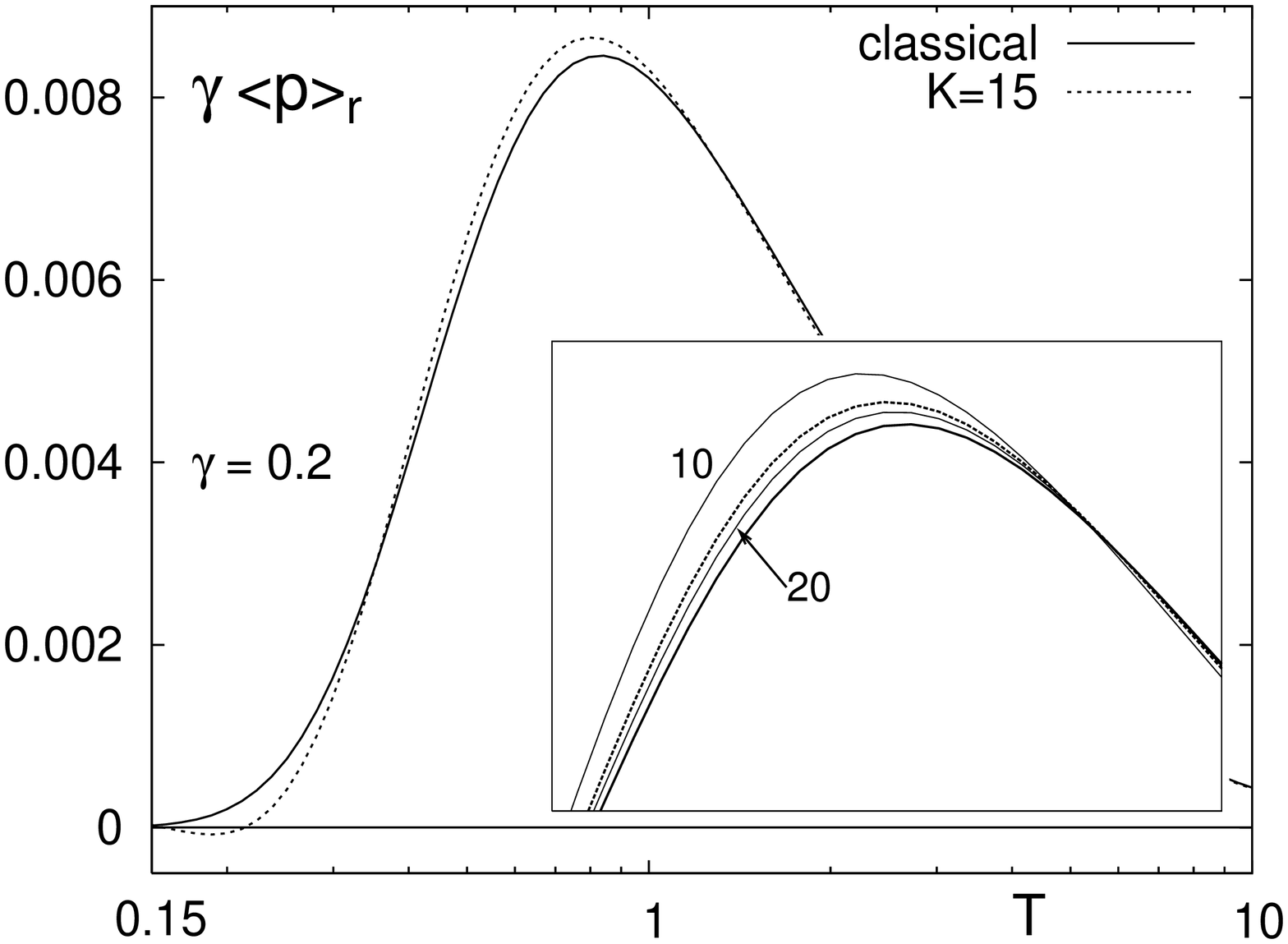}
\includegraphics[width=20.em]{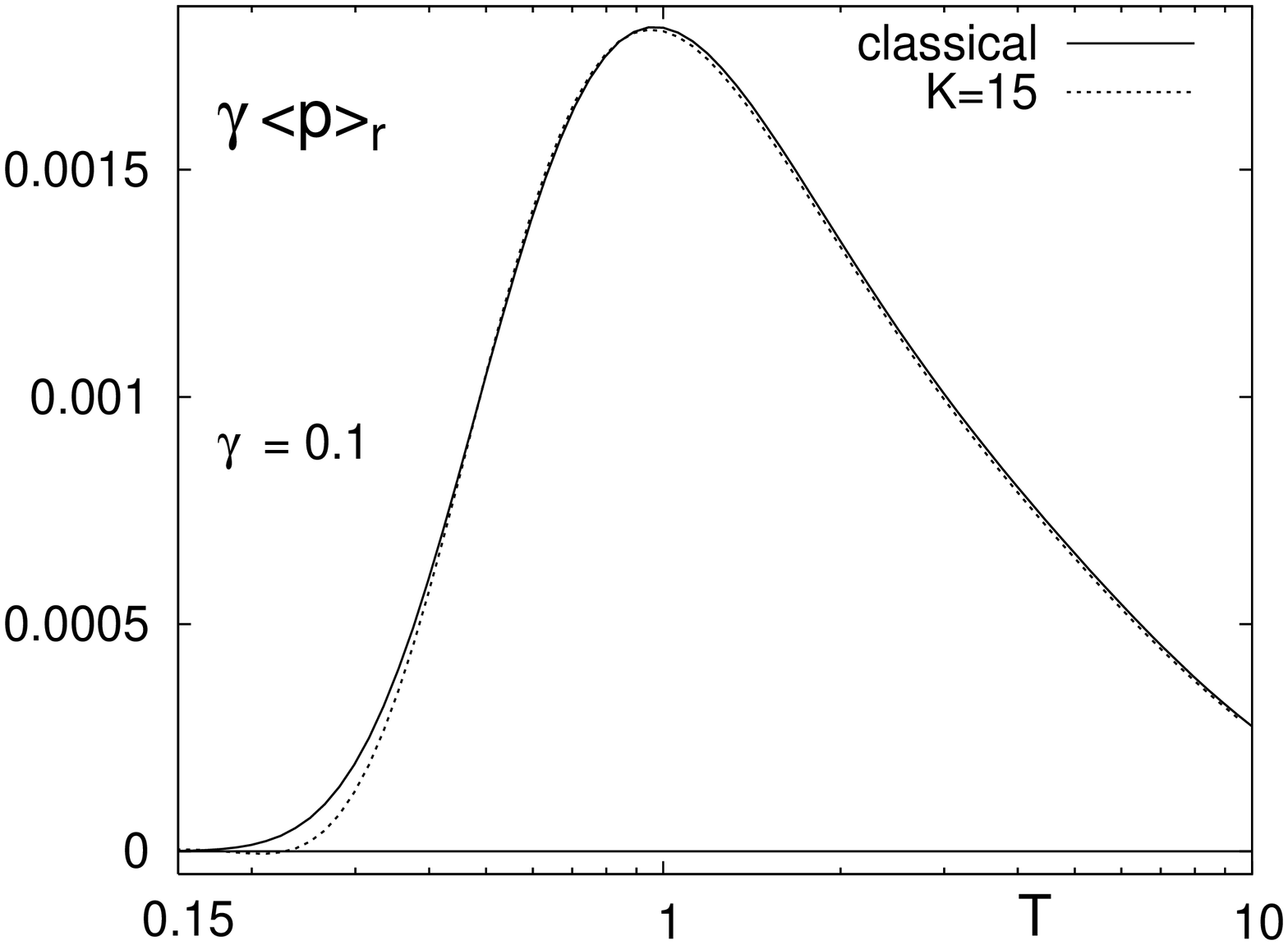}
\includegraphics[width=20.em]{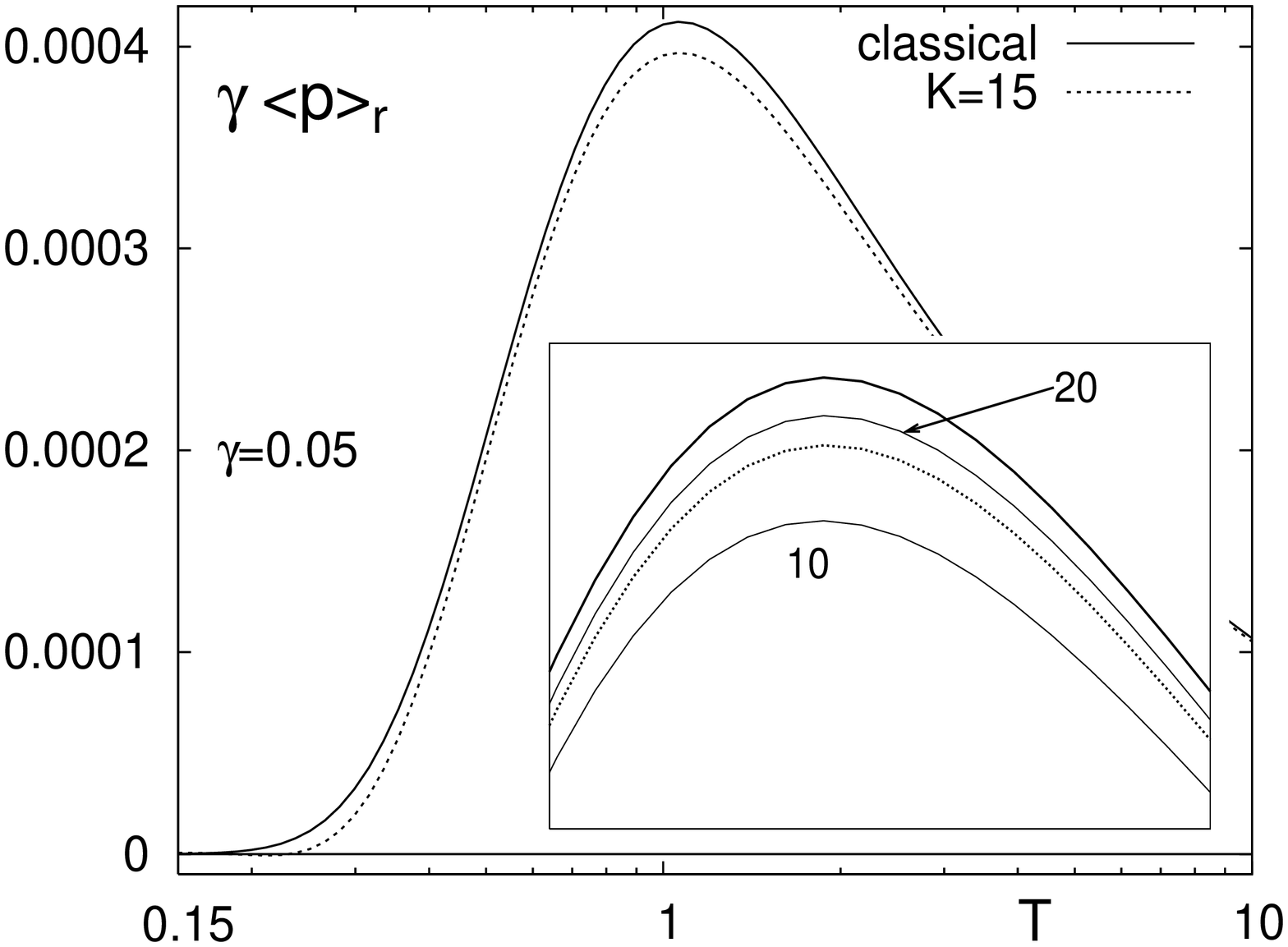}
\includegraphics[width=19.em]{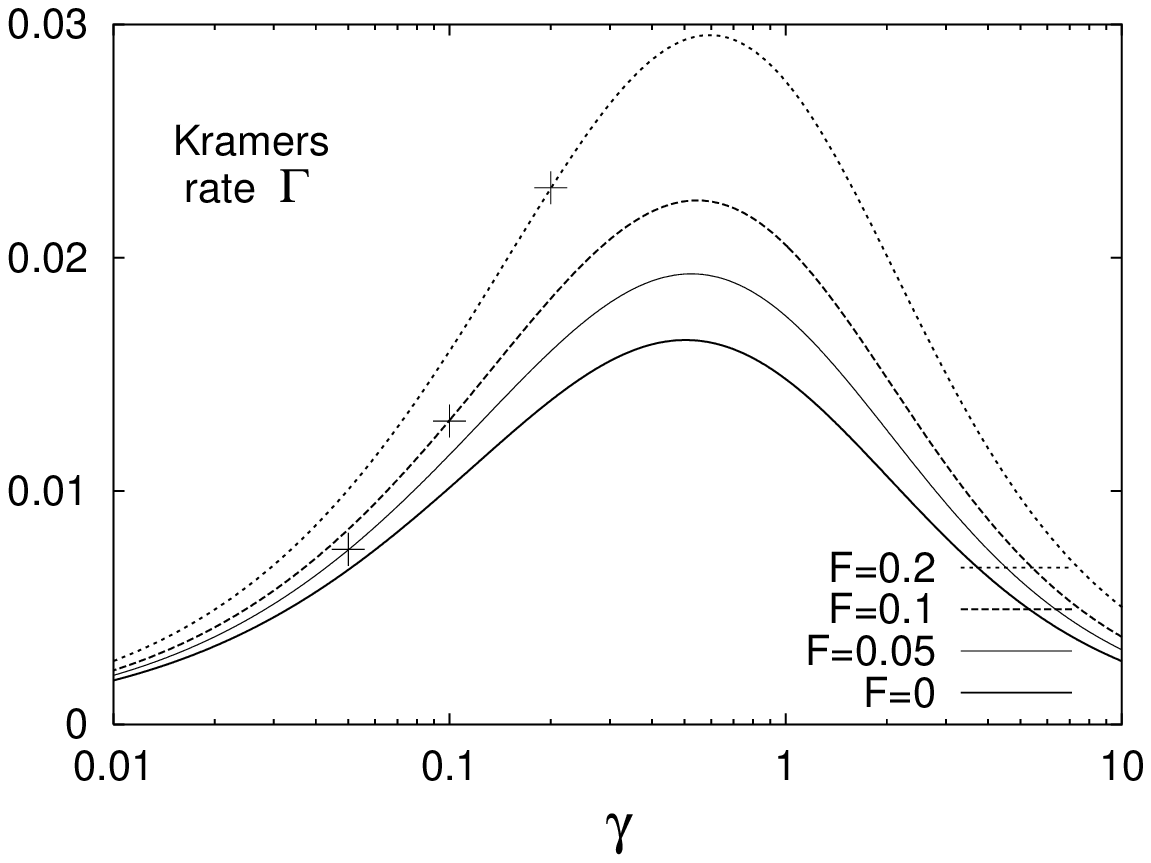}
\caption{
Rectified velocity vs.\ $T$ with $\gamma=0.2,~0.1,~0.05$ (using
$|F|/\gamma=1$) for $\kondobar=15$ and the reference classical curve
($\kondobar=1000$).
The insets of $\gamma=0.2$ and $0.05$ show enlargements of the peaks
(together with results for $\kondobar=20$ and $10$).
Bottom right panel: Kramers classical rate vs.\ $\gamma$ at various
forces (from Mel'nikov--Meshkov formula \cite{mel91}).
The crosses mark the points with $F/\gamma=1$.
}
\label{fig:Q-ratchet}
\end{figure}

After the mandatory exploration of the classical limit, we proceed to
make the system quantal, by decreasing $\kondobar$.
Figure~\ref{fig:Q-ratchet} shows the rectified current with
$\kondobar=15$ (with $7$ bands below the barrier; cf.\ figure
\ref{fig:bands}) for various $\gamma$, together with the reference
classical curves.
We have set $|F|/\gamma=1$ is all cases, to have roughly the same
amount of locked and running components in the solutions.
At high temperatures ($\kT\gg \Vo$) the classical and quantum curves
coincide, as expected.
However, decreasing $T$ down to $\kT\sim\Vo$, the ratchet effect gets
reduced or enhanced, with respect to the classical result, depending
on the dissipation.
Finally, at low temperatures ($\kT\ll\Vo$) the rectification is
always reduced.
%

To understand qualitatively the underlying physics we turn to some
semiclassical results for various related problems.
First, let us consider the {\em reflection/transmission coefficient\/}
for an asymmetric saw-tooth potential \cite[\S~50,~52]{lanlif3}.
The results depend on the energy of the incident quantum particle.
The reflection coefficient for energies higher than the barrier,
$\Refl$, and the underbarrier transmission coefficient, $\Tran$, read
\begin{equation}
\label{Rs}
\Refl
=
\frac{\hbar^{2}/2\sM}{(4\,E)^{3}}
(f_{2} - f_{1} )^{2}
\;,
\qquad
\Tran
=
\exp
\Big[
-
\frac{2}{\hbar}
\Big|
\int_{a}^{b}\!
\drm\x\,\p(\x)
\Big|
\Big]
\end{equation}
where $f_{1}$, $f_{2}$ are the slopes at both sides of the cusp, $E$
the particle energy, 
and $\p(\x)=\sqrt{2\sM[E-\V(\x)]}$.
Suppose now that we have an asymmetric potential, with an easy
direction to the right, and we deform it with forces $\pm F$,
computing the corresponding reflection coefficients $\Refl_{+}$ and
$\Refl_{-}$.
For energies higher than the barrier, the above $\Refl$ gives
$\Refl_{+}/\Refl_{-}<1$ (\ref{app:semicl}).
That is, {\em overbarrier wave reflection\/} is less intense when the
slope of the potential is smaller.
Thus overbarrier transmission (as thermal hopping) is favoured along
the easy direction for energies above the barrier \cite{linetal2002}.
In contrast, using $\Tran$ for energies below the barrier, we find
$\Tran_{+}/\Tran_{-}<1$, so that $\Refl_{+}/\Refl_{-}>1$.
That is, the reflection is more intense in the easy direction or,
equivalently, tunnelling events are favoured in the hard direction
\cite{reigrihan97}.

For a distribution of particle energies one would expect a competition
between these two types of quantum effects.
Then, the enhancement of the rectified velocity under $\pm F$ forcing
for the $\gamma=0.2$ curve follows from the increase of the thermal
escape with $\gamma$ in the intermediate-to-low damping range
$\gamma\leq0.5$ (figure~\ref{fig:Q-ratchet}).
At that damping more escape events occur and the particles launched
over the barrier are subjected to the discussed phenomenon of wave
reflection, which favours net motion along the easy side and hence
increases $\Ppro_{\rm r}$.
Lowering the damping, on the other hand, less escape events are
produced and wave-reflection becomes progressively dominated by tunnel
events (favouring the hard direction).
At $\kT\sim\Vo$ one would expect the tunnel to be thermally assisted
through the higher bands, which are wider.
A simple estimation of channel probabilities comes from multiplying a
band thermal population (using Kramers' rate) by the probability of
tunnel (given by the semiclassical transmission coefficient $\Tran$).
Here the exponential reduction of $\Gamma$ with $E$ compites with the
exponential increase of $\Tran$ for the upper bands
(\ref{app:semicl}).
At $\kT\sim\Vo$ and for $\kondobar=15$ we find that tunnel is indeed
favoured through the upper bands.

To check that the behaviours found at the maxima are systematic, we
increase $\kondobar$ to $\kondobar=20$ ($9$ bands below the barrier)
and reduce it to $\kondobar=10$ ($4$ bands; here the temperatures
around the peak are still inside the range of validity of the master
equation, small $\hbar\gamma/\kT$).
The results (insets) show clearly the tendency to amplify either the
enhancement or the reduction when the system becomes more quantal.

Finally, at low temperatures ($\kT\sim\Vo/5$) thermal activation
becomes ineficient for all dampings.
Then a reduction of the rectification is expected, due to dominance of
below-barrier energies and hence of tunnel events (through the lowest
bands).
Note that the decrease of $\Ppro_{\rm r}$ relative to the classical
results is comparable in the curves shown.
This suggests that we have similar crossover temperatures $T_{0}$ in
this damping range.
($T_{0}$, below which quantum effects dominate, is useful as a measure
of quantum corrections at a given $T$).
This quantity is known exactly in the problem of the escape from a
``quadratic-plus-cubic'' potential
$\V(\x)=\case{1}{2}\sM\wo^{2}\x^{2}\big[1-(2\x/3\xo)\big]$
\cite[Ch.~14]{weiss}
\begin{equation}
\label{T_0}
T_{0}
=
(\hbar\wo/2\pi k_{\rm B})
\big(\sqrt{1+a^{2}}-a\big)
\;,
\qquad
a
=
\gamma/2\wo
\;.
\end{equation}  
In our units
$T_{0}
\to
\big(
\sqrt{
1
+
\gamma^{2}/4
}
-
\gamma/2
\big)/\kondobar$
and in the damping range studied the factor in brackets is close to
one ($\sim0.9,~0.95,~0.975$).
This gives crossovers $T_{0}\sim1/\kondobar\sim0.07$ and hence
comparable reductions of the ratchet effect, as observed.
This reduction of the rectification is the precursor of the current
reversals ($\Ppro_{\rm r}<0$) found in these systems when tunnelling
completely dominates.
Although we find small $\Ppro_{\rm r}<0$ in some cases, the results
cannot be fully trusted.
For example, at $\gamma=0.2$, $\kondobar=15$ and $T=0.2$ we have
$\hbar\gamma/\kT\sim0.5$, which is bordering the limits of validity of
the quantum master equation employed here.


\section{
Periodic potentials: dynamical response
}
\label{sec:dynamics}

We conclude with the study of the (non-adiabatic) dynamical response
to oscillating forces $F(t)=\Delta F\cos(\omega t)$.
Then, the coefficients of the expansion (\ref{W:expansion:xp}) of the
Wigner function are periodic functions of $t$, so they can be Fourier
expanded as follows:
\begin{equation}
\label{C:fourier}
\ec(t)
=
\ec^{(0)}
+
\sum_{\iw=1}^{\infty}
\Big(\frac{\Delta F}{2}\Big)^{\iw}
\Big(
\ec^{(\iw)}
\e^{+\iu\,\iw\omega t}
+
\ec^{(-\iw)}
\e^{-\iu\,\iw\omega t}
\Big)
\;.
\end{equation}
To lowest order in the probing field, the static part of the
corresponding vectors [equation~(\ref{dcdt:vectors:matrices:x})] obeys
a recurrence relation of the type (vd.\ \ref{app:pert:ac})
\begin{equation}
\label{3RR:statics}
\mQ_{\ix}^{-}
\mc_{\ix-1}^{(0)}
+
\mQ_{\ix}
\mc_{\ix}^{(0)}
+
\mQ_{\ix}^{+}
\mc_{\ix+1}^{(0)}
=
0
\;.
\end{equation}
The equations for the harmonics have the form ($\mI$ is the identity
matrix)
\begin{equation}
\label{3RR:harmonics}
\mQ_{\ix}^{-}
\mc_{\ix-1}^{(k)}
+
(\iu\,\iw\omega\mI
+
\mQ_{\ix})
\mc_{\ix}^{(k)}
+
\mQ_{\ix}^{+}
\mc_{\ix+1}^{(k)}
=
-\mF_{\ix}
\end{equation}
with the ``forcing'' $\mF_{\ix}$ involving the previous order results.
Specifically, $\mF_{\ix}\sim\Delta\mQ_{\ix}\mc_{\ix}^{(k-1)}$, with
$\Delta\mQ_{\ix}$ the part of $\mQ_{\ix}$ corresponding to $\Delta F$,
namely
\begin{equation*}
\Delta\mQ_{\ix}
=
\left(
\begin{array}{ccccc}
0
\!&\!
\sqrt{1}\,\etap\Delta F
\!&\!
0
\!&\!
0
\!&\!
\ddots
\\[0.ex]
\sqrt{1}\,\etam\Delta F
\!&\!
0
\!&\!
\sqrt{2}\,\etap\Delta F
\!&\!
0
\!&\!
\ddots
\\[0.ex]
0
\!&\!
\sqrt{2}\,\etam\Delta F
\!&\!
0
\!&\!
\sqrt{3}\,\etap\Delta F
\!&\!
\ddots
\\[0.ex]
0
\!&\!
0
\!&\!
\sqrt{3}\,\etam\Delta F
\!&\!
0
\!&\!
\ddots
\\[0.ex]
\ddots
\!&\!
\ddots
\!&\!
\ddots
\!&\!
\ddots
\!&\!
\ddots
\end{array}
\right)
\end{equation*}
[vd.\ \ref{app:matrices} for the general $\mQ_{\ix\ixp}$; the
$\Delta\mQ_{\ix}$ actually entering in $\mF_{\ix}$ is divided by
$\Delta F$, cf.\ equations (\ref{EOM:generic}) and
(\ref{coefficients:a})].
Equations (\ref{3RR:statics}) and (\ref{3RR:harmonics}) can be solved
sequentially with the continued-fraction method.

\begin{figure}
\includegraphics[width=19.em]{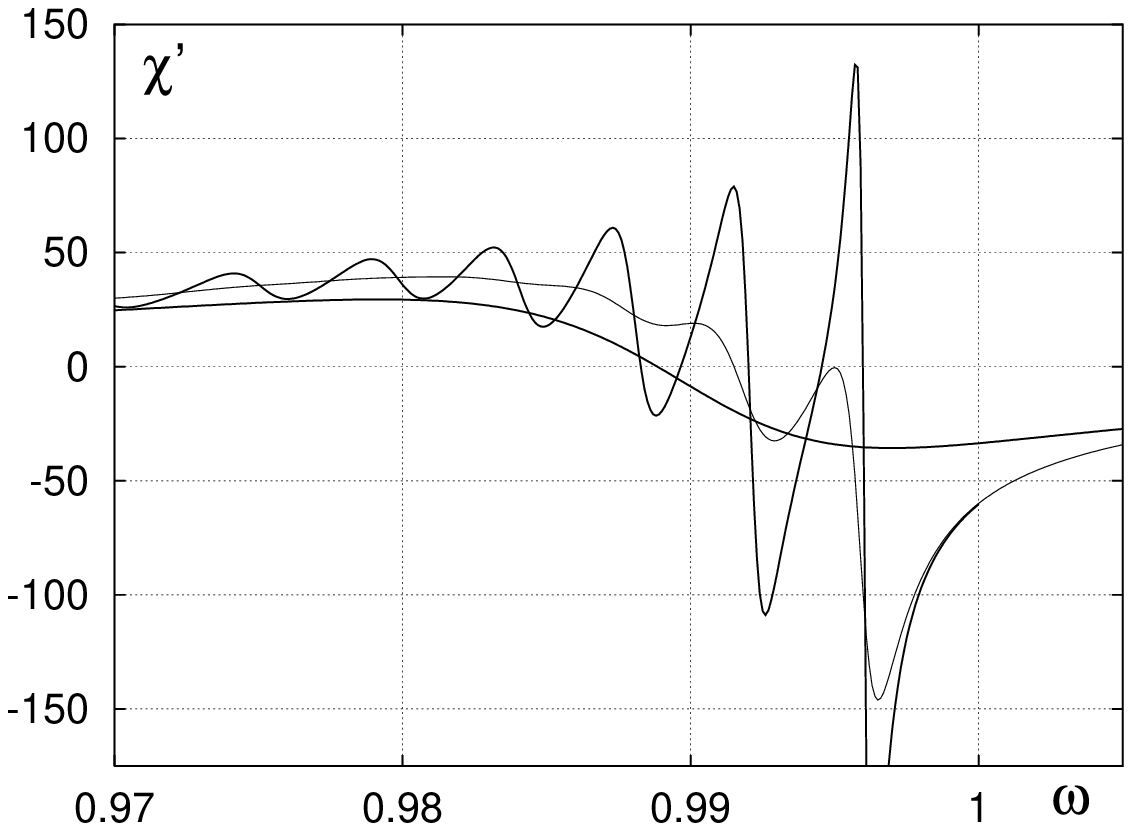}
\includegraphics[width=19.em]{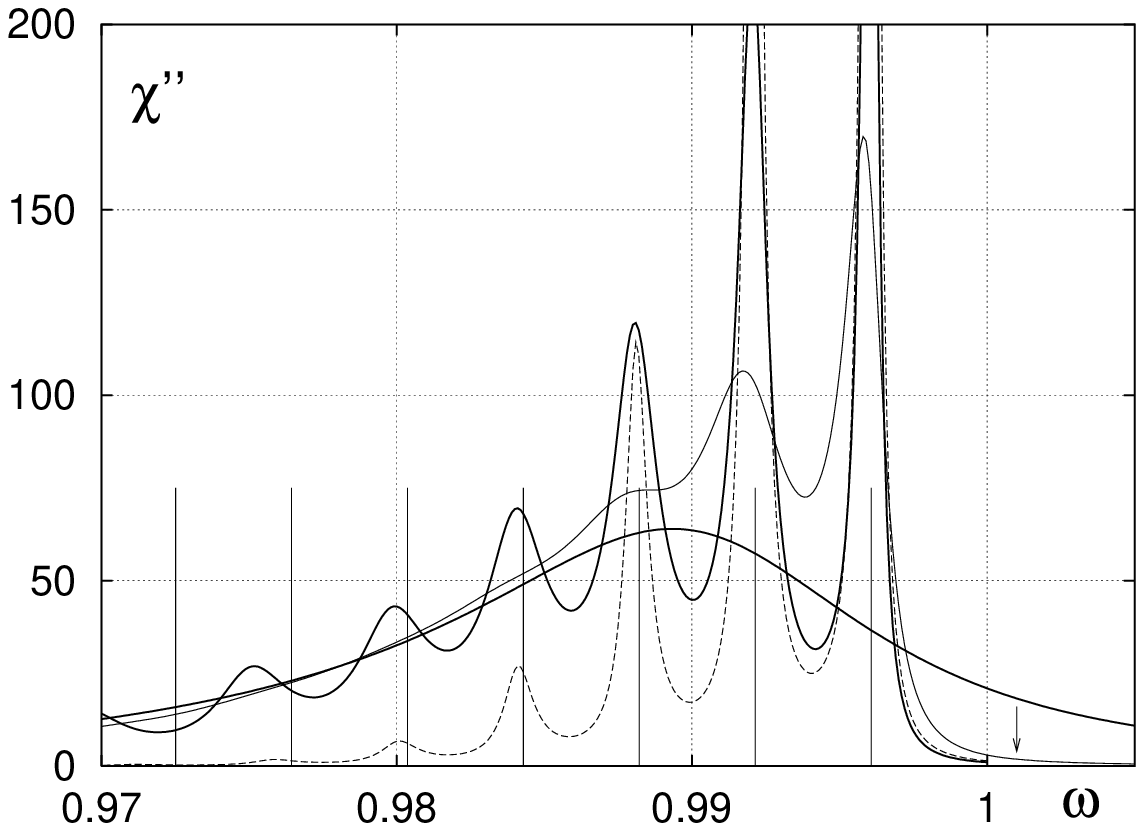}
\caption{
Linear dynamical susceptibility vs.\ frequency at $T=0.05$ and
$\kondobar=200$ with $\gamma=0.01$ ($\sim$ classical), $0.0003$ and
$0.0001$ (less to more peaked curves).
Left panel: real part.
Right panel: imaginary part (dashed line, $\gamma=0.0001$ but halving
$T=0.025$).
Vertical lines: loci of the transition frequencies of the associated
quadratic-plus-quartic anharmonic oscillator
[equation~(\ref{delta-pert-fey})].
}
\label{fig:suscep-w}
\end{figure}

Figure~\ref{fig:suscep-w} shows the linear susceptibility ($\iw=1$)
computed in this way for a particle in a {\em cosine\/} potential as a
function of $\omega$.
(Here we return to safe ground because
$\hbar\gamma/\kT\sim10^{-2}$--$10^{-4}$;
for related dielectric and Kerr relaxation curves of a quantum rotator
vd.\ \cite{tithou99} and references therein.)
For the largest damping the results correspond to the classical limit.
There the line-shape $\chi''(\omega)$ broadens and extends to
frequencies lower than the oscillation frequency near the bottom of
the potential wells ($\omega/\wo=1$ in our units).
For low damping, this classical {\em spreading of oscillation
frequencies\/} has an important contribution from the amplitude
dependence of the period of oscillation in anharmonic potentials
\cite{risken,jun93}.
This non-dissipative contribution to $\chi''(\omega)$ is also known
in classical spin problems, where the precession frequency depends on
$S_{z}$ due to the magnetic anisotropy \cite{gekh83e,garishpan90e}.

The quantum regime is approached by decreasing the Kondo parameter
$\kondo\sim\gamma\kondobar$ [equation~(\ref{kondos})].
Reducing $\gamma$ with a fixed $\kondobar$ we find that the
$\chi(\omega)$ curves develop a multipeaked structure.
These peaks could be vaguely seen as the mentioned nonlinear
oscillations becoming quantised.
To give content to this statement, we investigate perturbatively the
effect of the anharmonic terms of a cosine potential (those beyond
$\propto\x^{2}$) on the energy levels of the harmonic oscillator part.
Since the main contribution should come from the quartic term, we
consider the potential
$\V(\x)=\case{1}{2}\sM\wo^{2}\x^{2}+b\,\x^{4}$.
First-order perturbation theory gives the eigenvalues
\cite[Ch.~6.3]{feynman-statmech}
\begin{equation}
\label{pert-fey}
E_{m}
\simeq
\hbar
\wo 
\left(
m
+
\half
\right)
+
3b\,
\left (
\hbar/2\sM\wo
\right )^{2}
\left (
2m^{2}
+
2m
+
1
\right )
\;.
\end{equation}
Expanding the cosine potential $\V(\x)=-\Vo\cos(\x/\xo)$, we identify
$\sM\wo^{2}=\Vo/\xo^{2}$ and $b=-\Vo/(\xo^4\,4!)$ ($<0$, soft
anharmonicity).
Then we can compute the level spacing
\begin{equation}
\label{delta-pert-fey}
\Delta
E_{m+1,m}
=
\hbar\wo
\bigg[
1
-
\frac{\hbar\,(m+1)}{8\sM\wo\xo^{2}}
\bigg]
\end{equation} 
which has acquired a dependence on the level index $m$ absent in the
harmonic case.
The associated transition frequencies $E_{m+1,m}/\hbar$ are plotted in
figure~\ref{fig:suscep-w} [scaled as
$\omega_{m+1,m}/\wo=1-\pi(m+1)/4\kondobar$].
They agree well with the location of the main peaks of the quantum
$\chi''(\omega)$ curves, supporting their relation with the
nonlinearity of the potential.
Furthermore, the contribution from transitions between the higher
levels should be reduced when they are less populated.
Decreasing $T$, we indeed see that the corresponding peaks (located at
lower $\omega$) are substantially reduced.
In the suggestive language of nonlinear oscillations we would say that
lowering the temperature the large amplitude oscillations (having
longer period) become less probable.

The applicability of the quantum nonlinear oscillator picture is
grounded on the flattness of the lowest bands at the $\kondobar$
considered, which are well approximated by constant levels (vd.\
figure~\ref{fig:bands}).
We have computed the semiclassical width of the bands in a cosine
potential in \ref{app:semicl}.
An upper bound for the width [equation~(\ref{Wb:approx:top})] reads in
dimensionless units
$W_{\rm b}=(4/K)\exp\big[-\case{1}{2}\kondobar(1-\varepsilon)\big]$,
with $\varepsilon=E/\Vo$ the band mean energy relative to the
potential amplitude.
For moderate $\kondobar$ this bound gives in fact very flat bands,
specially the deepest ones ($E\to-\Vo$).
This also implies that the width of the $\chi''(\omega)$ peaks has a
negligible contribution from the intrinsic level width, being
dominated by the damping and thermal broadening.


\section{
Discussion
}

Traditionally, the continued-fraction method has been employed to
solve Fokker--Planck equations for few-variable classical systems
(translational and rotational problems).
When compared with numerical simulations of the Langevin equation,
the method has several {\em shortcomings\/}: (i) it is quite specific
of the problem to be solved (one needs to recalculate the recurrence
coefficients for each potential), (ii) its convergence and stability
depend on the parameters of the problem (and can be poor in some
ranges), and (iii) it does not return ``trajectories'', which in the
simulations provide helpful insight.
Nevertheless, when the method can be used its {\em advantages\/} are
most valuable: (i) it is free from statistical errors, (ii) is
essentially nonperturbative, (iii) specially apt to get stationary
solutions (static and dynamic), (iv) high efficiency (allowing to
explore parameter ranges out of the reach of the simulation), and (v)
it gives the distribution $\Wf$, which certainly compensates for the
lack of trajectories.

For these reasons, and the lack of quantum Langevin simulations, it
was worth to develope continued-fraction methods for quantum
dissipative systems.
This had been done for problems of spins in a thermal bath and in
quantum nonlinear optics.
In this article we have discussed in detail the adaptation of the
continued-fraction method to tackle quantum master equations in
phase-space problems, taking advantage of the parallels with the
classical case provided by the Wigner formalism.
We have seen that the quantum extension of the continued-fraction
method is more problem dependent (except for polynomial potentials),
due to the necessity of using the $\x$-recurrence of the quantum
hierarchies and finding suitable bases.
Nevertheless, it inherits from the classical methods most of the
abovementioned advantages, in particular, the obtaining of the Wigner
function, from which any observable can be computed.
Furthermore, the eigenvalue spectrum of the system, although helpful
in understanding the physics, is not necessarily required.
This is advantageous when dealing with non-bounded Hamiltonians,
continuous spectra, etc.
Finally, by changing the appropriate quantum parameter, the connection
with the classical results is attained in a natural way.

Note that ``continued-fraction'' is a quite generic term (like
``series expansion''), which can be found in many different contexts,
and in particular in works dealing with quantum dissipative systems
(see, e.g., \cite{balleetog2003} and references therein).
However, in those cases one usually obtains some continued-fraction
expression for a certain quantity (e.g., the linear susceptibility,
memory functions, etc.), whereas we have obtained the {\em complete
solution of the quantum master equation\/} by continued-fraction
methods.
The approach allows to study the interplay of quantum phenomena,
nonlinearity, thermal fluctuations, and dissipation in several
classes of systems.
This is important, because methods optimised for one of the sides
(e.g., nonlinearity or quantal behaviour) tend to perform poorly for
the others (e.g., fluctuations or dissipation).
In addition, the visualization of $\Wf(\x,\p)$ plus the knowledge of
the classical phase-space structure (orbits, separatrices, attractors,
etc.) can provide valuable insight in difficult problems.
Finally, the main limitations of the approach are those that the
starting master equations may have.
We have considered the celebrated Caldeira--Leggett quantum master
equation, but the method can be applied to a class of equations of
this type, and to possible generalisations.

We have implemented the method in the canonical problem of quantum
Brownian motion in periodic potentials.
We have considered both the cosine and ratchet potentials.
For the former we have recovered and extended a number of classic
results and interpreted them on the basis of the Wigner formalism.
Results for the dynamics under oscillating forcing have also been
obtained, illustrating the modification of the nonlinear effect of
spreading of oscillation frequencies by quantum phenomena.
For particles in ratchet potentials we have studied, under adiabatic
conditions, the effects of finite damping on the rectified velocity.
Taking into account the competition of thermal hopping, overbarrier
wave reflection and tunnel events, together with semiclassical
analytical results, has allowed us to understand in great detail the
physics of these systems.


\ack
This work was supported by DGES, project BFM2002-00113, and DGA, grant
no.\ B059/2003.
We acknowledge F.\ Falo, J.~J.\ Mazo, and A. P. Piazzolla for useful
discussions and valuable inspiration.


\appendix


\section{
Solving recurrence relations by continued fractions
}
\label{app:RR-CF}

Here we briefly discuss the solution of recurrence relations by
continued-fraction methods.
We first consider $3$-term recurrences and then differential
recurrences, vector-matrix cases, folding of larger-coupling
recursions into $3$-term ones, and the problem of the intial
conditions.
(For a brief description of the relation of continued fractions,
series expansions, recursions, and orthogonal polynomials see
\cite{benmil94}.)

Suppose we have the simplest case of a $3$-term recurrence relation of
the form
%
\begin{equation}
\label{RR}
\eQ_{\ir}^{-}\ec_{\ir-1}
+
\eQ_{\ir}\ec_{\ir}
+
\eQ_{\ir}^{+}\ec_{\ir+1}
=
-\eF_{\ir}
\;,
\qquad
\ir
=
0,\pm1,\pm2,\ldots
\end{equation}
with  $\eQ_{\ir}$ and $\eF_{\ir}$ given quantities.
To obtain the $\ec_{\ir}$ one inserts in (\ref{RR}) the ans\"atze
\cite{risken,junris85} (upper and lower signs for $\ir>0$ and $\ir<0$)
%
\begin{equation}
\label{C:risken:com}
\ec_{\ir}
=
\eS_{\ir}\ec_{\ir\mp1}+\ea_{\ir}
\end{equation}
getting the following ladder coefficients $\eS_{\ir}$ and shifts
$\ea_{\ir}$
%
\begin{equation}
\label{S:a}
\eS_{\ir}
=
-\frac
{\eQ_{\ir}^{\mp}}
{\eQ_{\ir}+\eQ_{\ir}^{\pm}\eS_{\ir\pm1}}
\;,
\qquad
\ea_{\ir}
=
-\frac
{\eF_{\ir}+\eQ_{\ir}^{\pm}\ea_{\ir\pm1}}
{\eQ_{\ir}+\eQ_{\ir}^{\pm}\eS_{\ir\pm1}}
\;.
\end{equation}
If the recurrence is finite or the $\ec_{\ir}$ decrease quickly with
$|\ir|$ (e.g., when they are coefficients of some judiciosly chosen
expansion), we can truncate at some large $|i|=\Itr$, by setting
$\eS_{\pm\Itr}=0$ and $\ea_{\pm\Itr}=0$.
Next we {\em iterate downwards\/} in (\ref{S:a}) getting $\eS_{\ir}$
and $\ea_{\ir}$, down to $\ir=0$.
Then, to get the $\ec_{\ir}$ from (\ref{C:risken:com}), we need
$\ec_{0}$, which is obtained from
\begin{equation}
\label{condini}
\big(
\eQ_{0}^{-}
\eS_{-1}
+
\eQ_{0}
+
\eQ_{0}^{+}
\eS_{1}
\big)
\,
\ec_{0}
=
-
\big(
\eF_{0}
+
\eQ_{0}^{-}\ea_{-1}
+
\eQ_{0}^{+}\ea_{1}
\big)
\end{equation}
combination of equation~(\ref{RR}) at $\ir=0$ and
$\ec_{\pm1}=\eS_{\pm1}\ec_{0}+\ea_{\pm1}$.
Thus, starting from $\ec_{0}$, we {\em iterate
$\ec_{\ir}=\eS_{\ir}\ec_{\ir\mp1}+\ea_{\ir}$ upwards}, getting the
solution of the recursion.
For $\eF_{\ir}\equiv0$ (homogeneous recurrences), we have
$\ea_{\ir}=0$ and the solution reads
$\ec_{\ir}^{\rm (h)}
=
\eS_{\ir}\eS_{\ir-1}\cdots\eS_{1}\ec_{0}^{\rm (h)}$.
Note that $\eS_{\ir}$ is given in terms of $\eS_{\ir+1}$ in the
denominator, which can in turn be written as a fraction with
$\eS_{\ir+2}$ in the denominator, and so on.
This has the structure of a {\em continued fraction}, naming the
method
%
\begin{equation}
\label{continued-fraction}
C
=
\frac{p_{1}}
{q_{1}
+
{\displaystyle
\frac{p_{2}}{q_{2}+\cdots}
}
}
\;.
\end{equation}

The iterative solution can be cast into explicit form as a series of
products of continued fractions \cite{cofkalwal94}.
The above form, however, has the virtue of being easy to implement in
a computer.
Convergence is checked increasing $\Itr\to2\Itr\to\cdots$ and
repeating the procedure.
Solving equation (\ref{RR}) in this way requires a computational
effort of order $\Itr$, instead of the order $\Itr^{2}$ of an
arbitrary linear algebra problem.
This reduction arises from the tridiagonal structure of the matrix
associated to (\ref{RR}).
(For analytical inversion of tridiagonal matrices see \cite{huamcc97}
and references therein.)

We can solve analogously a differential recurrence of the type
%
\begin{equation}
\label{dRR}
\drm\ec_{\ir}/\drm t
=
\eQ_{\ir}^{-}\ec_{\ir-1}
+
\eQ_{\ir}\ec_{\ir}
+
\eQ_{\ir}^{+}\ec_{\ir+1}
+
\eF_{\ir}
\;.
\end{equation}
Laplace transformation converts this equation into
[$\tilde{g}(s)\equiv\int_{0}^{\infty}\!\drm t\,\e^{-st}g(t)$]
\begin{equation}
\label{dRR:lap}
\eQ_{\ir}^{-}
\widetilde{\ec}_{\ir-1}
+
(\eQ_{\ir}-s)
\widetilde{\ec}_{\ir}
+
\eQ_{\ir}^{+}
\widetilde{\ec}_{\ir+1}
=
-
[\tilde{\eF}_{\ir}
+
\ec_{\ir}(0)]
\end{equation}
where $\tilde{\dot{g}}(s)=s\,\tilde{g}(s)-g(0)$ has been used.
Then, introducing $\eQ_{\ir}'=\eQ_{\ir}-s$ and
$\eF_{\ir}'=\tilde{\eF}_{\ir}+\ec_{\ir}(0)$, the above equation has
the structure of the ordinary recurrence (\ref{RR}).
%

The quantities involved in the above recurrences need not to be
scalars, but they can be $J$-vectors ($\ec_{\ir}$ and $\eF_{\ir}$) and
the coefficients $\eQ_{\ir}$ are $J\times J$-matrices.
Then we speak of {\em matrix continued fractions}.
The only change in the above solution is that the fraction bars then
mean matrix inversion (``from the left'', $A/B\;\rightarrow\;
B^{-1}\,A$).

This is important, because a recurrence relation involving more than
three coefficients can be ``folded'' into a $3$-term recurrence by
introducing vector and matrix quantities.
Let us show how to do so for the following $5$-term recursion
\begin{equation}
\label{RR:5}
\eQ_{\ir}^{--}\ec_{\ir-2}
+
\eQ_{\ir}^{-}\ec_{\ir-1}
+
\eQ_{\ir}\ec_{\ir}
+
\eQ_{\ir}^{+}\ec_{\ir+1}
+
\eQ_{\ir}^{++}\ec_{\ir+2}
=
-\eF_{\ir}
\;.
\end{equation}
Defining the $2$--vectors
\begin{equation*}
\mc_{\ir}
=
\left(
\begin{array}{c}
\ec_{2\ir}
\\
\ec_{2\ir+1}
\end{array}
\right)
\quad
\mbox{and}
\quad
\mF_{\ir}
=
\left(
\begin{array}{c}
\eF_{2\ir}
\\
\eF_{2\ir+1}
\end{array}
\right)
\end{equation*}
and the $2\times2$--matrices
\begin{equation*}
\fl
\mQ_{\ir}
=
\left(
\begin{array}{cc}
\eQ_{2\ir}
&
\eQ_{2\ir}^{+}
\\
\eQ_{2\ir+1}^{-}
&
\eQ_{2\ir+1}
\end{array}
\right)
\quad
\mQ_{\ir}^{-}
=
\left(
\begin{array}{cc}
\eQ_{2\ir}^{--}
&
\eQ_{2\ir}^{-}
\\
0
&
\eQ_{2\ir+1}^{--}
\end{array}
\right)
\quad
\mQ_{\ir}^{+}
=
\left(
\begin{array}{cc}
\eQ_{2\ir}^{++}
&
0
\\
\eQ_{2\ir+1}^{+}
&
\eQ_{2\ir+1}^{++}
\end{array}
\right)
\end{equation*}
it can be easily seen that equation~(\ref{RR:5}) for $2\ir$ and
$2\ir+1$ is equivalent to
\begin{equation}
\label{5to3:RR}
\mQ_{\ir}^{-}\mc_{\ir-1}
+
\mQ_{\ir}\mc_{\ir}
+
\mQ_{\ir}^{+}\mc_{\ir+1}
=
-\mF_{\ir}
\;.
\end{equation}
Then, insertion of $\mc_{\ir}=\mS_{\ir}\mc_{\ir\mp1}+\ma_{\ir}$ gives
the matrix version of (\ref{S:a}).
Note that the $\ec_{\ir}$ and $\eF_{\ir}$ can already be vectors and
the $\eQ_{\ir}$ matrices.
Then, the above $\mc_{\ir}$, $\mF_{\ir}$, and $\mQ_{\ir}$ are
``block'' vectors and matrices.

Finally, to iterate upwards
$\mc_{\ir}=\mS_{\ir}\mc_{\ir\mp1}+\ma_{\ir}$, the ``initial
condition'' $\mc_{0}$ is obtained from the matrix version of
(\ref{condini}) (which we write $\mA\,\mathbf{x}=\mathbf{b}$).
As in the absence of forcing ($\mF_{\ir}=0$) the general solution
involves an overall multiplicative constant, one adds to this system
an extra equation to fix it (e.g., a normalisation contidion on the
coefficients; vd.\ section~\ref{sec:obs}).
Alternatively, we can fix arbitrarily one element of $\mc_{0}$ and
rescale the solution at the end.
In either case, one has an extra equation which added to
$\mA\,\mathbf{x} = \mathbf{b}$ yields the extended $(J+1)\times J$
system
%
\begin{equation}
\label{condini:A*x=b*}
\mA^{\ast}\,\mathbf{x}=\mathbf{b}^{\ast}
\;,
\quad
\mA^{\ast}
=
\stackrel{\longleftarrow\;\;J\;\;\longrightarrow}
{
\left(
\begin{array}{c}
\mbox{\large$\mA$}
\\[0.ex]\hline\\[-2.ex]
\mbox{\,l.h.s.\ eq.\,}
\end{array}
\right)
}
\hspace{-0.5em}
\begin{array}{c}
\uparrow
\\
{\scriptstyle (J+1)}
\\
\downarrow
\end{array}
\quad
\mathbf{b}^{\ast}
=
\left(
\begin{array}{c}
b_{1}
\\[-1.5ex]
\vdots
\\[-1.5ex]
b_{J}
\\[0.ex]\hline\\[-2.5ex]
\mathrm{\!\!r.h.s.\!\!\!}
\end{array}
\right)
\end{equation}
This system can be solved by any method appropriate for cases with
more equations than unknowns (e.g., using $QR$- or SVD-decomposition
\cite{recipes}), getting $\mc_{0}$.


\section{
Derivation of the transformed evolution operator $\LFPb$
}
\label{app:Lb}

Here we derive $\LFPb$, the $\Wo^{-1}(\,\cdot\,)\Wo$ transform of the
operator $\LFP$ in the master equation with
$\Wo\propto\exp[-(\etab\,\p^{2}/2+\bVo)]$
[equation~(\ref{W:expansion})].
We split $\LFPb$ as $\LFPb=\Lirb+\Lkinb+\Lvb$, with $\Lir$, $\Lkin$,
and $\Lv$ given by (\ref{WKK:Lir:Lkin:Lv}).
For the ``product'' of operators we use that the transformation of the
product is equal to the product of the transformations:
\begin{equation}
\label{bar:product}
\overline{A\,B}
=
\overline{A}
\;
\overline{B}
\;,
\qquad
\overline{A^{m}}=\big(\overline{A}\big)^{m}
\end{equation}
which follows by sandwiching ``identities'' $\Wo\Wo^{-1}=1$ between
the factors.
%
Besides, the multiplication by functions of $(\x,\p)$ commutes with
$\Wo^{-1}(\,\cdot\,)\Wo$, so that only the differential terms need to
be transformed.
From these two properties we see that {\em only $\pxb$ and $\ppb$ need
to be calculated\/} (and then raised to some power or multiplied).


\subsection{
Calculation of $\pxb$ and $\ppb$
}

{\em Calculation of $\pxb$}.
In order to transform $\px$, we apply $\Wo^{-1}\px\Wo$, to an
arbitrary function $f(\x,\p)$.
The $\p$ dependent part of $\Wo$ cancels out and we obtain
%
\begin{equation}
\label{pxb}
\pxb
f
=
\e^{\bVo}
\left(
-\px\bVo\,
\e^{-\bVo}
f
+
\e^{-\bVo}
\px f
\right)
\;,
\quad
\leadsto
\quad
\pxb
=
\px
-
\bVo'
\end{equation}
which has the form of a ``displaced'' differential operator.

{\em Calculation of $\ppb$}.
Applying $\Wo^{-1}\pp\Wo$ to a function $f$, we similarly get $\ppb$:
%
\begin{equation}
\label{ppb:pre}
\fl
\ppb
f
=
\e^{\etab\,\p^{2}/2}
\left(
-\etab
\p\,
\e^{-\etab\,\p^{2}/2}
f
+
\e^{-\etab\,\p^{2}/2}
\pp
f
\right)
\;,
\quad
\leadsto
\quad
\ppb
=
\pp
-
\etab\,\p
\;.
\end{equation}
This $\ppb$ can be expressed in terms of the following creation and
annihilation operators
%
\begin{equation}
\label{bm-bp}
\bd
=
\pp
+
\case{1}{2}
\p
\;,
\qquad
\bp
=
-
\pp
+
\case{1}{2}
\p
\;.
\end{equation}
Thus, using the shifted $\etab$ parameters $\etapm=\etab\mp\case{1}{2}$
[equation~(\ref{etam:etap})] we arrive at
\begin{equation}
\label{ppb}
\ppb
=
-
\big(\etam\bp+\etap\bd\big)
\;.
\end{equation}
For $\etab=1/2$ (the choice in the classical case), $\etam=1$ and
$\etap=0$, so that $\ppb=-\bp$.


\subsection{
Calculation of $\Lirb=\Wo^{-1}\Lir\Wo$
}

Using the product property
$\overline{A\,B}
=
\overline{A}\,\overline{B}$,
we have $\Lirb=\gammaT\ppb\big(\p+\ppb\big)$.
Then, replacing $\pp$ by its ``bar'' form (\ref{ppb}) and writing
$\p=\bd+\bp$ [from~(\ref{bm-bp})], we find
\begin{equation*}
\fl
-\gammaT^{-1}
\Lirb
=
\etap(1-\etam)
+
\etam(1-\etam)
\bp\bp
+
2(\etab-\etam\etap)
\bp\bd
+
\etap(1-\etap)
\bd\bd
\;.
\end{equation*}
Here we have also used $\etam+\etap=2\etab$ and the commutation rule
$[\bd,\bp]=1$ to get a {\em normally ordered\/} form.
If we introduce some compact notation for the coefficients
%
\begin{equation*}
\begin{array}{rclcrcl}
\gammaoo
&=&
2\gammaT\,
(\etab-\etam\etap)
& &
\gammamm
&=&
\gammaT\,
\etam(1-\etam)
\\
\gammapm
&=&
\gammaT\,
\etap(1-\etam)
& &
\gammapp
&=&
\gammaT\,
\etap(1-\etap)
\end{array}
\end{equation*}
we can finally write
%
\begin{equation}
\label{Lirb}
\Lirb
=
-\big(
\gammapm
+
\gammamm
\bp\bp
+
\gammaoo\,
\bp\bd
+
\gammapp
\bd\bd
\big)
\;.
\end{equation}
(The choice $\etab=1/2$ gives the self-adjoint form
$\Lirb=-\gammaT\,\bp\bd$ \cite{risken}.)


\subsection{
Calculation of $\Lkinb=\Wo^{-1}\Lkin\Wo$
}

Using equations~(\ref{pxb}) and~(\ref{ppb}) for $\pxb$ and $\ppb$, as
well as $\p=\bd+\bp$ and the definition (\ref{DmDp}) of $\dplmi$ we
get for the kinetic part $\Lkinb=-\big(\p-\bDxp\ppb\big)\,\pxb$:
%
\begin{equation*}
\Lkinb
=
-
\bd
\dpl
\big(\px-\bVo'\big)
-
\bp
\dmi
\big(\px-\bVo'\big)
\;.
\end{equation*}
The coefficients of $\bd$ and $\bp$ are $\Dm$ and $\Dp$, also defined
in (\ref{DmDp}), and we have
%
\begin{equation}
\label{Lkinb}
\Lkinb
=
-
\left(
\bd\Dp
+
\bp\Dm
\right)
\;.
\end{equation}


\subsection{
Calculation of $\Lvb=\Wo^{-1}\Lv\Wo$
}

To obtain
$\Lvb=\sumiqo\qcoef^{(\iq)}\bV^{(2\iq+1)}\overline{\pp^{(2\iq+1)}}$
[vd.\ equation~(\ref{WKK:Lir:Lkin:Lv})], we use the product property
(\ref{bar:product}) to raise $\ppb$ to $2\iq+1$, getting
%
\begin{equation}
\label{Lvb}
\Lvb
=
-
\sumiqo
\qcoef^{(\iq)}
\,
\bV^{(2\iq+1)}
\,
\big(\etam\bp+\etap\bd\big)^{2\iq+1}
\;.
\end{equation}
To get later the matrix elements of $\Lvb$ between Hermite functions
it is convenient to cast (\ref{Lvb}) in a normally ordered form.
To this end we use Pathak's theorems \cite{pat2000} in the following
generalised form
%
\begin{equation}
\label{pathak:2:gral:odd:alt:rep}
\frac
{\big(\aplus+\aminus\big)^{2\iq+1}}
{(2\iq+1)!}
=
\sum_{q=0}^{\iq}
\frac
{\nol\big(\aplus+\aminus\big)^{2(\iq-q)+1}\nor}
{[2(\iq-q)+1]!}
\;
\frac{(\etapro/2)^{q}}{q!}
\;.
\end{equation}
Here
$\nol\big(\aplus+\aminus\big)^{m}\nor
=
\sum_{k=0}^{m}c_{k}^{m}(\aplus)^{m-k}\aminus^{k}$
with $c_{k}^{m}\equiv m!/[k!\,(m-k)!]$
($\nol f\nor$ is the result of moving the $\aplus$ to the left as if
they were scalars).
The generalisation resides in letting $\aminus$ and $\aplus$ have a
constant but different from one commutator.
In our case
%
\begin{equation}
\label{am-ap}
\aminus
=
\etap
\bd
\;,
\qquad
\aplus
=
\etam
\bp
\;,
\quad
\leadsto
\quad
[\aminus,\aplus]=\etapro
\end{equation}
follows from the commutator $[\bd,\bp]=1$ with $\etapro=\etam\etap$.
%
%
Note that $\aminus$ and $\aplus$ are not the adjoint of each other,
but the notation is symbolic.

With help from equation~(\ref{pathak:2:gral:odd:alt:rep}) and
$\qcoef^{(\iq)}=(-1)^{\iq}\ldB^{2\iq}\big/(2\iq+1)!$
[equation~(\ref{kappa:s})] we can write (\ref{Lvb}) as follows
%
\begin{equation}
\label{Lvb:deriv}
\Lvb
=
-
\sum_{\iq=0}^{\infty}
\sum_{q=0}^{\iq}
A_{\iq,q}
\,
\hat{X}^{2(\iq-q)+1}
\end{equation}
where we have introduced the notations
%
%
\begin{equation*}
A_{\iq,q}
=
\frac{(-1)^{\iq}\ldB^{2\iq}}{[2(\iq-q)+1]!}
\frac{(\etapro/2)^{q}}{q!}
\bV^{(2\iq+1)}
\;,
\qquad
\hat{X}^{\ell}
=
\nol\big(\aplus+\aminus\big)^{\ell}\nor
\;.
\end{equation*}
For a given $\iq$ different powers of $\hat{X}$ appear.
We take a common factor $\hat{X}^{\ell}$ and add coefficients, e.g.,
$A_{00}+A_{11}+A_{22}+\cdots$ with $\hat{X}$,
$A_{10}+A_{21}+A_{32}+\cdots$, with $\hat{X}^{3}$, etc.
%
\begin{equation*}
\fl
\sum_{\iq=0}^{\infty}
\sum_{q=0}^{\iq}
A_{\iq,q}
\,
\hat{X}^{2(\iq-q)+1}
=
\sum_{\ell=0}^{\infty}
\big(
\sum_{\iq=\ell}^{\infty}
A_{\iq,\iq-\ell}
\Big)
\,
\hat{X}^{2\ell+1}
=
\sum_{\ell=0}^{\infty}
\big(
\sum_{q=0}^{\infty}
A_{q+\ell,q}
\Big)
\,
\hat{X}^{2\ell+1}
\;.
\end{equation*}
For $s=q+\ell$ in $A_{\iq,q}$ we write
$\bV^{[2(q+\ell)+1]}=\px^{2q}\bV^{(2\ell+1)}$ and introduce the
operator $\dlpxx$
%
\begin{equation}
\label{Aqlq}
A_{q+\ell,q}
=
\frac{(-\dlpxx/2)^{q}}{q!}
\;
\frac{(-1)^{\ell}\ldB^{2\ell}}{(2\ell+1)!}
\bV^{(2\ell+1)}
\;,
\qquad
\dlpxx
=
\etapro\ldB^{2}\px^{2}
\;.
\end{equation}
Now the $q$ and $\ell$ dependences factorise, so we can sum the series
of $q$ in $E_{\ell}\equiv\sum_{q=0}^{\infty}A_{q+\ell,q}$, getting an
exponential.
Then, recalling the definition of $\qcoef^{(\iq)}$ we finally obtain
the normally ordered form for $\Lvb$ we were looking for
%
\begin{equation}
\label{Lvb:norord:El}
\fl
\Lvb
=
-
\sum_{\ell=0}^{\infty}
E_{\ell}
\,
\nol\big(\aplus+\aminus\big)^{2\ell+1}\nor
\;,
\qquad
E_{\ell}
=
\qcoef^{(\ell)}
\exp(-\dlpxx/2)
\;
\bV^{(2\ell+1)}
\;.
\end{equation}
The operator $\exp(-\dlpxx/2)$ does not act on what could appear after
$\V(\x)$, since $\dlpxx$ was introduced as some abreviated notation
for the high-order derivatives of the potential.


\section{
The different contributions to
$\dptT\ec_{\ip}=\sum_{\ipp}\Q_{\ip\ipp}\ec_{\ipp}$
}
\label{app:Qnm}

Here we calculate the matrix elements
$\Q_{\ip\ipp}
=
\int\!\drm\p\,\psi_{\ip}\LFPb\psi_{\ipp}$,
with the transformed operator $\LFPb=\Lirb+\Lkinb+\Lvb$ obtained in
\ref{app:Lb}.
We split in parts the calculation by introducing the corresponding
decomposition:
$\Q_{\ip\ipp}^{\rm ir}=\langle\ip|\Lirb|\ipp\rangle$,
$\Q_{\ip\ipp}^{\rm kin}=\langle\ip|\Lkinb|\ipp\rangle$, and
$\Q_{\ip\ipp}^{\rm v}=\langle\ip|\Lvb|\ipp\rangle$,
where we have used the custom ``bra''--``ket'' notation.
The calculation is simplified because we have all operators $\LFPb$
expressed as normally ordered forms of $\bd$ and $\bp$.
Then we can take advantage of the ladder actions
$\bp|\ip\rangle=\sqrt{\ip+1}\,|\ip+1\rangle$,
$\bd|\ip\rangle=\sqrt{n}\,|\ip-1\rangle$,
and the number property $\bp\bd|\ip\rangle=\ip\,|\ip\rangle$.


\subsection{
Calculation of $\sum_{\ipp}\Q_{\ip\ipp}^{\rm ir}\ec_{\ipp}$
}
\label{sec:Qir}

From~(\ref{Lirb}) for $\Lirb$ and the orthonormality of the
$|\ipp\rangle$, we find for $\Q_{\ip\ipp}^{\rm
ir}=\langle\ip|\Lirb|\ipp\rangle$
%
\begin{equation*}
\fl
-\Q_{\ip\ipp}^{\rm ir}
=
\gammamm
\sqrt{(\ip-1)\ip}
\,
\delta_{\ip-2,\ipp}
+
(\ip\,\gammaoo+\gammapm)
\,
\delta_{\ip\ipp}
+
\gammapp
\sqrt{(\ip+1)(\ip+2)}
\,
\delta_{\ip+2,\ipp}
\;.
\end{equation*}
Then, we introduce the notation $\gammao_{\ip}=\ip\,\gammaoo+\gammapm$
for the coefficient of the diagonal term [cf.\
equation~(\ref{gammas})], multiply by $\ec_{\ipp}$ and sum over
$\ipp$, getting
%
\begin{equation}
\label{sum:Qir:c}
\fl
-
\sum_{\ipp}
\Q_{\ip\ipp}^{\rm ir}
\ec_{\ipp}
=
\gammamm
\sqrt{(\ip-1)\ip}
\,
\ec_{\ip-2}
+
\gammao_{\ip}
\,
\ec_{\ip}
+
\gammapp
\sqrt{(\ip+1)(\ip+2)}
\,
\ec_{\ip+2}
\;.
\end{equation}


\subsection{
Calculation of $\sum_{\ipp}\Q_{\ip\ipp}^{\rm kin}\ec_{\ipp}$
}
\label{sec:Qrev}

Using equation~(\ref{Lkinb}) for $\Lkinb$ and the ``ladder'' action of
$\bd$ and $\bp$, we have
%
\begin{equation*}
\fl
\Q_{\ip\ipp}^{\rm kin}
=
-
\langle\ip|
\bd\Dp
+
\bp\Dm
|\ipp\rangle
=
-
\left(
\sqrt{\ip+1}\;
\Dp
\delta_{\ip+1,\ipp}
+
\sqrt{\ip}\;
\Dm
\delta_{\ip-1,\ipp}
\right)
\end{equation*}
where we have used the Kronecker delta to interchange $\ip$ and
$\ipp\mp1$.
Then, multiplying by $\ec_{\ipp}$ and summing over $\ipp$, one obtains
%
\begin{equation}
\label{sum:Qrev:c}
\sum_{\ipp}
\Q_{\ip\ipp}^{\rm kin}
\ec_{\ipp}
=
-
\left(
\sqrt{\ip}\;
\Dm
\ec_{\ip-1}
+
\sqrt{\ip+1}\;
\Dp
\ec_{\ip+1}
\right)
\;.
\end{equation}


\subsection{
Calculation of $\sum_{\ipp}\Q_{\ip\ipp}^{\rm v}\ec_{\ipp}$
}
\label{sec:Qq}

This will take longer.
First, the operator $\nol(\aplus+\aminus)^{2\ell+1}\nor$ appearing in
equation (\ref{Lvb:norord:El}) for $\Lvb$ can be put in explicit form
by using the binomial formula
\begin{equation}
\label{sum:pathak}
\nol\big(\aplus+\aminus\big)^{2\ell+1}\nor
=
\sum_{k=0}^{2\ell+1}
c_{k}^{2\ell+1}
(\aplus)^{2\ell+1-k}
\aminus^{k}
\;.
\end{equation}
The $\aminus$ and $\aplus$ are proportional to $\bd$ and $\bp$
[equation~(\ref{am-ap})], so the contribution of each summand to
$\langle\ip|\Lvb|\ipp\rangle$ involves
$\langle\ip|
(\bp)^{2\ell+1-k}\bd^{k}|\ipp\rangle
\propto
\langle\ip-(2\ell+1)+k|\ipp-k\rangle$.
From orthonormality only one term of the sum (\ref{sum:pathak}) will
contribute, namely $2k=(\ipp-\ip)+2\ell+1$, while $\ipp-\ip$ will be
odd, $\ipp=\ip\mp(2\iq+1)$ with $\iq\ge0$ (reflecting the odd powers
of $\pp$ in $\Lv$).
Then, in terms of $\iq$, the index $k$ is restricted to
%
\begin{equation}
\label{k:restriction}
k
=
\left\{
\begin{array}{lll}
\ell-\iq
\;,
&
\mbox{for}
&
\ipp=\ip-(2\iq+1)
\\
\ell+\iq+1
\;,
&
\mbox{for}
&
\ipp=\ip+(2\iq+1)
\end{array}
\right.
\;.
\end{equation}


\paragraph{
Relating the results for $\ipp=\ip\mp(2\iq+1)$.
}

The two cases can be easily related.
First, the combinatorial coefficients are equal
$c_{\ell+\iq+1}^{2\ell+1}=c_{\ell-\iq}^{2\ell+1}$, because
$c_{k}^{m}=c_{m-k}^{m}$.
Second, the $\etab$ factors, coming from the proportionality of the
$\aminus,\aplus$, to $\bd,\bp$, are simply obtained from each other by
exchanging pluses and minuses ($\etapro=\etam\etap$):
%
\begin{equation}
\begin{array}{rcl}
k=\ell-\iq
\quad
\leadsto
&
\etam^{\ell+\iq+1}
\etap^{\ell-\iq}
=
\etapro^{\ell-\iq}
\etam^{2\iq+1}
&
\\
k=\ell+\iq+1
\quad
\leadsto
&
\etam^{\ell-\iq}
\etap^{\ell+\iq+1}
=
\etapro^{\ell-\iq}
\etap^{2\iq+1}
&
\end{array}
\end{equation}
Third and last, the matrix elements are also relatable.
For $\ipp=\ip-(2\iq+1)$, we have
$\langle\ip|(\bp)^{\ell+\iq+1}\bd^{\ell-\iq}|\ip-(2\iq+1)\rangle$,
while for $\ipp=\ip+(2\iq+1)$
%
\begin{equation*}
\langle\ip|
(\bp)^{\ell-\iq}
\bd^{\ell+\iq+1}
|\ip+(2\iq+1)\rangle
=
\langle\ipp|
(\bp)^{\ell+\iq+1}\bd^{\ell-\iq}
|\ipp-(2\iq+1)\rangle
\end{equation*}
which is formally identical to the former by replacing
$\ip\to\ip'=\ipp$ and $\ipp\to\ipp'=\ip$.

Thus, from now on we consider only the case $\ipp=\ip-(2\iq+1)$
(factor $\etapro^{\ell-\iq}\etam^{2\iq+1}$).
The result for $\ipp=\ip+(2\iq+1)$, will be readily obtained from this
by replacing
%
\begin{equation}
\label{rule}
\ip'=\ipp
\;,
\quad
\ipp'=\ip'-(2\iq+1)
\quad
\mbox{and}
\quad
\etam^{2\iq+1}
\to
\etap^{2\iq+1}
\;.
\end{equation}
From all these considerations, we can write the matrix elements of
$\nol(\aplus+\aminus)^{2\ell+1}\nor$ as
\begin{equation*}
\big\langle\ip\big|
\nol\big(\aplus+\aminus\big)^{2\ell+1}\nor
\big|\ipp\big\rangle
=
\etam^{2\iq+1}
c_{\ell-\iq}^{2\ell+1}
\,
\etapro^{\ell-\iq}
\;
\big\langle\ip\big|
(\bp)^{\ell+\iq+1}\bd^{\ell-\iq}
\big|\ipp\big\rangle
\end{equation*}
where we omit a Kronecker delta ensuring $\ipp=\ip-(2\iq+1)$.
The rest of the calculation consists of multiplying this by $E_{\ell}$
and sum over all $\ell$ to get $\langle\ip\big|\Lvb\big|\ipp\rangle$
[equation~(\ref{Lvb:norord:El})].


\paragraph{
Restrictions on the sum on $\ell$.
}

The contribution of the infinite sum (\ref{Lvb:norord:El}) to
$\langle\ip\big|\Lvb\big|\ipp\rangle$ is fortunately cut-off.
First, the lower index in $c_{\ell-\iq}^{2\ell+1}$ should be positive
and, introducing the shifted index $\ell'=\ell-\iq$, we obtain (we
rename $\ell'\to\ell$ at the end)
%
\begin{equation}
\label{Lvb:matele:1}
\fl
\big\langle\ip\big|
\Lvb
\big|\ipp\big\rangle
=
-
\etam^{2\iq+1}
\sum_{\ell=0}^{\ipp}
E_{\ell+\iq}
\;
c_{\ell}^{2(\ell+\iq)+1}
\,
\etapro^{\ell}
\;
\big\langle\ip\big|
(\bp)^{(2\iq+1)+\ell}\bd^{\ell}
\big|\ipp\big\rangle
\;.
\end{equation}
Here we have already used that the sum is also cut-off from above at
$\ell_{\rm max}=\ipp$, since $\bd^{\ell}|\ipp\rangle\equiv0$, for
$\ell>\ipp$.
On the other hand, for $E_{\ell+\iq}$ [equation~(\ref{Lvb:norord:El})]
we have
\begin{equation}
E_{\ell+\iq}
=
\e^{-\dlpxx/2}
\;
\frac{(2\iq+1)!}{[2(\ell+\iq)+1]!}
(-\ldB^{2}\px^{2})^{\ell}
\frac{(-1)^{\iq}\ldB^{2\iq}}{(2\iq+1)!}
\bV^{(2\iq+1)}
\;.
\end{equation}
Combining this with $c_{\ell}^{2(\ell+\iq)+1}$, gathering
$(\ldB^{2}\px^{2})^{\ell}$ with $\etapro^{\ell}$ to form the operator
$\dlpxx$ [equation~(\ref{Aqlq})], and recalling the
definition~(\ref{kappa:s}) of $\qcoef^{(\iq)}$, we obtain
%
\begin{equation*}
\fl
\big\langle\ip\big|
\Lvb
\big|\ipp\big\rangle
=
-
\etam^{2\iq+1}
\qcoef^{(\iq)}\,
\e^{-\dlpxx/2}
\sum_{\ell=0}^{\ipp}
\frac{(2\iq+1)!}{[(2\iq+1)+\ell]!}
\;
\big\langle\ip\big|
(\bp)^{(2\iq+1)+\ell}\bd^{\ell}
\big|\ipp\big\rangle
\;
\frac{(-\dlpxx)^{\ell}}{\ell!}
\;
\bV^{(2\iq+1)}
\;.
\end{equation*}
Note that the quotient of factorials equals $1/(2\iq+2)_{\ell}$, with
the Pochhammer symbol defined by \cite[Ch.~13.5]{arfken}
%
\begin{equation}
\label{pochhammer}
(a)_{\ell}
=
a(a+1)\cdots(a+\ell-1)
=
(a+\ell-1)!\big/(a-1)!
\;.
\end{equation}


\paragraph{
Computation of the matrix element.
}

To conclude, we only need the matrix element
$\langle\ip|(\bp)^{(2\iq+1)+\ell}\bd^{\ell}|\ipp\rangle$
in (\ref{Lvb:matele:1}).
We first pass $2\iq+1$ times the action of $\bp$ to $|\ip\rangle$ by
taking its adjoint $\bd$ and then we use repeteadly its downward
ladder action, getting
\begin{equation}
\bd^{2\iq+1}\big|\ip\big\rangle
=
\sqrt{\ip(\ip\!-\!1)\cdots(\ip\!-\!2\iq)}\;
\big|\ip\!-\!(2\iq\!+\!1)\big\rangle
=
\sqrt{\ip!/\ipp!}\;
\big|\ipp\big\rangle
\end{equation}
with $\ipp=\ip-(2\iq+1)$.
Then, in terms of the Pochhammer symbol (\ref{pochhammer}), we can
write
%
\begin{equation*}
\fl
\big\langle\ip\big|
(\bp)^{(2\iq+1)+\ell}\bd^{\ell}
\big|\ipp\big\rangle
=
\sqrt{\ip!/\ipp!}\;
\big\langle\ipp\big|
(\bp)^{\ell}\bd^{\ell}
\big|\ipp\big\rangle
=
\sqrt{\ip!/\ipp!}\;
(-1)^{\ell}
(-\ipp)_{\ell}
\;.
\end{equation*}
Now, taking into account that the factor $(-1)^{\ell}$ cancels that of
$(-\dlpxx)^{\ell}$ and using the modified coefficient
$\qcoef^{(\iq)}_{\ip}=\qcoef^{(\iq)}\sqrt{\ip!/\ipp!}$
[equation~(\ref{Gamma:qcoeff})], we arrive at
%
\begin{equation*}
\big\langle\ip\big|
\Lvb
\big|\ipp\big\rangle
=
-
\etam^{2\iq+1}
\qcoef^{(\iq)}_{\ip}
\,\e^{-\dlpxx/2}
\;
\bigg[
\sum_{\ell=0}^{\ipp}
\frac{(-\ipp)_{\ell}}{(2\iq+2)_{\ell}}
\;
\frac{\dlpxx^{\ell}}{\ell!}
\bigg]
\bV^{(2\iq+1)}
\;.
\end{equation*}
Comparing with the series definition of the confluent hypergeometric
(Kummer) function $\kum(a,c\,;z)$ \cite[Ch.~13.6]{arfken} we get
%
\begin{equation*}
\fl
\big\langle\ip\big|
\Lvb
\big|\ipp\big\rangle
=
-
\etam^{2\iq+1}
\qcoef^{(\iq)}_{\ip}
\big[
\e^{-\dlpxx/2}
\,
\kum(-\ipp,2\iq\!+\!2\,;\dlpxx)
\big]
\bV^{(2\iq+1)}
\quad
\kum(a,c\,;z)
=
\sum_{\ell=0}^{\infty}
\frac{(a)_{\ell}}{(c)_{\ell}}
\frac{z^{\ell}}{\ell!}
\;.
\end{equation*}
The operator in front of $\bV(\x)$ is simply $\G_{\ip}^{\iq,-}$
[equation~(\ref{Gamma:qcoeff})], so that we can finally write
%
\begin{equation}
\label{Lvb:matele:kummer}
\Q_{\ip\ipp}^{\rm v}
=
\big\langle\ip\big|
\Lvb
\big|\ipp\big\rangle
=
-
\G_{\ip}^{\iq,-}
\,
\bV^{(2\iq+1)}(\x)
\;,
\quad
\ipp=\ip-(2\iq+1)
\;.
\end{equation}
Note that $\kum(-\ipp,2\iq\!+\!2\,;z)$ is a polynomial because the
first negative index cuts the upper Pochhammer symbol.
Indeed the ``integer-plus-one'' second argument $(2\iq+1)+1$ tells us
that it is an associated Laguerre polynomial \cite[Ch.~13]{arfken}
%
\begin{equation}
\label{assoc:laguerre:poly:kummer}
L_{\ell}^{k}(z)
=
\frac{(\ell+k)!}{k!\,\ell!}
\kum(-\ell,k+1\,;z)
\;.
\end{equation}
In our case $\ell=\ipp$ and $k=\ip-\ipp=2\iq+1$, so that
$\kum(-\ipp,2\iq+2\,;z)
=
L_{\ipp}^{2\iq+1}(z)/c_{\ip}^{\ipp}$.


\paragraph{
Final expressions for $\sum_{\ipp}\Q_{\ip\ipp}^{\rm v}\ec_{\ipp}$.
}

Once we have computed $\Q_{\ip\ipp}^{\rm v}$, the matrix elements of
$\Lvb$, we multiply by $\ec_{\ipp}$ and sum over $\ipp$ (i.e., over
$\iq$), to get
%
\begin{equation*}
\sum_{\ipp<\ip}
\Q_{\ip\ipp}^{\rm v}
\ec_{\ipp}
=
-\sumiqo
\big[
\G_{\ip}^{\iq,-}
\,
\bV^{(2\iq+1)}
\big]
\,
\ec_{\ip-(2\iq+1)}
\;.
\end{equation*}
Note that this $\iq$-sum is restricted because $\ip-(2\iq+1)\geq0$.
This yields the upper limit $\iq_{\rm max}=[(\ip-1)/2]$, with $[a]$
the integer part of $a$.
The contribution from $\ipp>\ip$ is easily obtained recalling the
transformation rules (\ref{rule}) (here the sum is not restricted)
%
\begin{equation*}
\sum_{\ipp>\ip}
\Q_{\ip\ipp}^{\rm v}
\ec_{\ipp}
=
-\sumiqo
\big[
\G_{\ip+(2\iq+1)}^{\iq,+}
\,
\bV^{(2\iq+1)}
\big]
\,
\ec_{\ip+(2\iq+1)}
\end{equation*}
with $\G_{\ip}^{\iq,+}$ given by (\ref{Gamma:qcoeff}).
We have enclosed into square brackets the action of the operators
$\G_{\ip}^{\iq,\pm}$ to recall that they only act on the
$\x$-dependence of the potential.

For $\etab=1/2$, the choice in the classical case, $\etam=1$,
$\etap=0$, and $\etapro=0$.
Then,
$\G_{\ip}^{\iq,-}(\etapro=0)=\qcoef^{(\iq)}_{\ip}$
and
$\G_{\ip}^{\iq,+}(\etapro=0)=0$,
so that only the $\ec_{\ip-(2\iq+1)}$ terms survive.
The chain of equations then adquires an unbalanced structure (like in
the old hierarchies \cite{plorho85,joh98,burmol2002}), which may
result in a poor stability when the quantum terms are important.
In constrast, $\etab\sim0$ gives $|\etapm|\sim1/2$, and hence the
weight of the terms at both sides of $\ip$ ($\sim$ above and below the
diagonals in the matrices $\mQ_{\ix\ixp}$) is similar, improving the
stability when going into the deep quantum regime.


\section{
The auxiliary integrals $\Ial_{\ix}$ and $\Knl_{\ip}^{(\ell)}$
}
\label{app:auxint}

For special $\etab$ and $\bVo$, the integrals $\Ial_{\ix}$ and
$\Knl_{\ip}^{(\ell)}$ [equation~(\ref{Ial:Knl})] appearing in the
expressions (\ref{J:Ial:Knl}) for the observables, are easily done.
For $\etab=1/2$, we simply have
$\Knl_{\ip}^{(0)}=\delta_{\ip,0}$,
$\Knl_{\ip}^{(1)}=\delta_{\ip,1}$,
and
$\Knl_{\ip}^{(2)}=\delta_{\ip,0}+\sqrt{2}\,\delta_{\ip,2}$.
If $\bVo=0$ (i.e., $\epspot=0$), we get
$\Ial_{\ix}=\Ial_{0}\delta_{\ix,0}$.
Then the first moments reduce to $1=\Ial_{0}\ec_{0}^{0}$
(normalisation) and to
$\langle\,\p\,\rangle=\Ial_{0}\ec_{1}^{0}$
and
$\langle\p^{2}\rangle
=
\Ial_{0}(\ec_{0}^{0}+\sqrt{2}\ec_{2}^{0})$.
In the general case, one can derive recurrences relating the
$\Knl_{\ip}^{(\ell)}$ with lower order (in $\ell$) ones.
Thus, only $\Knl_{\ip}^{(0)}$ is needed, which can be found
analytically.
Concerning the $\Ial_{\ix}$, we mostly use $\epspot=0$, and hence
$\Ial_{\ix}=\Ial_{0}\delta_{\ix,0}$.
Anyway, for periodic potentials one can derive recurrence relations
for them which can also be solved by continued fractions.


\subsection{
The integrals $\Knl_{\ip}^{(\ell)}$ (arbitrary potentials)
}
\label{sec:Knl}

To get the recurrence for the $\Knl_{\ip}^{(\ell)}$ we start from
their definition (\ref{Ial:Knl}) with index $\ell+1$, express the last
$\p$ as $\p=\bd+\bp$ [equation~(\ref{bm-bp})], and then use
$\p\,\psi_{n}=\sqrt{\ip}\,\psi_{\ip-1}+\sqrt{\ip+1}\,\psi_{\ip+1}$,
getting
%
\begin{equation}
\label{Knl:RR}
\Knl_{\ip}^{(\ell+1)}
=
\sqrt{\ip}\,
\Knl_{\ip-1}^{(\ell)}
+
\sqrt{\ip+1}\,
\Knl_{\ip+1}^{(\ell)}
\;.
\end{equation}
This recurrence can be used to get the $\Knl_{\ip}^{(\ell+1)}$ from
the $\Knl_{\ip}^{(\ell)}$.

We can get analytical expressions for the first few
$\Knl_{\ip}^{(\ell)}$ with help from the tabulated integral
\cite[equation~(7.374-4)]{gradshteyn-ryzhik}
%
\begin{equation}
\label{hermite:func:int:1}
\fl
\int_{-\infty}^{\infty}\!\drm x\,
\e^{-x^{2}}\,
H_{m}(x)
H_{2n+m}(ax)
=
\sqrt{\pi}
\,
\frac{(2n+m)!}{n!}
\,
(2a)^{m}
\,
(a^{2}-1)^{n}
\;.
\end{equation}
For $\Knl_{\ip}^{(0)}=r_{0}\int\!\e^{-\etab\,\p^{2}/2}\,\psi_{\ip}$, we
use this result with $m=0$, obtaining
%
\begin{equation}
\label{Knl:0}
\fl
\Knl_{2\ip+1}^{(0)}
=
0
\;,
\qquad
\Knl_{2\ip}^{(0)}
=
\frac{\sqrt{(2\ip)!}}{\ip!}
\,
\etam^{-1/2}
\;
\Lambda^{\ip}
\;,
\qquad
\Lambda=-\etap/2\etam
\;.
\end{equation}
Although $\Knl_{\ip}^{(1)}$ and $\Knl_{\ip}^{(2)}$ can now be obtained
from (\ref{Knl:RR}), they can also be done with help from
equation~(\ref{hermite:func:int:1}), getting: $\Knl_{2\ip}^{(1)}=0$,
$\Knl_{2\ip+1}^{(2)}=0$, and
%
\begin{equation}
\label{Knl:1}
\fl
\Knl_{2\ip+1}^{(1)}
=
\frac{\sqrt{(2\ip+1)!}}{\ip!}
\,
\frac{1}{\etam^{3/2}}
\;
\Lambda^{\ip}
\;,
\qquad
\Knl_{2\ip}^{(2)}
=
\frac{\sqrt{(2\ip)!}}{\ip!}
\,
\frac{1}{\etam^{3/2}}
\,
\Big(
\frac{\ip}{\etam}
+
\Lambda
\Big)
\,
\Lambda^{\ip-1}
\;.
\end{equation}


\subsection{
Recursions for the integrals $\Ial_{\ix}$ (periodic potentials)
}
\label{sec:Vper:Ial}

For a general periodic potential we can derive a recurrence relation
for the integrals
$\Ial_{\ix}
=
\int_{0}^{2\pi}\!\drm\x\,\exp[-\epspot\bV(\x)]\,u_{\ix}(\x)$
[equation~(\ref{Ial:Knl})].
We integrate by parts the expression for $\iu\ix\,\Ial_{\ix}$ and use
$\V'=\sum_{\ix}\V_{\ix}'\,\e^{\iu\ix\x}$ with
$u_{\ix}=\e^{\iu\ix\x}/\sqrt{2\pi}$, getting
%
\begin{equation}
\label{RR:Ial:Vper}
\iu\ix\,
\Ial_{\ix}
=
\epspot
\sum_{\ixp}
\V_{\ixp}'
\,
\Ial_{\ix+\ixp}
\;.
\end{equation}
For a finite number of harmonics ($\V_{\ixp}'\equiv0$, if $|\ixp|>B$),
the recurrence (\ref{RR:Ial:Vper}) has a finite coupling range.
For instance, for the ratchet potential (\ref{Vrat}), we have
$\V_{\pm1}'=-\Vo/2\kT$ and $\V_{\pm2}'=-\rat\Vo/2\kT$ and the above
recurrence reduces to
%
\begin{equation}
\label{RR:Ial:Vrat}
\fl
\iu\ix\,
\Ial_{\ix}
=
-\epspotT
\left[
\left(\Ial_{\ix-1}+\Ial_{\ix+1}\right)
+
\rat\,
\left(\Ial_{\ix-2}+\Ial_{\ix+2}\right)
\right]
\;,
\qquad
\epspotT=\epspot\,\Vo/2\kT
\;.
\end{equation}
With $\Ial_{0}$ as input (obtained numerically, e.g., by Simpson rule
\cite[App.~2]{arfken}), this recursion can be solved by continued
fractions (\ref{app:RR-CF}).


\section{
The matrices $\mQ_{\ix\ixp}$ for general periodic potentials
}
\label{app:matrices}

Here we write expliticly the matrices $\mQ_{\ix\ixp}$ for arbitrary
periodic potentials.
We shall give the results for a slightly generalised basis, obtained
by shifting the origin of momenta by an amount $\pshift$ [cf.\
equation~(\ref{W:expansion:xp})]
\begin{equation}
\Wf(\x,\p)
=
\Wo(\x,\p-\pshift)
\sum_{\ip,\ix}
\ec_{\ip}^{\ix}\;
u_{\ix}(\x)\,\psi_{\ip}(\p-\pshift)
\;.
\end{equation}
This momentum shift is convenient to ``catch'' solutions centred far
from zero $\p$, where the ordinary basis would need many
$\psi_{\ip}(\p)$ to reconstruct the distribution.
(A useful choice is, when generating a curve, to set $\pshift=\Ppro$
of the precedent point.)
The momentum shift is handled as a change of variable
$\p\to\p_{\pshift}=\p-\pshift$.
Then, $\partial_{\p}=\partial_{\p_{\pshift}}$, but the term $\p\,\px$
makes the force to enter in the combination (scaled units)
%
\begin{equation}
\label{Fshift}
\bFsh
=
\bF-\gammaT\,\pshift
\;.
\end{equation}
As mentioned in section~\ref{sec:Vper}, we extract the force $\bF$
from $\V(\x)$ and set the auxiliary potential $\bVo$ proportional to
the periodic part $\bVo=\epspot\bV$.

The $\mQ_{\ix\ixp}$ are obtained by inserting the formulae for
$\Dpm_{\ix\ixp}$, and $[\G_{\ip}^{\iq,\pm}\bV^{(2\iq+1)}]_{\ix\ixp}$
[equation~(\ref{DmDpVper:matr-elem})], into the general matrices of
section~\ref{sec:MQBH}.
Instead of giving a single expression, we write separately the
matrices for $\ixp=\ix$ and $\ixp=\ix\pm\qab$.
The central matrix $\mQ_{\ix\ix}$ is determined by the terms involving
$\delta_{\ix\ixp}$, and reads
\begin{equation*}
\fl
\hspace*{-3.em}
-\mQ_{\ix\ix}
=
\left(
\begin{array}{cccccc}
\gammao_{0}
\!\!+\!\!
\pshift\,\iu\ix
\!\!\!\!&\!\!\!\!
\sqrt{1}(\iu\,\dpl\ix\!-\!\etap\bFsh)
\!\!\!\!&\!\!\!\!
\sqrt{1\cdot2}\,
\gammapp
\!\!\!\!&\!\!\!\!
0
\!\!\!\!&\!\!\!\!
0
\!\!\!\!&\!\!
\ddots
\\[0.ex]
\sqrt{1}(\iu\,\dmi\ix\!-\!\etam\bFsh)
\!\!\!\!&\!\!\!\!
\gammao_{1}
\!\!+\!\!
\pshift\,\iu\ix
\!\!\!\!&\!\!\!\!
\sqrt{2}(\iu\,\dpl\ix\!-\!\etap\bFsh)
\!\!\!\!&\!\!\!\!
\sqrt{2\cdot3}\,
\gammapp
\!\!\!\!&\!\!\!\!
0
\!\!\!\!&\!\!
\ddots
\\[0.ex]
\sqrt{1\cdot2}\,
\gammamm
\!\!\!\!&\!\!\!\!
\sqrt{2}(\iu\,\dmi\ix\!-\!\etam\bFsh)
\!\!\!\!&\!\!\!\!
\gammao_{2}
\!\!+\!\!
\pshift\,\iu\ix
\!\!\!\!&\!\!\!\!
\sqrt{3}(\iu\,\dpl\ix\!-\!\etap\bFsh)
\!\!\!\!&\!\!\!\!
\sqrt{3\cdot4}\,
\gammapp
\!\!\!\!&\!\!
\ddots
\\[0.ex]
0
\!\!\!\!&\!\!\!\!
\sqrt{2\cdot3}\,
\gammamm
\!\!\!\!&\!\!\!\!
\sqrt{3}(\iu\,\dmi\ix\!-\!\etam\bFsh)
\!\!\!\!&\!\!\!\!
\gammao_{3}
\!\!+\!\!
\pshift\,\iu\ix
\!\!\!\!&\!\!\!\!
\sqrt{4}(\iu\,\dpl\ix\!-\!\etap\bFsh)
\!\!\!\!&\!\!
\ddots
\\[0.ex]
0
\!\!\!\!&\!\!\!\!
0
\!\!\!\!&\!\!\!\!
\sqrt{3\cdot4}\,
\gammamm
\!\!\!\!&\!\!\!\!
\sqrt{4}(\iu\,\dmi\ix\!-\!\etam\bFsh)
\!\!\!\!&\!\!\!\!
\gammao_{4}
\!\!+\!\!
\pshift\,\iu\ix
\!\!\!\!&\!\!
\ddots
\\[0.ex]
\ddots
\!\!\!\!&\!\!\!\!
\ddots
\!\!\!\!&\!\!\!\!
\ddots
\!\!\!\!&\!\!\!\!
\ddots
\!\!\!\!&\!\!\!\!
\ddots
\!\!\!\!&\!\!
\ddots
\end{array}
\right)
\end{equation*}
with $\gamma_{\pm}$ and $\gammao_{\ip}$ given by (\ref{gammas}),
$\etapm=\etab\mp1/2$, and $\dplmi=1+\etapm\bDxp$.
Here the periodic potential does not appear; only $\bF$ enters in the
matrices $\mQ_{\ix\ix}$.
Besides, they do depend on the index $\ix$ (cf.\ below).

The matrices $\mQ_{\ix,\ix\pm\qab}$ are determined by the terms
involving $\V_{\ix-\ixp}'$ in equation~(\ref{DmDpVper:matr-elem}) and
read
\begin{equation*}
\fl
\hspace*{-3.em}
-\mQ_{\ix,\ix\pm\qab}=\bV_{\mp\qab}'
\left(
\begin{array}{ccccccc}
-\pshift\,\epspot
\!\!\!&\!\!\!
\etap \Gs_{1}^{0}
\!-\!
\sqrt{1}\,\epspot
\!\!\!&\!\!\!
0
\!\!\!&\!\!\!
\etap^{3}
\Gs_{3}^{1}
\!\!\!&\!\!\!
0
\!\!\!&\!\!\!
\etap^{5}
\Gs_{5}^{2}
\!\!\!&\!\!\!
\ddots
\\[0.ex]
\etam \Gs_{1}^{0}
\!-\!
\sqrt{1}\,\epspot
\!\!\!&\!\!\!
-\pshift\,\epspot
\!\!\!&\!\!\!
\etap \Gs_{2}^{0}
\!-\!
\sqrt{2}\,\epspot
\!\!\!&\!\!\!
0
\!\!\!&\!\!\!
\etap^{3}
\Gs_{4}^{1}
\!\!\!&\!\!\!
0
\!\!\!&\!\!\!
\ddots
\\[0.ex]
0
\!\!\!&\!\!\!
\etam \Gs_{2}^{0}
\!-\!
\sqrt{2}\,\epspot
\!\!\!&\!\!\!
-\pshift\,\epspot
\!\!\!&\!\!\!
\etap \Gs_{3}^{0}
\!-\!
\sqrt{3}\,\epspot
\!\!\!&\!\!\!
0
\!\!\!&\!\!\!
\etap^{3}
\Gs_{5}^{1}
\!\!\!&\!\!\!
\ddots
\\[0.ex]
\etam^{3}
\Gs_{3}^{1}
\!\!\!&\!\!\!
0
\!\!\!&\!\!\!
\etam \Gs_{3}^{0}
\!-\!
\sqrt{3}\,\epspot
\!\!\!&\!\!\!
-\pshift\,\epspot
\!\!\!&\!\!\!
\etap \Gs_{4}^{0}
\!-\!
\sqrt{4}\,\epspot
\!\!\!&\!\!\!
0
\!\!\!&\!\!\!
\ddots
\\[0.ex]
0
\!\!\!&\!\!\!
\etam^{3}
\Gs_{4}^{1}
\!\!\!&\!\!\!
0
\!\!\!&\!\!\!
\etam \Gs_{4}^{0}
\!-\!
\sqrt{4}\,\epspot
\!\!\!&\!\!\!
-\pshift\,\epspot
\!\!\!&\!\!\!
\etap \Gs_{5}^{0}
\!-\!
\sqrt{5}\,\epspot
\!\!\!&\!\!\!
\ddots
\\[0.ex]
\etam^{5}
\Gs_{5}^{2}
\!\!\!&\!\!\!
0
\!\!\!&\!\!\!
\etam^{3}
\Gs_{5}^{1}
\!\!\!&\!\!\!
0
\!\!\!&\!\!\!
\etam \Gs_{5}^{0}
\!-\!
\sqrt{5}\,\epspot
\!\!\!&\!\!\!
-\pshift\,\epspot
\!\!\!&\!\!\!
\ddots
\\[0.ex]
\ddots
\!\!\!&\!\!\!
\ddots
\!\!\!&\!\!\!
\ddots
\!\!\!&\!\!\!
\ddots
\!\!\!&\!\!\!
\ddots
\!\!\!&\!\!\!
\ddots
\!\!\!&\!\!\!
\ddots
\end{array}
\right)
\end{equation*}
where $\Gs_{\ip}^{\iq}$ is evaluated at $z=-\etapro\qab^{2}\ldB^{2}$,
with $\etapro=\etam\etap$.
These matrices enjoy properties opposite to those of $\mQ_{\ix\ix}$.
The substrate potential does appear (multiplying all elements in front
of the matrices) while the force is absent.
Besides, the matrices $\mQ_{\ix,\ix\pm\qab}$ are independent of the
index $\ix$, which permits to write them simply as $\mQ^{\pm\qab}$.
The only technical complication with respect to the common choice
$\etab=1/2$ is that we need to compute the Kummer functions
$\kum(a,c\,;z)$ included in $\Gs_{\ip}^{\iq}$, which in our case are
simple polynomials [equation~(\ref{assoc:laguerre:poly:kummer})].
Note that to compute the observables we must eventually undo the
momentum shift $\p_{\pshift}\to\p=\p_{\pshift}+\pshift$.


\section{
Semiclassical analytical results
}
\label{app:semicl}

In this appendix we derive some semiclassical formulae discussed in
the main text.


\subsection{
Reflection and transmission for a saw-tooth barrier
}

\begin{figure}
\includegraphics[width=10.em]{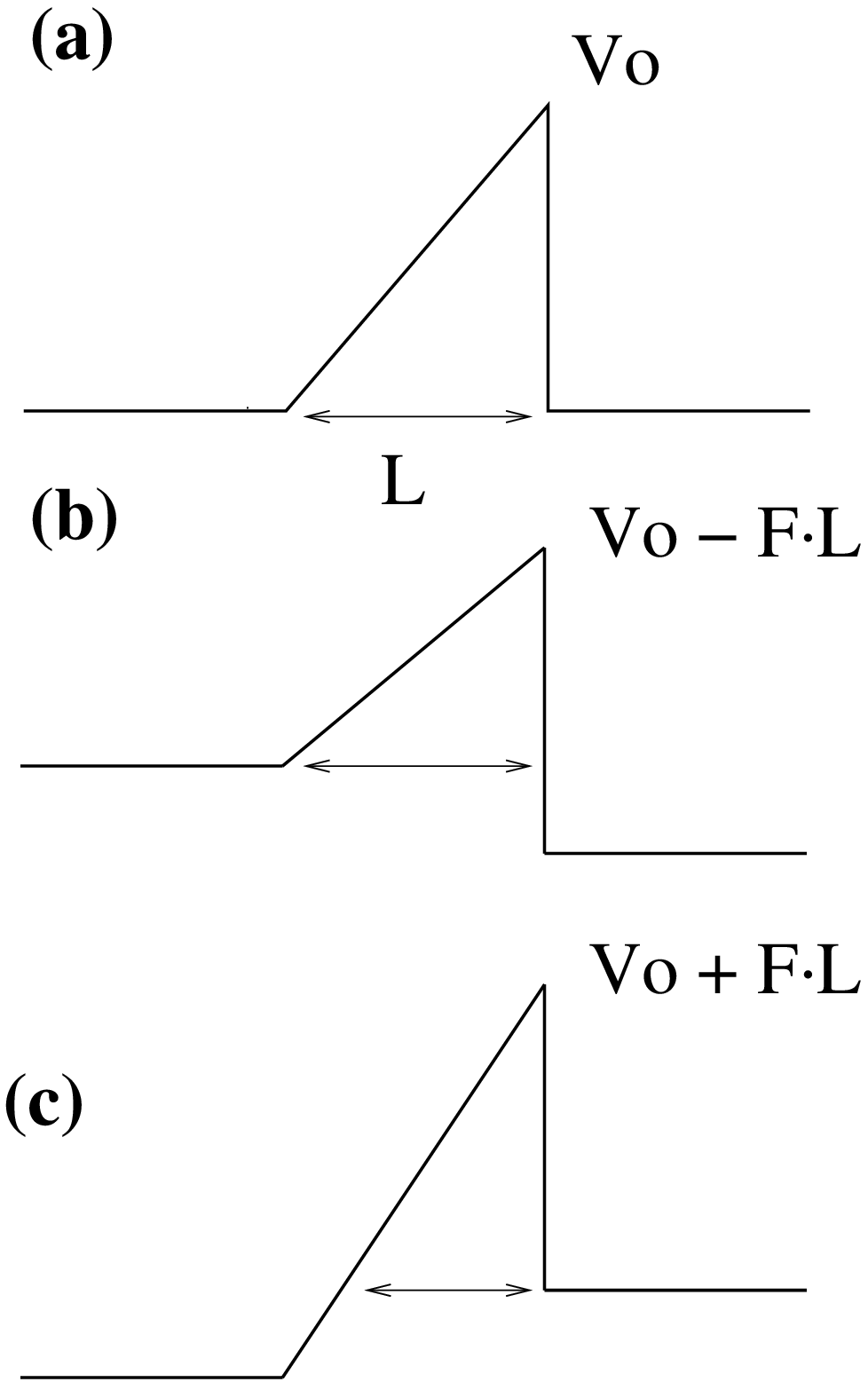}
\qquad
\qquad
\includegraphics[width=23.em]{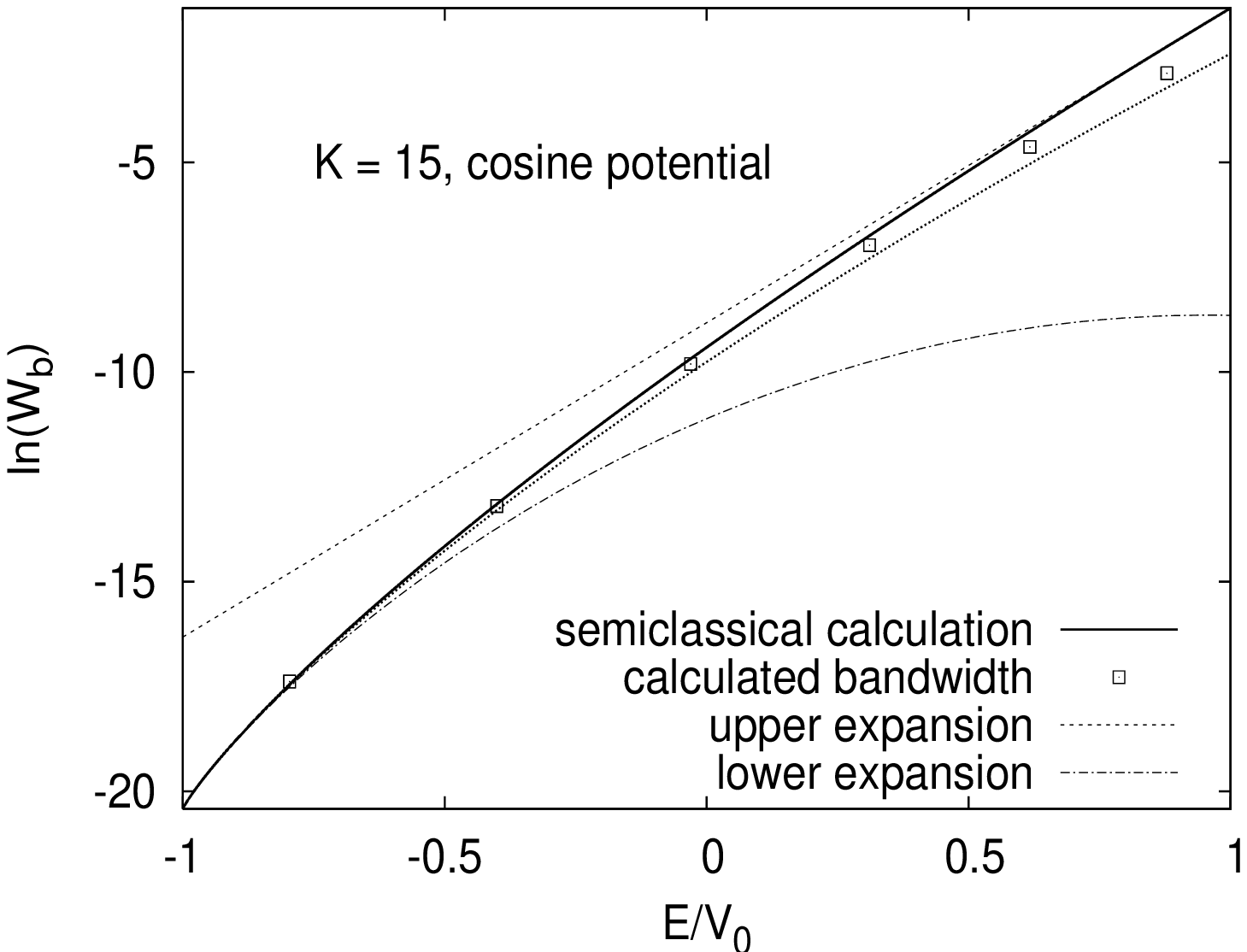}
\caption{
Left panel: profile of the saw-tooth potential barrier for
non-deformed (a) and deformed by $-F\x$ (b) and by $+F\x$ (c).
%
%
Right panel: width of the energy bands $W_{\rm b}$ in a cosine
potential (exact and various approximations).
}
\label{fig:saw-width}
\end{figure}

In section~\ref{ratQBM} we invoked the current through a saw-tooth
barrier deformed by $\pm F\x$ (figure \ref{fig:saw-width}).
We introduced the reflexion coefficients $\Refl_{+}$ and $\Refl_{-}$
for $\pm F$.
One has to analyse separately the cases of incident particles with
energies above or below the barrier.
In the first case, the reflexion coefficient by a potential with a
cusp (with slopes $f_{1}$ and $f_{2}$) is given by the first
equation~(\ref{Rs}) \cite[\S~52]{lanlif3}.
Owing to the right slope in figure~\ref{fig:saw-width} is infinity, we
make it finite and finally take the vertical limit, getting
\begin{equation}
\frac{\Refl_{+}}{\Refl_{-}}
=
\frac{\Vo - F\,L}{\Vo +F\,L}
<
1
\;.
\end{equation}
Next we consider the case of tunnelling through the barrier.
The semiclassical transmission coefficient is also written in
(\ref{Rs}).
Defining $A=|\int_{a}^{b}\!\drm\x\,\p(\x)|$, we calculate $A_{+}$ and
$A_{-}$ for positive and negative forces.
After simple integrals, we find
\begin{equation}
\frac{A_{+}}{A_{-}}
=
\frac{\Vo +FL}{\Vo-FL}
\left (
\frac{\Vo -FL - E}{\Vo - E}
\right )^{3/2}
\end{equation}
with $E$ the particle energy (cf.\ reference~\cite{rei2002}).
For moderate forces, $F \leq 0.618\,\Vo/L$, we get for the
transmission coefficients $\Tran=\exp[-(2/\hbar)A]$
\begin{equation}
A_{+}/A_{-}
>
1
\quad
\leadsto
\quad
\Tran_{+}/\Tran_{-}
<
1
\;.
\end{equation}
Thus, for $E$ below the barrier $\Refl_{+}/\Refl_{-}>1$, while
$\Refl_{+}/\Refl_{-}<1$ for $E$ above the barrier.


\subsection{
Transmision coefficient and energy bands for a cosine potential
}

In order to get the semiclassical transmision coefficient through a
potential $\V(\x)=-\Vo\cos(\x/\xo)$, we need to do the integral
\begin{equation}
A
=
\Big|
\int_{a}^{b}\!\drm\x\,
\sqrt{2\sM[E + \Vo \cos (\x/\xo)]}
\Big|
\end{equation}
with turning points obeying $\cos(a/\xo)=\cos(b/\xo)=-E/\Vo$.
Due to $E+\Vo\cos(\x/\xo)<0$ below the barrier, we change the sign
inside the square root extracting an $\iu$ which is absorbed by the
modulus.
Simple tranformations bring the integral into a tabulated form
\cite[equation~(2.576)]{gradshteyn-ryzhik}, and we obtain
\begin{equation}
\fl
A
=
4
\So
\;
\big[
2\,{\bf E}(s)
-
(1+\varepsilon)\,
{\bf K}(s)
\big]
\;,
\qquad
\varepsilon
=
E/\Vo
\;,
\quad
s=\sqrt{\half(1-\varepsilon)}
\end{equation}
where $\So=\sqrt{\sM\Vo\xo^{2}}$ while ${\bf K}(s)$ and ${\bf E}(s)$
the first and second complete elliptic integrals
\cite[Ch.~5.8]{arfken}.
Insertion in $\Tran=\exp[-(2/\hbar)A]$ gives the transmision
coefficient through a cosine potential.
This $\Tran$ also gives the {\em semiclassical energy bands}
\cite[\S~55]{lanlif9}
\begin{equation}
E_{\q}
=
E_{0}
\pm
(\hbar \wo/\pi)
\sqrt{\Tran}
\cos(2 \pi \xo \q)
\;.
\end{equation}
with $\wo$ the oscillation frequency in the wells.
The width of the bands is therefore
\begin{equation}
\fl
W_{\rm b}
= 
2(\hbar\wo/\pi)
\exp
\big\{
-4\,
\sqrtK
\;
\big[
2\,{\bf E}(s)
-
(1+\varepsilon)\,
{\bf K}(s)
\big]
\big\}
\;,
\qquad
\varepsilon
=
E/\Vo
\;.
\end{equation}
The parameter $s$ goes from $1$ when $\varepsilon=-1$ (potential
bottom) to $0$ at the barrier top $\varepsilon=1$.
For $s\sim1$ we can expand the elliptic integrals, getting
\begin{eqnarray}
\label{Wb:approx:bottom}
\fl
W_{\rm b}
\simeq
2(\hbar\wo/\pi)
\exp
\Big[
-\sqrtK
\;
\Big(
8
&
-
(1+\varepsilon)
\Big\{
1
+
\ln
\big[
\case{1}{32}
(1+\varepsilon)
\big]
\Big\}
\nonumber\\
&
+
(1+\varepsilon)^{2}
\Big\{
1
+
\case{1}{8}
\ln
\big[
\case{1}{32}
(1+\varepsilon)
\big]
\Big\}
\Big)
\Big]
\;.
\end{eqnarray}
At the potential bottom $\varepsilon=-1$ this gives
$W_{\rm b}=\exp[-8\,\sqrtK]$.
The band width increases exponentially with the energy.
Near the top, $\varepsilon\sim1$, one has $s\sim 0$, and we can use
the series expansion of the elliptic integrals, obtaining
\begin{equation}
\label{Wb:approx:top}
W_{\rm b}
\simeq
2(\hbar\wo/\pi)
\exp
\big[
-\pi\sqrtK
\;
(1-\varepsilon)
\big]
\;.
\end{equation}
Figure~\ref{fig:saw-width} shows that equations
(\ref{Wb:approx:bottom}) and (\ref{Wb:approx:top}) approximate well
the exact results in their respective ranges (the later also serves
as un upper bound).
If we disregard the terms with $(1+\varepsilon)^2$ in
(\ref{Wb:approx:bottom}), the result happens to fit remarkably
$W_{\rm b}$ in all the range.


\section{
Perturbative solution for periodic forcing
}
\label{app:pert:ac}

In this last appendix we shall solve the following generic
differential equation
\begin{equation}
\label{EOM:generic}
\tau
\frac{\drm \mc}{\drm t}
+
\rel\,\mc
=
\dF(t)
\,
\Vpert\,\mc
\;,
\qquad
\dF(t)
=
\dFp
\e^{+\iu\omega t}
+
\dFm
\e^{-\iu\omega t}
\end{equation}
perturbatively in the forcing $\dF$.
(For cosine forcing $\dFp=\dFm=\dF/2$.)
Comparing with section~\ref{sec:dynamics}, we see that $\mc$
corresponds to the vectors $\mc_{\ix}$, $\rel$ to the static part of
the matrices $\mQ_{\ix\ixp}$, and the perturbation $\Vpert$ to
$\Delta\mQ_{\ix\ix}$, the part of $\mQ_{\ix\ix}$ involving $\Delta F$.

A periodic solution can be Fourier expanded as
\begin{equation}
\label{ansatz:X}
\mc
=
\ec^{(0)}
+
\sum_{\iw=1}^{\infty}
\big(
\dFp^{\iw}\,
\ec^{(\iw)}
\e^{+\iu\omega_{\iw} t}
+
\dFm^{\iw}\,
\tilde{\ec}^{(\iw)}
\e^{-\iu\omega_{\iw} t}
\big)
\;,
\quad
\omega_{\iw}=\iw\,\omega
\end{equation}
with $\ec^{(0)}$ the unperturbed response, while
$\tilde{\ec}^{(\iw)}\neq\ec^{(\iw)\ast}$ for complex $\mc$.
Separating terms oscillating with $\e^{+\iu\omega_{\iw} t}$ and
$\e^{-\iu\omega_{\iw} t}$, the left-hand side of (\ref{EOM:generic})
reads
\begin{eqnarray*}
\tau
\frac{\drm \mc}{\drm t}
+
\rel\,\mc
=
\rel\,
\ec^{(0)}
&+&
\sum_{\iw=1}^{\infty}
\dFp^{\iw}\,
\big[
\quad
\iu(\omega_{\iw}\tau)\ec^{(\iw)}
+
\rel\,\ec^{(\iw)}
\big]
\e^{+\iu\omega_{\iw} t}
\\
&+&
\sum_{\iw=1}^{\infty}
\dFm^{\iw}\,
\big[
-\iu(\omega_{\iw}\tau)\tilde{\ec}^{(\iw)}
+
\rel\,\tilde{\ec}^{(\iw)}
\big]
\e^{-\iu\omega_{\iw} t}
\;.
\end{eqnarray*}
To obtain the right-hand side, we use
$\omega_{\iw}\pm\omega
=\iw\omega\pm\omega
=\omega_{\iw\pm1}$,
redefine the indices (keeping the same names) to get all the
oscillating factors at $\pm\omega_{\iw} t$, and introduce the
definition $\tilde{\ec}^{(0)}\equiv\ec^{(0)}$, arriving at
\begin{eqnarray*}
\fl
\dF(t)
\,
\Vpert\,\mc
=
\dFp\dFm
\big(
\Vpert\,\ec^{(1)}
+
\Vpert\,\tilde{\ec}^{(1)}
\big)
&+&
\sum_{\iw=1}^{\infty}
\dFp^{\iw}
\big[
\Vpert\,\ec^{(\iw-1)}
+
\left(\dFp\dFm\right)
\Vpert\,\ec^{(\iw+1)}
\big]
\e^{+\iu\omega_{\iw} t}
\\
&+&
\sum_{\iw=1}^{\infty}
\dFm^{\iw}
\big[
\Vpert\,\tilde{\ec}^{(\iw-1)}
+
\left(\dFp\dFm\right)
\Vpert\,\tilde{\ec}^{(\iw+1)}
\big]
\e^{-\iu\omega_{\iw} t}
\;.
\end{eqnarray*}
Equating terms with the same oscillating factors at both hand sides
({\em uniqueness\/} of the Fourier expansion), we obtain
\begin{equation}
\label{coeffs:per}
\fl
\begin{array}{rclcl}
\rel\,
\ec^{(0)}
&=&
\Vpert\,
\big[\;
0
&+&
\left(\dFp\dFm\right)
(\ec^{(1)}+\ec^{(-1)})
\big]
\\[1.ex]
\iu(\omega_{\iw}\tau)\ec^{(\iw)}
+
\rel\,\ec^{(\iw)}
&=&
\Vpert\,
\big[\;
\ec^{(\iw-1)}
&+&
\left(\dFp\dFm\right)
\ec^{(\iw+1)}
\big]
\end{array}
\end{equation}
The $\tilde{\ec}^{(k)}$ are included as
$\ec^{(-k)}\equiv\tilde{\ec}^{(k)}$ with
$\omega_{-\iw}=-\omega_{\iw}$.

The contribution $\Vpert\ec^{(\iw-1)}$ comes from products of
``rotating" terms $\e^{\pm\iu\omega t}\times\e^{\pm\iu\omega_{\iw}
t}$.
For instance, the oscillating factor $\e^{+\iu\omega_{\iw-1} t}$ in
$\ec^{(\iw-1)}$ when multiplied by the part $\e^{+\iu\omega t}$ of the
field raises the harmonic from $\iw-1$ to $\iw$.
The term $\Vpert\ec^{(\iw+1)}$ comes from products of
``counter-rotating" terms $\e^{\pm\iu\omega
t}\times\e^{\mp\iu\omega_{\iw} t}$.
For example, the product of the oscillating factor
$\e^{+\iu\omega_{\iw+1} t}$ and the $\e^{-\iu\omega t}$ part of the
field lowers the order from $\iw+1$ to $\iw$.
This contribution (the ``contamination" from higher harmonics) has an
order higher in $\dF$ than the contribution of $\ec^{(\iw-1)}$, which
is therefore the leading one.
Indeed, expanding the $\ec^{(\iw)}$ in powers of $\dF$ we get the
leading terms at each harmonic
\begin{equation}
\label{coefficients:a}
\rel\,
\ec^{(0)}
=
0
\;,
\qquad
\iu(\omega_{\iw}\tau)\ec^{(\iw)}
+
\rel\,\ec^{(\iw)}
=
\Vpert\,\ec^{(\iw-1)}
\;.
\end{equation}
The $\iw=0$ equation gives the static response [corresponding to
(\ref{3RR:statics})], $\iw=1$ the linear dynamical response,
etc. [corresponding to equation~(\ref{3RR:harmonics})].


\section*{References}



\end{document}